\documentclass[sigconf]{acmart}

\usepackage{acmart-taps}

\AtBeginDocument{%
  \providecommand\BibTeX{{%
    \normalfont B\kern-0.5em{\scshape i\kern-0.25em b}\kern-0.8em\TeX}}}

\copyrightyear{2023} 
\acmYear{2023} 
\setcopyright{acmlicensed}\acmConference[ASIA CCS '23]{ACM ASIA Conference on Computer and Communications Security}{July 10--14, 2023}{Melbourne, VIC, Australia}
\acmBooktitle{ACM ASIA Conference on Computer and Communications Security (ASIA CCS '23), July 10--14, 2023, Melbourne, VIC, Australia}
\acmPrice{15.00}
\acmDOI{10.1145/3579856.3582827}
\acmISBN{979-8-4007-0098-9/23/07}

\usepackage{tikz}
\usepackage{amsmath}
\usepackage{amsfonts}
\usepackage{filecontents}
\usepackage{graphicx}
\usepackage[indention=0pt]{subcaption}
\usepackage{caption}
\usepackage{wrapfig}
\usepackage{tabularx}
\usepackage{enumitem}
\usepackage{dirtytalk}

\newcolumntype{?}{!{\vrule width 1pt}}
\definecolor{islamicgreen}{rgb}{0.0, 0.56, 0.0}

\usepackage{xcolor}
\usepackage{ifthen}

\newboolean{showcomments}
\setboolean{showcomments}{false}
\newcommand{\yf}[1]{\ifthenelse{\boolean{showcomments}}
{ \textcolor{red}{(#1)}}{}}
\newcommand{\ad}[1]{\ifthenelse{\boolean{showcomments}}
{ \textcolor{green}{(#1)}}{}}
\newcommand{\dk}[1]{\ifthenelse{\boolean{showcomments}}
{ \textcolor{cyan}{(#1)}}{}}
\newcommand{\ry}[1]{\ifthenelse{\boolean{showcomments}}
{ \textcolor{blue}{(#1)}}{}}
\newcommand{\SY}[1]{\ifthenelse{\boolean{showcomments}}
{ \textcolor{magenta}{(#1)}}{}}

\newcommand{\name}{EMShepherd }

\usepackage{tabularx}
\makeatletter
\def\hlinewd#1{%
\noalign{\ifnum0=`}\fi\hrule \@height #1 %
\futurelet\reserved@a\@xhline}
\makeatother
\usepackage{multirow}

\begin{document}
\title{EMShepherd: Detecting Adversarial Samples via Side-channel Leakage} 

\author{Ruyi Ding}
\affiliation{%
  \institution{Northeastern University}
  \city{}
  \state{}
  \country{}
}
\email{ding.ruy@northeastern.edu}

\author{Cheng Gongye}
\affiliation{%
  \institution{Northeastern University}
  \city{}
  \state{}
  \country{}
}
\email{gongye.c@northeastern.edu}

\author{Siyue Wang}
\affiliation{%
  \institution{Northeastern University}
  \city{}
  \state{}
  \country{}
}
\email{wang.siy@northeastern.edu}

\author{Aidong Adam Ding}
\affiliation{%
  \institution{Northeastern University}
  \city{}
  \state{}
  \country{}
}
\email{a.ding@northeastern.edu}

\author{Yunsi Fei}
\affiliation{%
  \institution{Northeastern University}
  \city{}
  \state{}
  \country{}
}
\email{y.fei@northeastern.edu}




\begin{abstract}
Deep Neural Networks (DNN) are vulnerable to adversarial perturbations --- small changes crafted deliberately on the input to mislead the model for wrong predictions. 
Adversarial attacks have disastrous consequences for deep learning empowered critical applications.  
Existing defense and detection techniques both require extensive knowledge of the model, testing inputs and even execution details. 
They are not viable for general deep learning implementations where the model internal is unknown, a common `black-box' scenario for model users. 
Inspired by the fact that electromagnetic (EM) emanations of a model inference are dependent on both operations and data and may contain footprints of different input classes, we propose a framework, EMShepherd, to capture EM traces of model execution, perform processing on traces and exploit them for adversarial detection. 
Only benign samples and their EM traces are used to train the adversarial detector: a set of EM classifiers and class-specific unsupervised anomaly detectors.  
When the victim model system is under attack by an adversarial example, the model execution will be different from executions for the known classes, and the EM trace will be different.  
We demonstrate that our air-gapped EMShepherd can effectively detect different adversarial attacks on a commonly used FPGA deep learning accelerator for both Fashion MNIST and CIFAR-10 datasets. 
It achieves a $100\%$ detection rate on most types of adversarial samples, which is comparable to the state-of-the-art `white-box' software-based detectors. 

\end{abstract}



\maketitle
\section{Introduction} \label{sec: introduction}
\ry{In this paragraph, I want to introduce the background, emphasize the harm of adversarial on machine learning models, especially on the medical applications.}
Recent advances in deep learning have revolutionized many application domains, including computer vision \cite{forsyth2011computer} and natural language processing \cite{manning1999foundations}.
Society has benefited significantly from technological developments in AI-empowered authentication and access control~\cite{parkhi2015deep}, medical diagnosis~\cite{foster2014machine, richens2020improving}, autonomous driving~\cite{bojarski2016end}, etc.
However, deep-learning models in critical applications face serious security threats, including the most common adversarial attacks~\cite{goodfellow2014explaining}.
An adversary can manipulate the input to DNN models for inference with carefully selected perturbation to result in misclassification or misdetection.
The disastrous consequences of adversarial examples include access right escalation (e.g., to critical industrial control systems or nuclear plants),  fraudulent medical claims, and driving accidents.  

\ry{talk about existing adversarial detection and defense methods}
A large body of work has been developed for both defense and detection against adversarial attacks~\cite{li2017adversarial, bhagoji2017dimensionality, srivastava2014dropout, tao2018attacks, feinman2017detecting}. 
Defense techniques harden the DNN models through adversarial training~\cite{ganin2016domain} or stochastic methods~\cite{liu2018towards, wang2018defensive, wang2019protecting}. 
However, the adversarial examples are provided by certain adversarial generation methods, and the model retrained may not be resilient to other unknown adversarial attacks, potentially stronger with different feature characteristics. 
Model retraining also has privacy implications as it requires to be iterative to keep up with new attacks~\cite{wang2019beyond}.
Furthermore, these protection methods have been circumvented recently by the most sophisticated attack~\cite{carlini2017towards}.
Another line of work is to detect the adversarial examples during model execution, which can be external to the model, therefore, more agile, general, and robust.
Existing detection methods rely on observing intermediate execution features or model behavior~\cite{xu2017feature,ma2019nic} or input statistics~\cite{ma2018characterizing, grosse2017statistical}, 
and leveraging them for adversarial detection, a `white-box' scenario with the model internal and run-time execution details known.\yf{what is the difference between execution features or behavior? I merged}

\ry{This paragraphy is for motivation: what is a `black-box' machine learning model, key points: privacy, parameters invisible and can't be often updated.}
However, there are plenty of cases where the model users have limited access to the model intermediate parameters or testing images, which we also called it a `black-box' system.
For example, machine learning models in the healthcare system may be kept confidential due to their values and privacy concerns, 
where the model suppliers tend to offer model users only limited interfaces so as to prevent reverse engineering or membership inference attacks~\cite{shokri2017membership}.
\ry{Further clarify the `black-box' setting, pointing out the urgent requirements of a new detection method. Why current methods are not suitable on `black-box' setting}
Current detection strategies do not suit such ``black-box" systems, and they require direct access to the model structure and parameters, execution details such as activations, testing images, or model intermediate outputs, including model logits. 
With privacy concerns, we target building an adversarial detector without access to both testing inputs and model execution details (i.e., feature maps). 

\ry{Introduce the EM analysis briefly}
Inspired by the simple Electromagnetic analysis (SEMA), which associates the EM emanation patterns of a computer platform with the operations it is running and data processing~\cite{emsca}, we leverage side-channel EM leakage for detecting adversarial examples. The intuition is that different classes of images will activate the network model differently and yield different patterns in the EM traces. 
For adversarial examples that impose perturbations on a source class and fool the victim model into misclassifying it as a target class, 
the semantic information leaked from the EM trace may present a discrepancy from that of the target class (learned with prior training), i.e., revealing an anomaly.

\begin{figure}[t]
    \centering
    \includegraphics[width=0.9\linewidth]{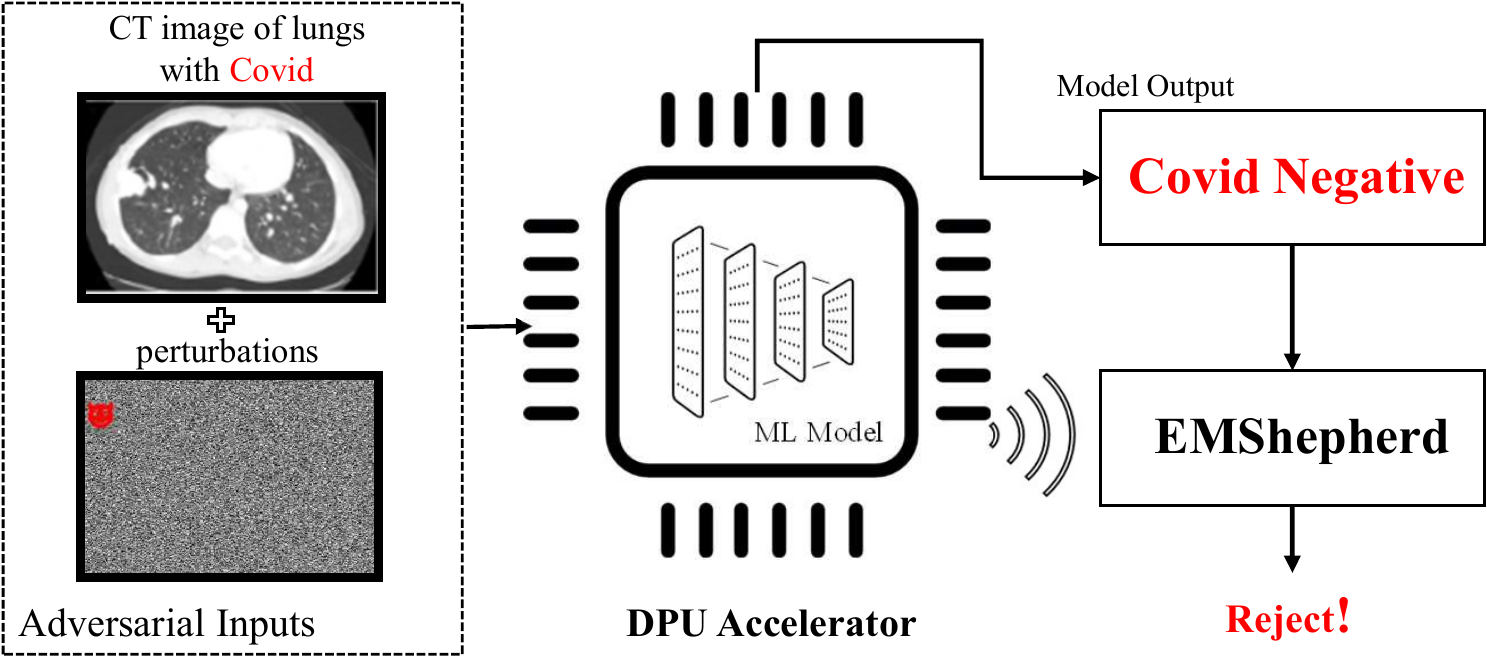}
    \caption{EMShepherd detection framework: the medical machine learning model misdiagnoses an attacked CT lung image with Covid, captured by \name}
    \Description{An example of detecting adversarial examples for AI chips with EM emanations.}
    \label{fig: task explanation}
    \vspace{-5mm}
\end{figure}

\ry{Detection Flow of EM side channel attack, given an example of melanoma detection.}
In this paper, we propose EMShepherd, a framework of adversarial sample detection via EM side-channel leakage, 
which treats the victim model as a `black-box' without probing it for execution details.
The victim model is deployed on a physical device that the user can access, e.g., an IoT edge device or a local inspection station.  
Note remote access to DNN models in the cloud, a so-called Machine-learning-as-a-Service (MLaaS) scenario,  is out of scope for our work. 
The DNN implementation on the device can be software running on a CPU or GPU or a hardware accelerator running on FPGA, NPU, or TPU.
Fig.~\ref{fig: task explanation} shows an example setup of EMShepherd, 
where the deep learning inference system is attacked by samples with malicious perturbations, and the additional air-gapped \name fends off the adversarial sample at run-time with correct detection. 
For running the adversarial detector, neither the victim model (both static model internals and dynamic execution details) nor the inputs are needed. 
Note our work uses EM side-channel as an example, while the framework is generally applicable to power side-channel as well, collected either by equipment like an oscilloscope or through on-chip power sensors~\cite{intelrapl,Lipp2021Platypus} where the measuring resolution has to be commensurate with the detection goal.

This work makes several contributions as follows:
\begin{itemize}[leftmargin=*]
    \item We leverage side-channel EM leakage, for the first time, for detecting adversarial attacks under a ``black-box" scenario. 
    We propose \name adversarial detector, which requires no prior knowledge of the victim models, adversarial attacks, intermediate execution details, and the model inputs.
   \item Our novel \name framework consists of scripts for EM measurements, a novel data processing method to tame the EM traces for follow-on class-specific feature extraction and learning, and an unsupervised anomaly detector. 
    \item We evaluate the framework on $5$ different adversarial attacks on the Fashion MNIST dataset running on a common FPGA deep learning accelerator. 
   Our results show that the EM-based detector can effectively detect all attacks 
    with over $90\%$ detection accuracy and an acceptable false positive rate (less than $10\%$).
    \name is also applied on a large VGG neural network accelerator with the CIFAR-10 dataset. 
    \item We further evaluate its performance on a robust retrained model with adversarial examples, and \name demonstrates high accuracy and low false positive rate as well.
    \item We compare our method with the state-of-the-art white-box software-based detection methods.
    The results show that the performance of our adversarial detector is comparable to the prior methods. 
\end{itemize}

\section{Background} \label{sec: background}
This section presents relevant background on adversarial attacks, protection, and EM side-channel.

\subsection{Adversarial Attacks on DNNs}\label{sec:adversarial}
DNN is an artificial neural network with multiple layers to represent a function, $F:X\rightarrow Y$, 
with parameters $\omega_F$ such as weights, kernels, and biases, where $X$ denotes the input space and $Y$ the output space.
In the training phase, a DNN is trained with a dataset of input-output pairs to arrive at optimal values of $\omega_F$, 
to minimize the loss function $J_F$, which is a distance measurement between the model predicted result $F(x)$ and the ground truth $y*$.
The widely-used optimizers include stochastic gradient descent (SGD)~\cite{sgd} and Adam~\cite{adam}.
Taking image classification as an example, the DNN model runs inference on the unknown input, $x_t$, and predicts a class, out of $m$ classes, with the largest probability.
\begin{equation}
    y = F(x_t) = \text{softmax}(Z(x_t))
    \label{eq: m-class classifier}
\end{equation}
where the vector $Z(x_t)$ is known as logits.
Our detection method assumes that the defender can only query the model and know the output $y$ while the logits $Z(x)$ are unavailable.

DNN model is vulnerable to adversarial attacks.  
An adversarial sample ($x'$) is a carefully crafted sample, which has a human-imperceivable difference from the original benign sample ($x$), 
but causes the DNN to misclassify it to a different class $F(x') \neq y$.
If $F(x')$ is an arbitrary class except for $y$, $x'$ is an untargeted adversarial sample.
A more restrictive and harmful case is the targeted adversarial sample, where $F(x')=l \neq y$, a specific target class. 
In this paper, we consider both untargeted and targeted attacks.
The difference between the adversarial example and benign example can be measured by $L_p$, defined as $\Delta_p = \sum_{i=0}^n(|x_i-x'_i|^p)^{\frac{1}{p}}$.
Common choices of $L_p$ include: $L_0$, the number of pixels changed; $L_1$, the Manhattan norm; $L_2$, the Euclidean distance norm; $L_{\inf}$ the largest absolute change of any pixels.
The adversarial attack can be viewed as an optimization problem:
\begin{equation}
    \textbf{minimize} \quad \Delta_p(x, x') \quad\textbf{s.t.} \quad F(x') \neq y
    \label{eq: adversarial attack}
\end{equation}

GoodFellow et al.~\cite{goodfellow2014explaining} first proposed the concept of the adversarial sample and introduced Fast Gradient Sign Method (FGSM) to generate adversarial samples.
Madry et al. introduced the Projected Gradient Descent (PGD)~\cite{madry2017towards} attack to improve the attack efficiency of the Basic Iterative Method (BIM)~\cite{kurakin2016adversarial}. 
Other learning-based attacks include DeepFool~\cite{moosavi2016deepfool} and Carlini and Wagner Attack (CW)~\cite{carlini2017towards}, 
where CW attack is proven to be one of the strongest attacks~\cite{carlini2017towards, carlini2017adversarial} at the cost of generation speed.

\subsection{Existing Software Detection Methods}
The existing adversarial detection methods, all software-based, can be classified into three categories.

\textit{Distributional Detection:}
These detectors perform some statistical analysis on the inputs or intermediate values of model execution (e.g., activation values) to find adversarial samples. 
Grosse \cite{grosse2017statistical} used Maximum Mean Discrepancy test (MMD) to determine whether the benign and adversarial inputs have the same underlying distribution or not.
Feinman \cite{feinman2017detecting} use Kernel Density Estimation to measure the distance of distributions. 
However, these detection methods are ineffective on more complex datasets \cite{carlini2017adversarial}.
Ma \cite{ma2018characterizing} introduced Local Intrinsic Dimensionality (LID) to characterize adversarial regions of the model. However, LID is proven to perform poorly on a number of attacks~\cite{ma2019nic}.

\textit{Latent Space Detection:}
The second type of detector employs a pre-processing step to reduce variation. 
Grosse \cite{grosse2017statistical} found that adversarial examples tend to place a higher weight on larger principal components, narrowing down the targets for detection. 
Some approaches train denoisers to reconstruct the inputs by removing the adversarial noise added by the attacker, such as
auto-encoders used in MagNet \cite{meng2017magnet}  and the mean blur method used in \cite{li2017adversarial}.
Most of them work for simple attack methods such as FGSM, but cannot resist the state-of-the-art CW attack.

\textit{Inconsistency Detection:}
This approach focuses on the model misbehavior during inference of adversarial examples.
Feinman et al.~\cite{feinman2017detecting} proposed Bayesian neural network uncertainty to measure the uncertainty of a DNN under a given input.
By introducing some randomness (e.g., Dropout \cite{srivastava2014dropout}) during the inference, the DNN model tends to give the same outputs for benign inputs but different outputs for adversarial ones. 
The Feature Squeezing approach \cite{xu2017feature}
reduces the color depth 
and observe that
adversarial samples are likely to induce different classification results while benign inputs are not.
Tao et al.~\cite{tao2018attacks} introduced the Attacks meet Interpretability structure (AmI),
which measures the inconsistency of the victim DNN with another neural network enhanced with human perceptible attributes under adversarial examples.
In the Network Invariant Checking (NIC) work proposed by Ma \cite{ma2019nic}, 
the key idea is during model execution, there are class-dependent
provenance channels (the distribution of activated neurons in the network) and activation value channels (value distributions of activated neurons). 
It employs a one-class SVM to determine outliers.
NIC shows promising results against a broader range of attacks, including the CW attack.
Our approach generally falls into the type of inconsistency detection, 
in a black-box victim system scenario. 

\subsection{Class Activation Map in Adversarial Detection} \label{sec: CAM}
\begin{figure}[t]
    \centering
    \small
    \begin{tabularx}{\linewidth}{XXXX}
    \begin{tabular}[x]{@{}X@{}} \textbf{Benign:}\\\textbf{``Trouser''}\end{tabular} &
    \includegraphics[width=\linewidth]{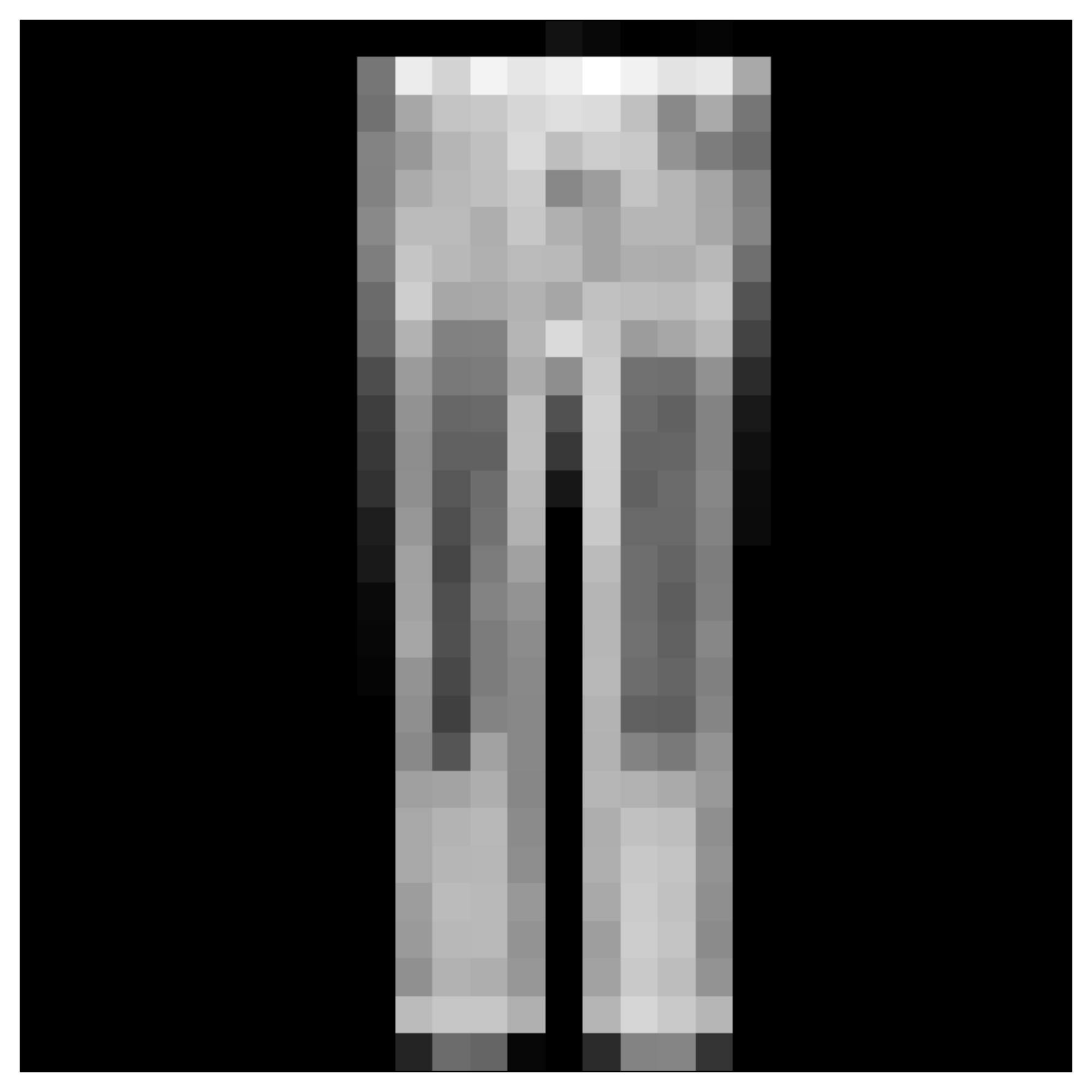} &
    \includegraphics[width=\linewidth]{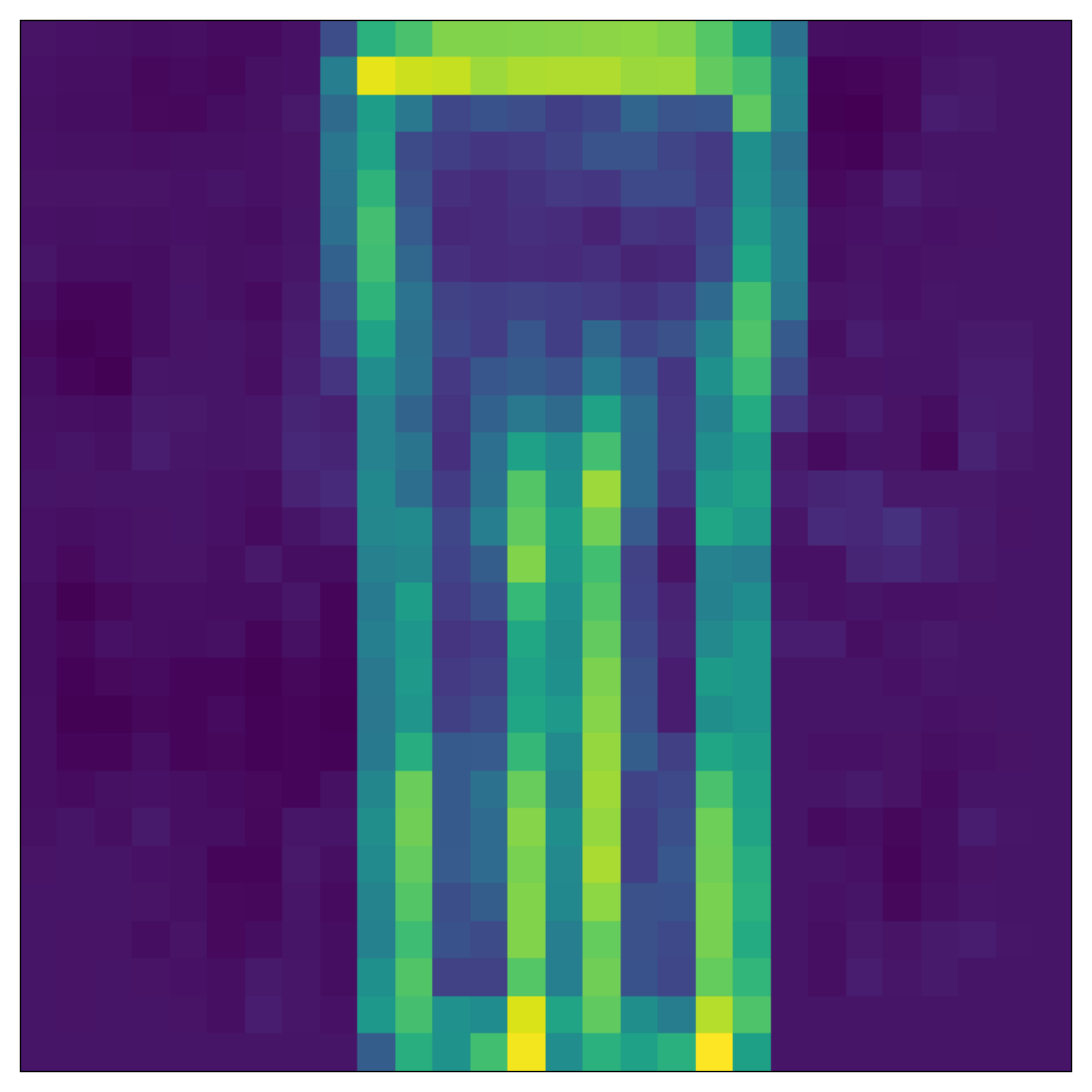}&
    \includegraphics[width=\linewidth]{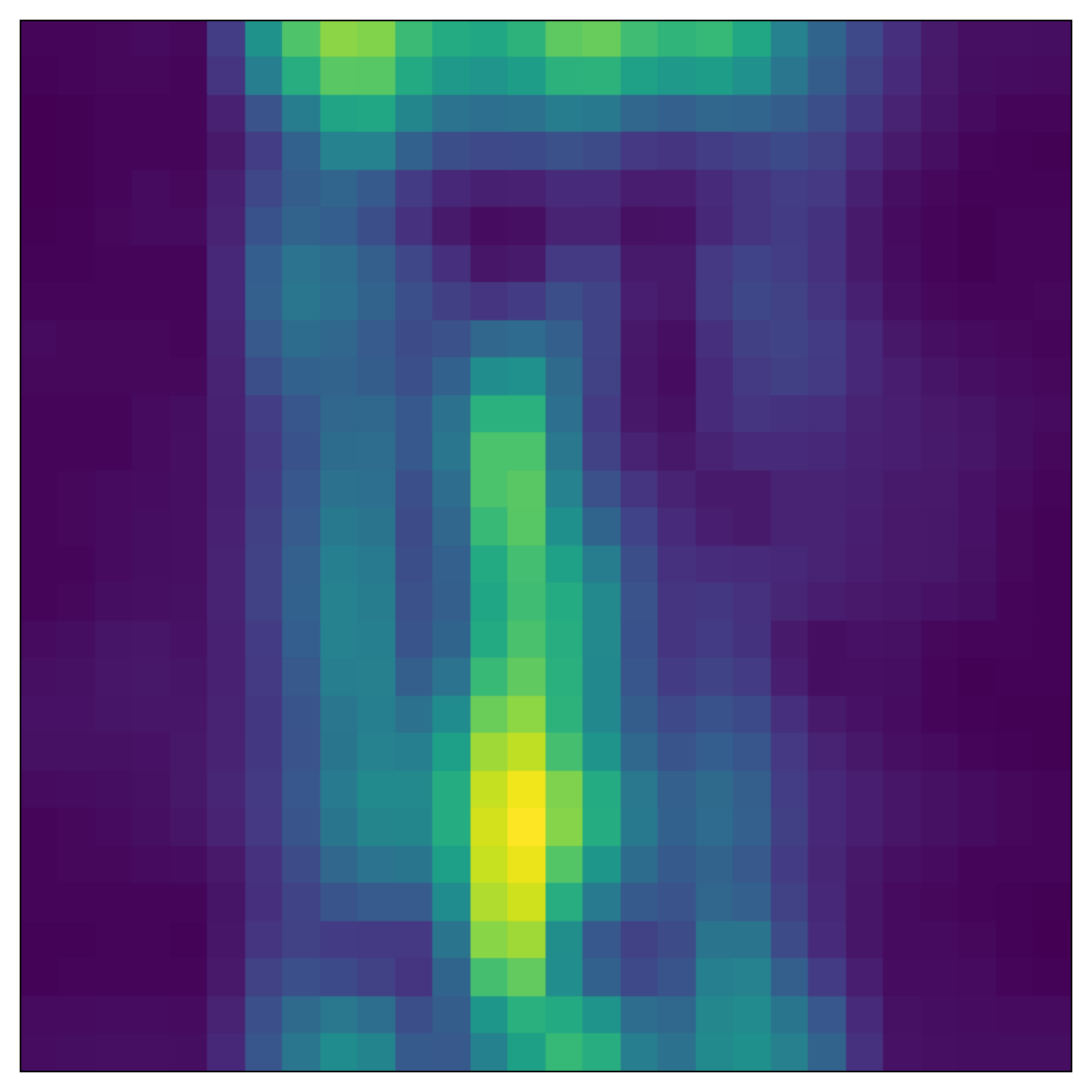}\\
    \begin{tabular}[x]{@{}X@{}} \textbf{Benign:}\\\textbf{``Sandal''}\end{tabular}&
    \includegraphics[width=\linewidth]{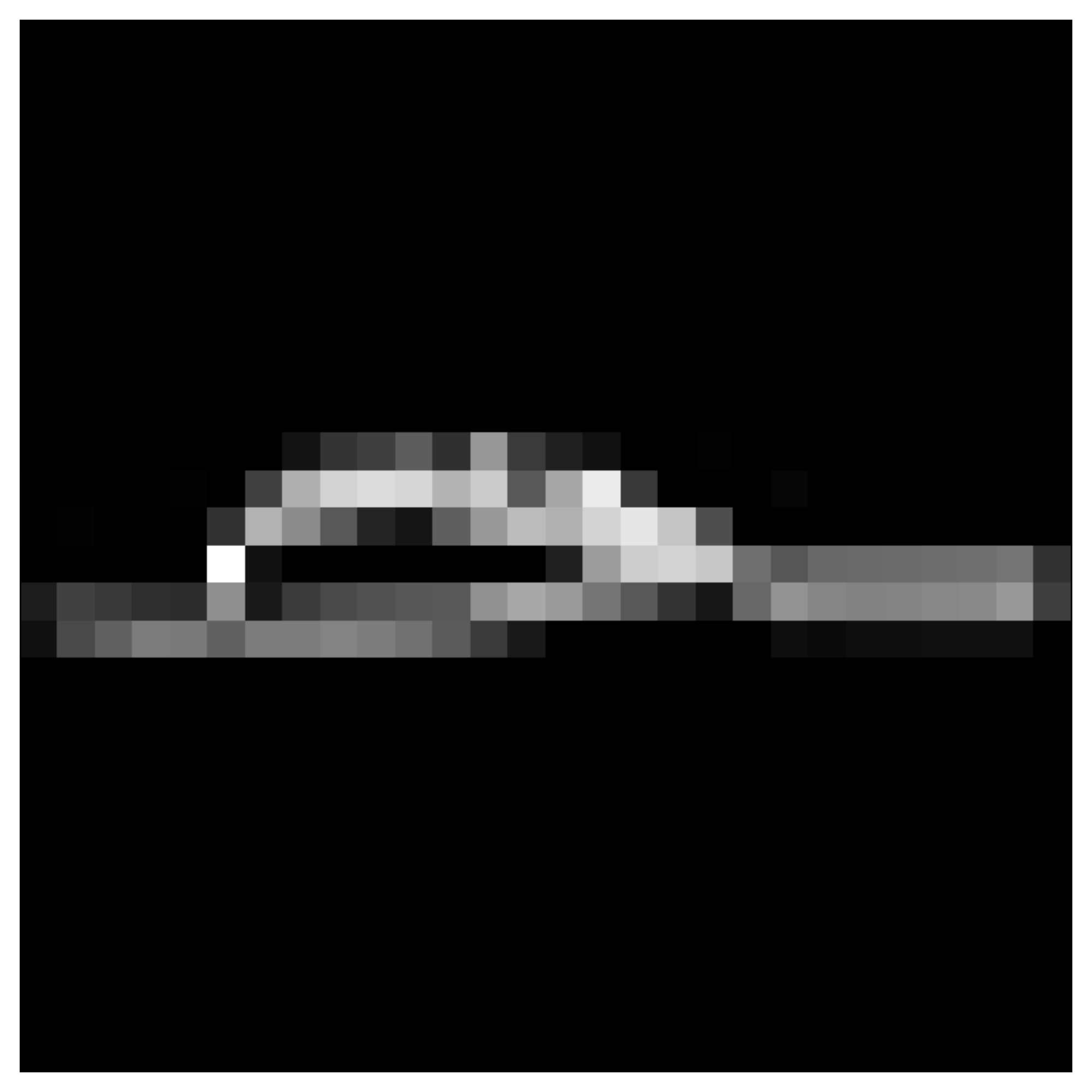}&
    \includegraphics[width=\linewidth]{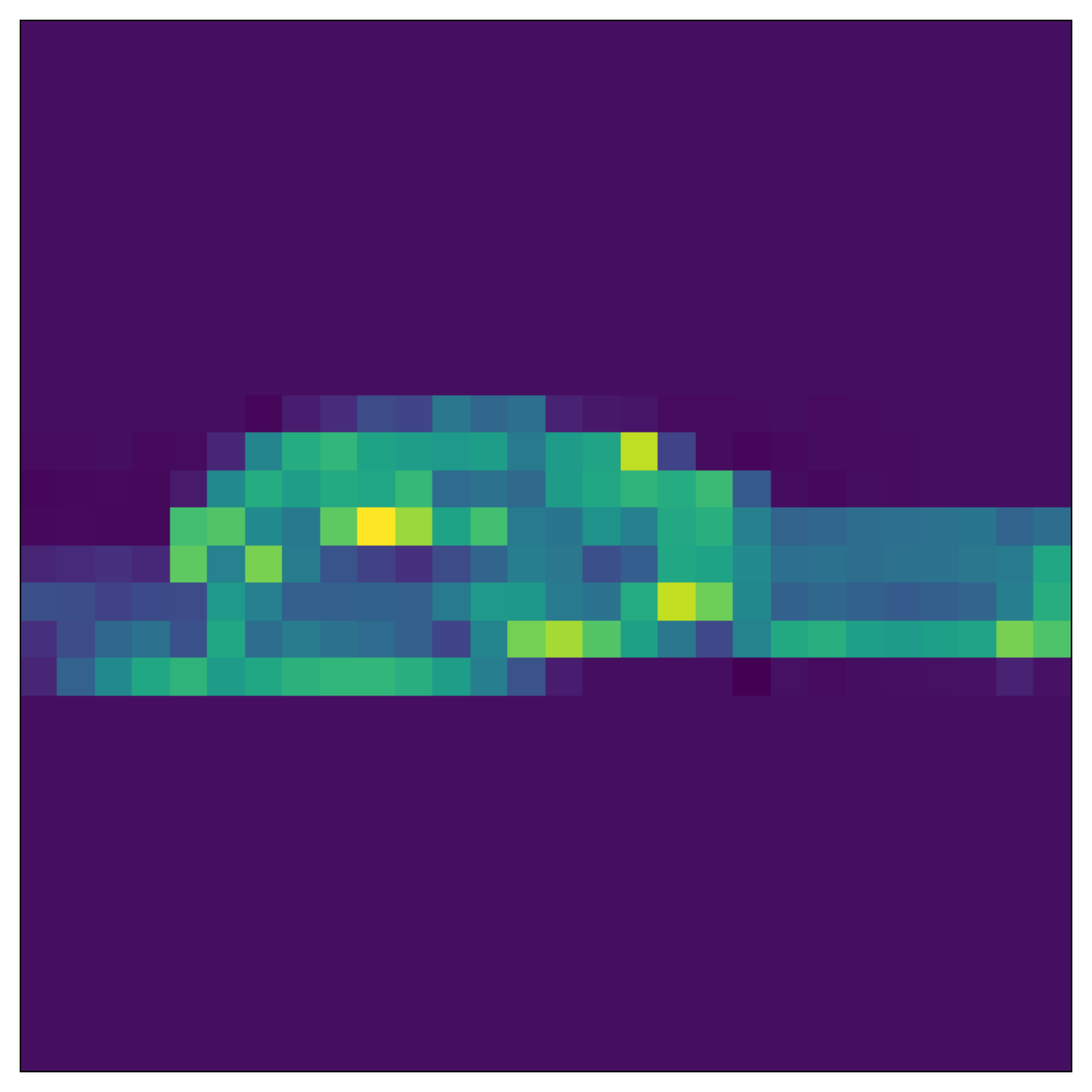}&
    \includegraphics[width=\linewidth]{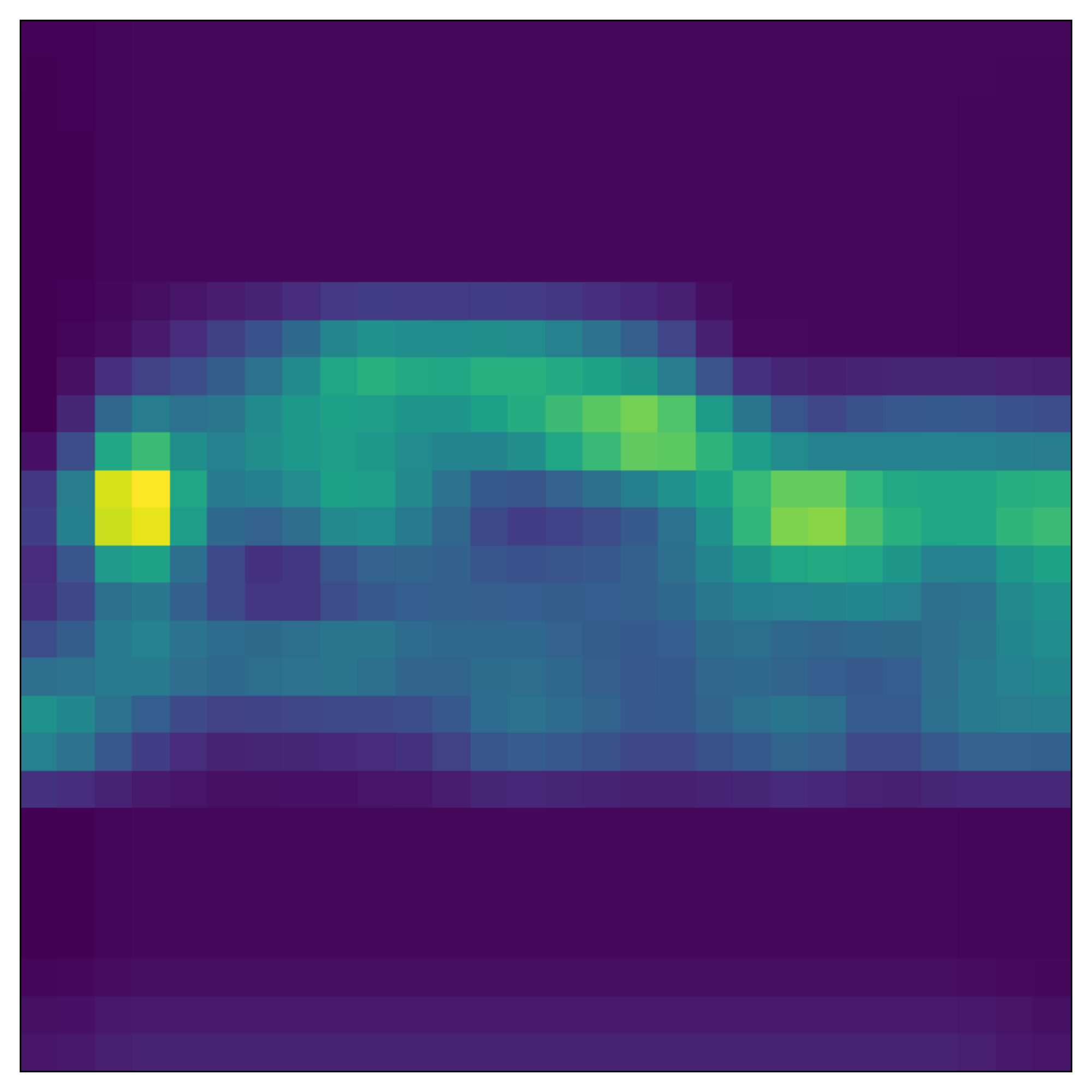}\\
    \begin{tabular}[x]{@{}X@{}} \textbf{Adversarial:}\\\textbf{``Sandal''}->\textbf{``Trouser''}\end{tabular} &
    \includegraphics[width=\linewidth]{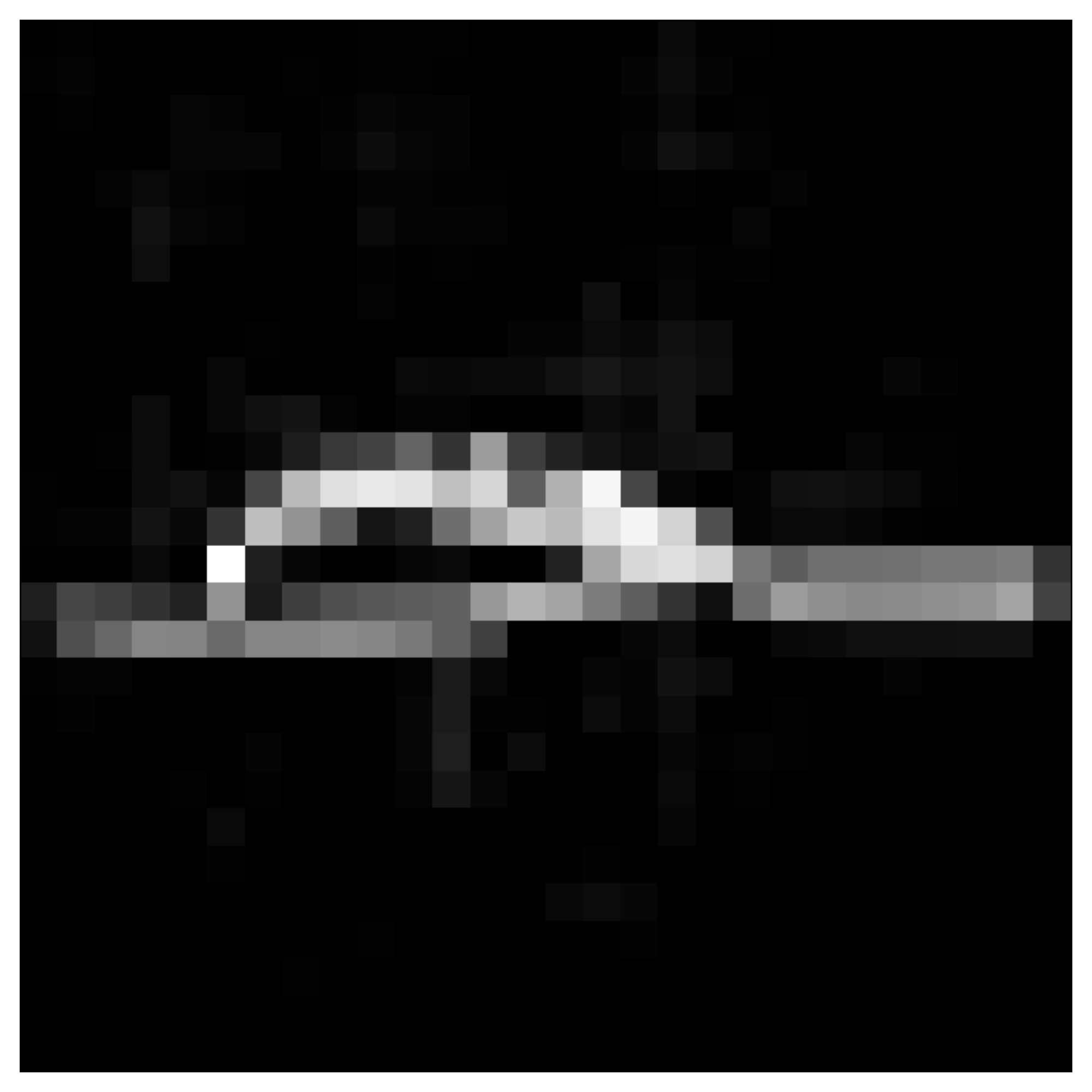}&
    \includegraphics[width=\linewidth]{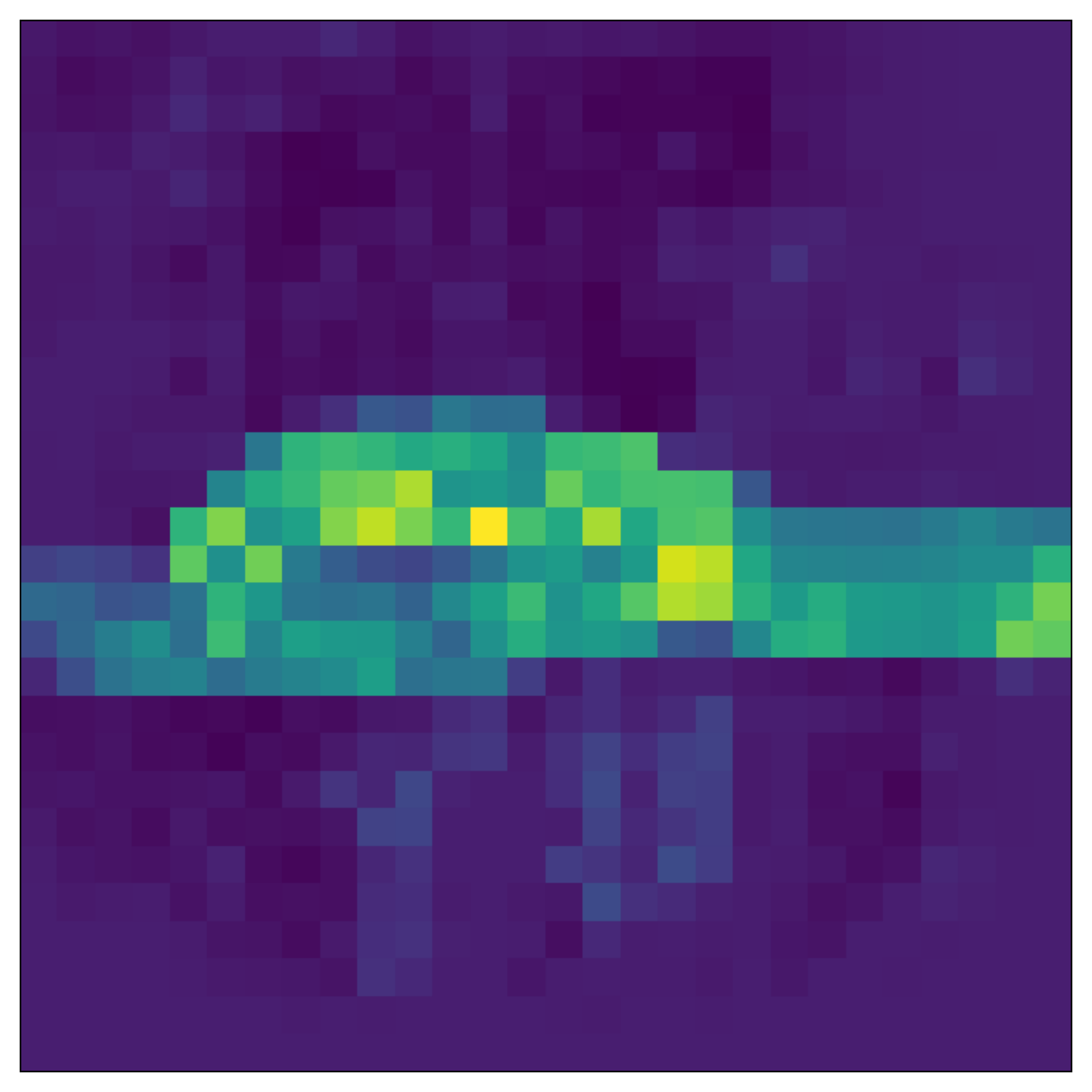}&
    \includegraphics[width=\linewidth]{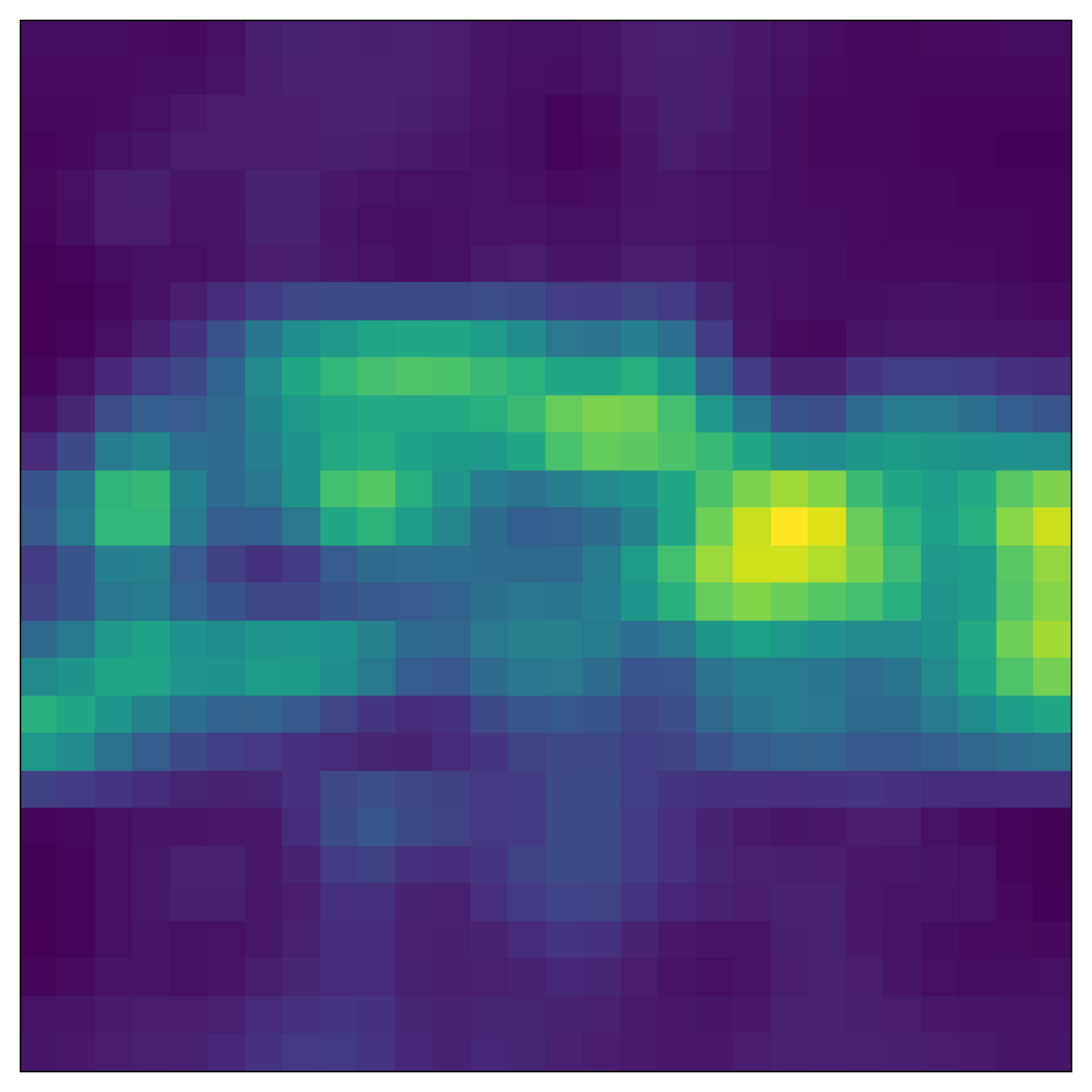}\\
    \end{tabularx}
    \Description{GradCAM of Adversarial sample}
    \caption{GradCAM illustration of adversarial attacks. The malicious noise in adversarial sample (the third row) increases with the model depth, which finally causes it's misclassification to the target.}
    \label{fig: adv-gradcam}
    \vspace{-5mm}
\end{figure}

\ry{In this paragraph, I want to propose adversarial example detection using CAM, which leverages detection using semantic and discriminative information.}
Class Activation Map (CAM)~\cite{selvaraju2016grad, jiang2021layercam} is commonly used to explain the behavior of deep neural networks, 
\ry{semantic information is the CAM from benign samples, deterministic information is the real CAM used by adversarial sample leading to wrong results.}
showing how the network progressively (with more layers) identifies the important region of the input (features) that leads to the class prediction. \yf{the previous line needs rephrasing. Verify if my edits are correct}
For benign samples, CAMs can represent the images' semantic information~\cite{zhou2021removing, vinogradova2020towards}. 
However, the adversarial perturbations can impact the focus of neural networks, which leads to wrong predictions. \yf{complete the previous line.} 
For example, in Fig.~\ref{fig: adv-gradcam}, we present an example originally comes from Class ``Sandal'' and is classified as ``Trouser'' via CW attack, 
together with samples from source and target class, followed by their class activation maps of the first two convolutional layers, respectively. 
We conclude that:
\begin{itemize}[leftmargin=*]
    \item For benign samples, the class activation maps (semantic information) visually show features that lead to the classification result. \yf{where are discriminative features? never mentioned. how did you arrive at the first conclusion.}
    \item The CAMs of adversarial samples do not resemble those of benign examples that represent the target class. 
    And because of adversarial noise, their CAMs diverge from the source class to some anomalies gradually by the model depth.
    \item CAMs of different layers vary, and the impact of adversarial perturbations will be amplified by the network depth. 
\end{itemize}
Therefore, one way for adversarial detection is to train an out-of-distribution detector to figure out the mismatch between CAMs from benign samples and ones from adversarial samples.


\subsection{Electromagnetic and Power Side-Channel} \label{sec: EM and power side channel attack}
\ry{Background about EM traces}
Both EM emanations and power consumption of a computer system depend on the circuit operations and data~\cite{agrawal2002side}.
Such side-channels have been extensively analyzed to retrieve the secret key of cryptographic algorithms ~\cite{kocher1999differential, chari2002template, das2019x}, and recently have been utilized to infer deep neural network model information. 
Yu \cite{yu2020deepem} proposed a SEMA to retrieve the topology of the victim model.
Batina \cite{batina2019csi} applied differential EM analysis to recover simple MLP model parameters of microcontroller implementations.
Zhang \cite{zhang2021stealing} successfully extracted the structure of a network running on an FPGA via power side-channel.
Chmielewski \cite{chmielewski2021reverse} targeted GPU DNN implementations and recovered the model structure with EM side-channel information.
All the prior work focuses on reverse engineering partial \textit{model} information, while our work associates the EM emanation patterns with input sample classes.

\ry{Different methods to analyze EM power traces.}
There are multiple strategies to analyze EM/power side channel information, such as traditional statistical way and modern learning-based methods.
Statistical analysis requires alignment of the EM/power measurements with the computation processes and relies on certain power models, such as Hamming Weight and Hamming Distance model, or mutual information between distributions to discern the secret. 
When leveraging EM side-channel leakage of DNN model execution for classification and adversarial detection, the model structure is complex, the execution is computation-intensive, and the hardware platform supports highly parallel operations, and therefore learning-based methods are more suitable for coping with the misalignment, noise, and feature extraction, etc.

\subsection{Target Platform} \label{sec: DPU}
\ry{Are we going to put the target platform part here or in the experiment part.}
We choose Xilinx® Deep Learning Processing Unit (DPU) as our platform.  DPU is a popular configurable hardware neural network accelerator on FPGA
and achieves the best throughput for DNN inference~\cite{9069951}.
DPU supports common CNNs such as VGG\cite{simonyan2015deep}, ResNet\cite{he2015deep}, GoogLeNet\cite{szegedy2014going}, YOLO\cite{redmon2016look}, and MobileNet\cite{howard2017mobilenets}.
Xilinx provides Vitis AI~\cite{BibEntry2022Jan}, a development stack to compile neural networks software trained with generic DNN platforms such as ~\cite{BibEntry2021Dec},  onto a DPU accelerator.
\section{Threat Model and Advantages of Our Hardware-based Adversarial Detector} \label{sec: motivation}
\ry{In this part, I want to highlight the comparison between hardware and software attacks}
\subsection{Threat Model} \label{sec: threat model}
\ry{This section is threat model: attack is `white-box', detector is `black-box'}
The victim is a DNN classifier, which is pre-trained with a public dataset. The testing dataset may be kept private.
We assume the strongest `white-box' attack model, where the attacker has full knowledge of the victim model and training dataset in order to generate adversarial samples with minimum perturbations. 
On the contrary, the detection system assumes the most limited scenario, under a `black-box' view of the victim, without access to the victim's inputs, parameters, and intermediate outputs or execution details. 
The only information available to the detector to distinguish adversarial samples is the EM side-channel measurement and the victim model's prediction class.
For training the adversarial detector with EM traces, a public benign dataset is used. 

\if false 
\ry{In this part, we discuss more settings of the detector especially the data used in two phases.}
In general, the detecting process can be summed up into two phases, training phase and detecting phase.
To begin with, we train an Out-of-Distribution(OOD) detector on a public benign dataset of the same classification task, which should be distinct from the victim's training dataset.
For each query, the detector will obtain the classification result and an EM trace along with the model execution to fit its EM classifiers and anomaly detectors.  
During the detection phase, the victim model is in operation and under attack when the pre-trained detector decides whether the current input is adversarial or not, only based on the victim model output and its EM trace.
\fi 

\subsection{Advantages}
Compared to software-based adversarial detection methods, our hardware-based detector, EMShepherd, has three distinct advantages: privacy-preserving, portability, and robustness.

\begin{itemize}[leftmargin=*]
    \item \ry{Add a new motivation here. The motivation is that using \name can help the user protect their privacy.} 
    \name protects the DNN model user's data privacy as it is agnostic to the model's inputs, which instead are always required by prior reconstruction-based detection methods~\cite{meng2017magnet, yang2022you}. 
    The sensitive inputs should not be shared with \textit{third-party detectors}. 
    Our design only requires the output class labels and the EM signals, which are passively leaked to common acquisition equipment. 
    \item \ry{The second motivation is still related to privacy. This time we consider model privacy when the model structure or parameters should be kept private.}
   \name also protects the model confidentiality.  No model information, including 
   hyper-parameters, parameters, and logits, is needed, in stark contrast to the previous software-based detection methods~\cite{ma2019nic,feinman2017detecting}.
    The EM data processing and the adversarial detector training process are both victim model-agnostic. 
    Therefore, our method has more general usage, applicable to closed-source DNN applications, which are pervasive in edge devices where the user only queries the models for the final prediction output. 
    \item \ry{The third motivation is portability.}  
    Owing to the model-agnostic feature, EMShepherd can be easily ported for wide-range hardware devices with different DNN implementations for diverse applications. It can be used as a `plug and play' (PnP) device, aside from the target system, to work automatically without user intervention or contact with the victim system. 
    \item \ry{The last motivation is about adaptive attacks, we should propose that EM signal is hard to imitate, so it is hard for adaptive attacks to generate sample fraud both detector and victim.} 
    Adaptive attack~\cite{adaptive} is a threat to most software defense methods where the attacker adjusts the adversarial perturbations to mislead both the victim models and defense systems.
   However, due to the high complexity and non-explicit dependency of the EM signals on computations and data, 
   it is extremely hard to have an adaptive attack on our detection method, 
   i.e., adversarial examples whose EM signals are deliberately controlled to evade the EM-based detector. 
\end{itemize}


\section{\name Design} \label{sec: methods}

We next present the design rationales for the \name framework, the composition of the adversarial detector and salient functions.

\subsection{EM Emanations of DNN}\label{sec: EM-inspiration}
\begin{figure}[t]
    \centering
    \subcaptionbox{CNN Feature Space}[.53\linewidth][c]{%
    \includegraphics[width=\linewidth]{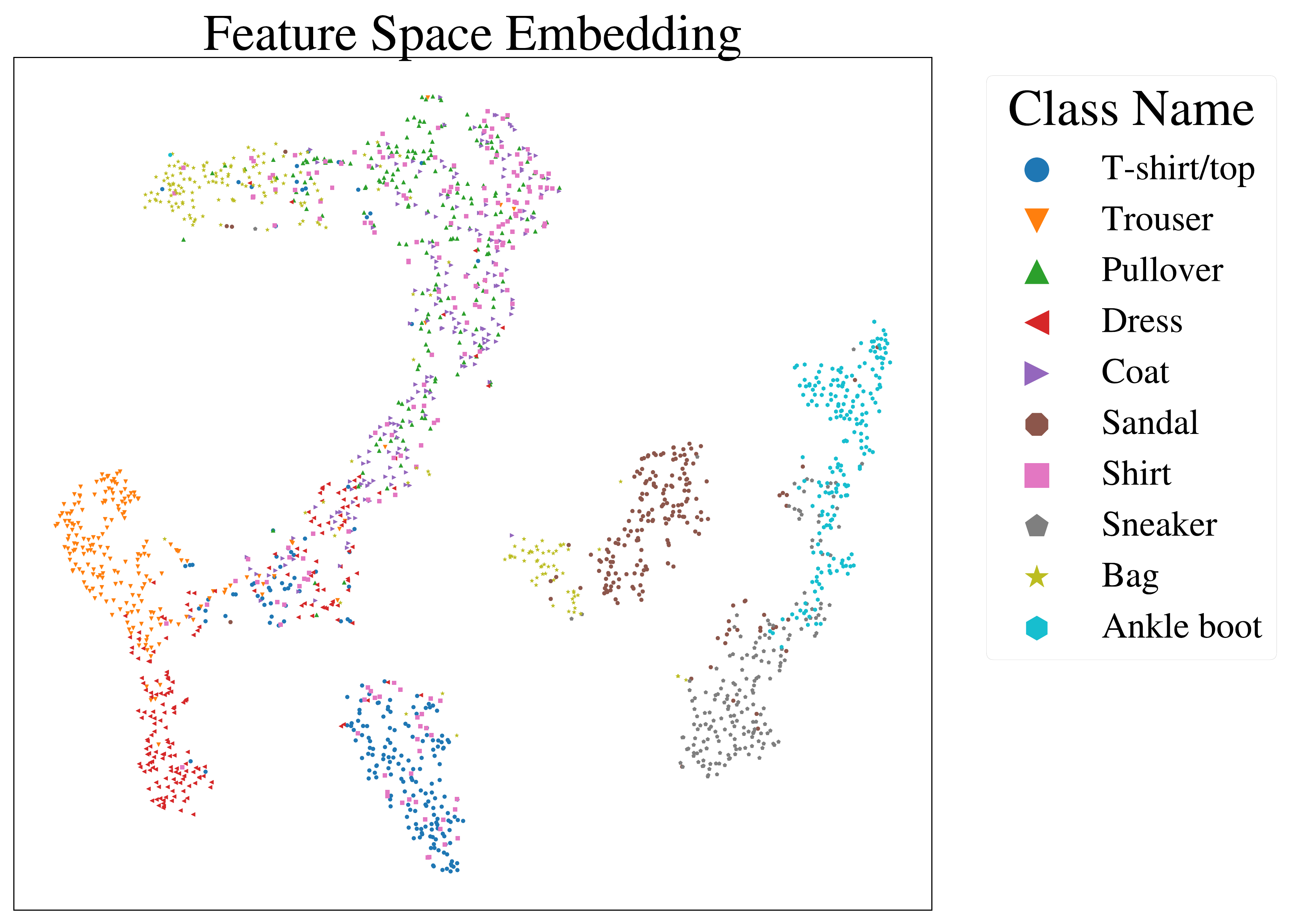}}\quad
    \subcaptionbox{Confusion Matrix}[.4\linewidth][c]{%
    \includegraphics[width=\linewidth]{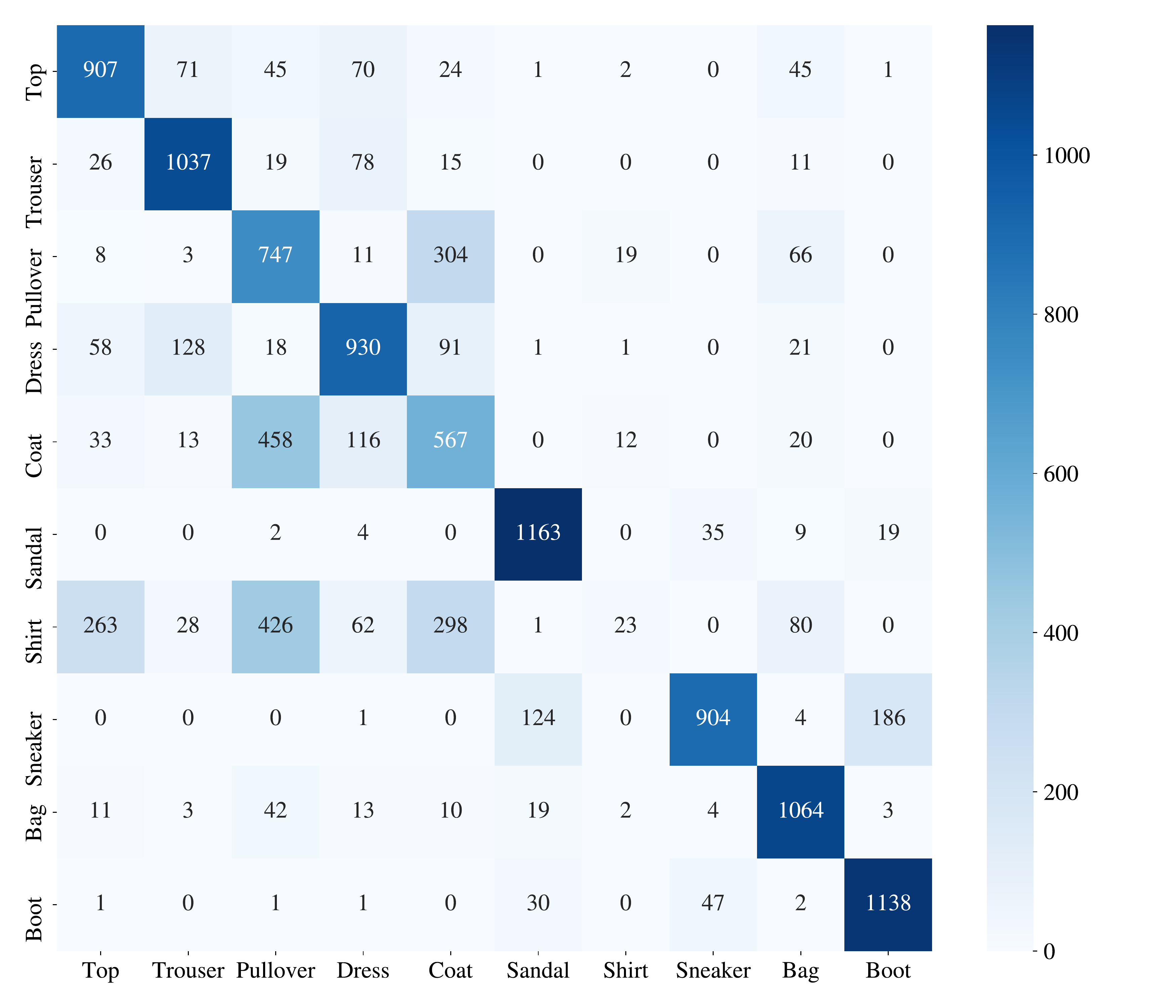}}\quad
    \Description{Confusion matrix on EM classifiers}
    \caption{Feature Space Representation and Confusion Matrix of EM signals}
    \label{fig: em-features}
\end{figure}

\ry{In this section, I want to mention the findings of EM signals about class, and there is semantic information in it.}
As mentioned in Section~\ref{sec: CAM}, semantic information can be used in adversarial detection, but how can we get it under a `black-box' setting?
We leverage EM side-channel leakage to characterize the semantic computational information for benign inputs.
Model inference is a highly computation-intensive task, involving multiple stages of parallel computation, making the EM signals complex and their dependency on computations hard to model. 
Learning-based methods can tackle these noisy EM signals well to extract class-specific features.
We build convolutional neural network (CNN) classifiers based on an EM dataset collected from a target system running on Fashion MNIST (with 10 classes). 
\yf{Here needs more context: What CNN? - classification? verify if the line I added is correct}
Figure~\ref{fig: em-features}(a) presents the feature space embedding of a CNN classification model on the testing EM dataset using a commonly-used visualization tool, T-distributed stochastic neighbor embedding (TSNE)~\cite{van2008visualizing}.\yf{what is the function of TSNE? when you use a tool you have to explain very high-level of the tool.} 
Figure~\ref{fig: em-features}(b) shows the confusion matrix of the CNN model prediction results.
We notice that,
\begin{itemize}[leftmargin=*]
    \item The CNN classifier can extract class-related features from the EM signals. 
    Some classes' features are distinct from others, while some overlap with others.
    \item The embeddings of the classes with similar semantic information are located near each other, 
    such as the cluster of (Sandal, Sneaker, Ankle Boot) (all shoes) and the cluster of (Shirt, Coat, Pullover) (all tops).
\end{itemize}
Based on these findings, we will discuss our design of \name, which leverages this semantic information in the EM signals with the model outputs to detect adversarial samples.
The design will also overcome the low prediction accuracy for some classes by further exploiting the feature space with anomaly detectors.



\subsection{Overview of the Detection System}\label{sec: design-overview}

\begin{figure}
    \centering
    \includegraphics[width=0.95\linewidth]{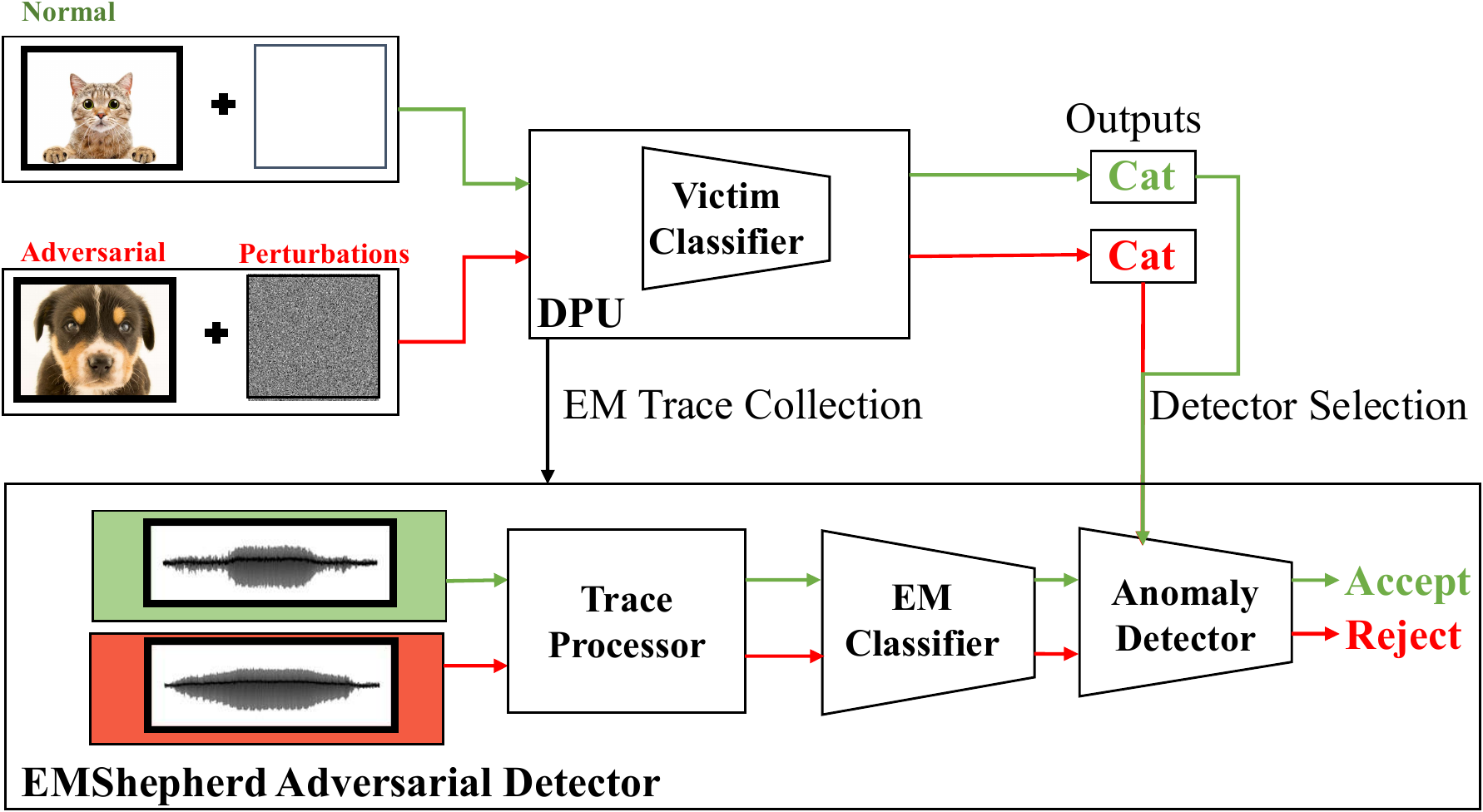}
    \Description{The Flow chart for training and detecting the framework}
    \caption{Overview of the \name detection flow}
    \label{fig: flow-chart}
    \vspace{-3mm}
\end{figure}

\ry{In this paragraph, we will explain the main framework of the detection system as Fig.4.}
Figure~\ref{fig: flow-chart} shows an overview of the \name detection framework. 
The victim model is an image classifier running on a Xilinx DPU, and an EM trace is collected for each model execution. 
During the training phase, the model is queried with a benign training dataset and corresponding a training EM trace dataset is collected. 
These traces will be used to train EM classifiers whose outputs are utilized to fit a set of class-specific anomaly detectors.
After this phase, all the trainable components are fitted and the parameters are fixed, an \name detector is generated. 
In the detection phase, the pre-trained \name takes in the EM trace collected during an image inference, processes it, and feeds it to the follow-on EM classifiers and anomaly detector, to accurately detect adversarial examples guided by the output label from the victim classifier. 


\if false
The adversarial detector is based on two DNN models: the EM classifier and the Adversarial Detector.
Shown in Figure \ref{fig: flow-chart} (a), the framework takes benign examples as a training dataset to train these DNN models.
Firstly, the EM leakage are collected and processed via the trace processor.
Then the preprocessed EM traces will train a number of EM classifiers.
Finally, the classification results will train a class-specific adversarial detector for the future detection.
When it comes to inferring an adversarial example (an adversarial image from the class Cat targeted to Dog),
the original model is fooled and predicts the image as Dog.
However, the well-trained EM classifier and the adversarial detector
will notice the attack and claim that the output should not be Dog but adversarial.

In details, our method uses the EM leakage during model execution.
It is passive side-channel information which won't impact the original model so that there will be no performance reduction for the normal execution.
Moreover, there is no direct information exchange between the original model and the detector such as weights, activation values, logits and model structures.
We can monitor the adversarial behaviours independently with normal inference.
As we only use the benign training dataset as inputs to train our technique, our method is not specific to a certain attack.
Ideally, our protection method is suitable for those packed edge devices with limited interfaces.
\fi

\subsection{Notation and Definition}
We next define notations used along with our system design. 
The victim model $\mathcal{M}_v$ is a pre-trained $N$-class classification model running on a device, facing adversarial attacks.
One input image to the victim model is denoted $Im_i \in \mathcal{I}_v$, and the corresponding output label is $y_i$.  
The corresponding EM trace for the execution collected is $\mathbf{T}_i \in \mathcal{T}_v$, each with $P$ number of points.
Every $\mathbf{T}_i$ can be partitioned into multiple computation segments $\mathbf{T}_i = concat(\{ \mathbf{B}_1, \mathbf{B}_2,..., \mathbf{B}_M\}_i)$, 
where the number of segments, $M$, depends on both the structure of $\mathcal{M}_v$ and its implementation on the victim device. Note that $M$ may be larger than the number of layers, as off-chip communications can happen within a layer due to the on-chip resource constraints.  

EM segments are 1-D time series, and are preprocessed by Short-Time-Fourier-Transform (STFT) to generate 2-D EM spectrograms (details will be given in Section~\ref{sec: preprocessor}), 
$\mathbf{B}_m \rightarrow \mathbf{S}_m(t, f)$ in both time and frequency dimensions, where $m=1, 2, \dots, M$ denotes the segment index. 
Correspondingly one EM classifier is built on one EM segment, denoted $\mathcal{C}_m$.  
Given an input image $Im_i$, the logits of EM classifier $m$ is denoted as $\{\mathbf{L}_m\}_i$.
Then logits of all $M$ EM classifiers are concatenated into one vector $\boldsymbol{l}_i= concat(\{\mathbf{L}_m\}_i)$, 
providing a holistic view of the victim model's internal processing for input image $Im_i$.

The class-specific anomaly detectors are built for all $N$ classes, denoted $\{\mathcal{D}_n\}$, where $n=1, 2,...,N$, one for each class.
$\mathcal{D}_n$ is trained with $\boldsymbol{l}_i$ of the benign training samples with $y_i=n$, and a threshold is selected for each class.
For a testing input $Im_j$ with the prediction label $n$, the loss of the anomaly detector, denoted as  $\ell(Im_i)$, is compared with the corresponding class threshold, $L_T$, to detect whether the input is adversarial. 
More details will be illustrated in Section~\ref{sec: anomaly detection}.

\begin{figure}[t]
    \centering
    \includegraphics[width=0.7\linewidth]{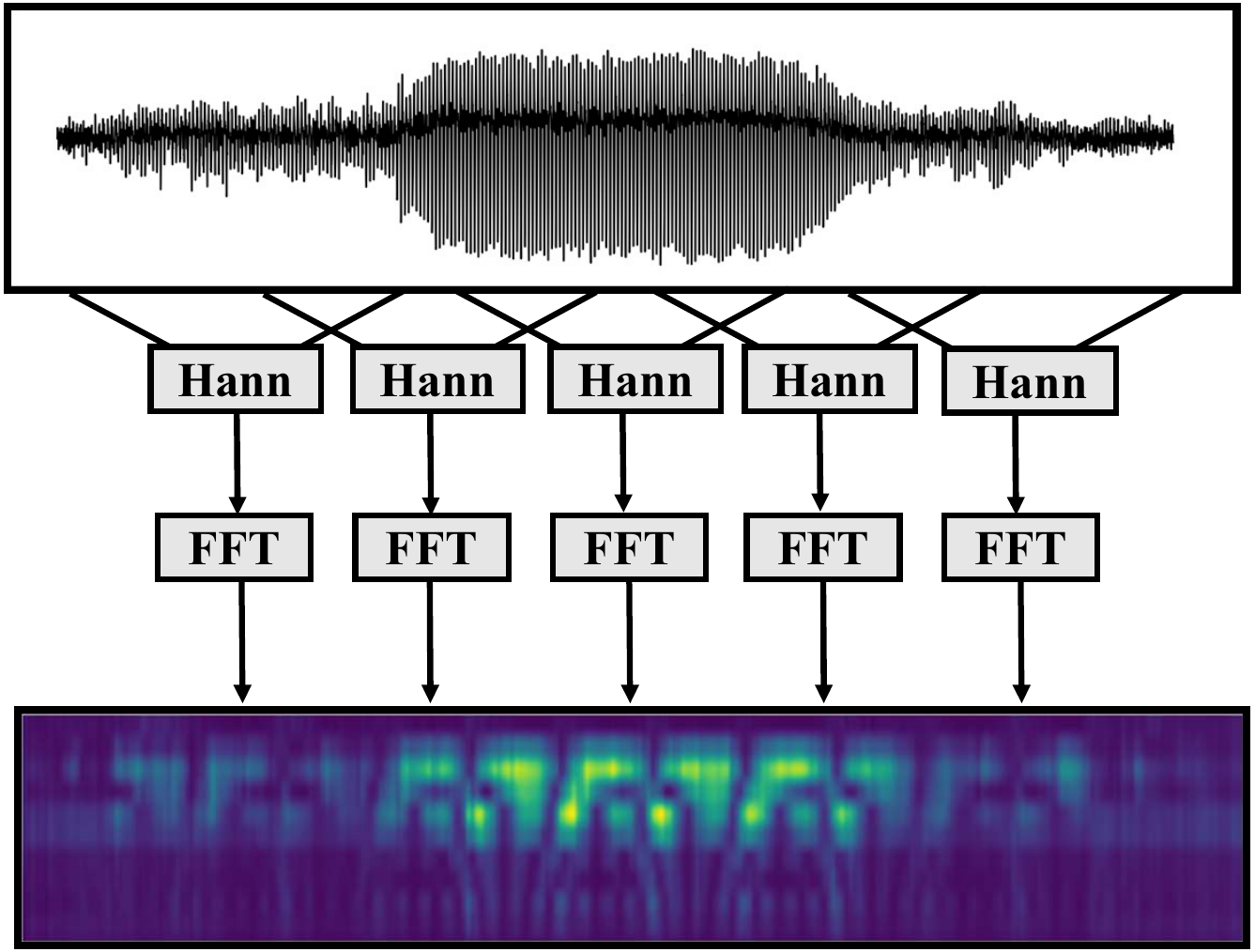}
    \Description{Illustration of STFT.}
    \caption{Short-Time Fourier Transform}
    \label{fig: stft}
    \vspace{-5mm}
\end{figure}

\subsection{Data Collection and Preprocessing} \label{sec: preprocessor}
Fig.~\ref{fig: flow-chart} shows that the raw EM traces, $\mathbf{T}_i \in \mathcal{T}_v$, will first go through a trace processor for denoising and transformation.
The trace processor performs two tasks:  extracting high-energy segments from raw traces and converting each EM segment into a spectrogram.
We analyze the trace profile and find that different input images will result in different amplitudes on the EM traces, but they all have the same number of segments due to the hardware accelerator structure.
We pick local maximum leakage points and split each trace into multiple segments.
Note that we do not need to align the segments in the time domain as we will use Short-Time Fourier Transformation to do time-frequency domain conversion.
Fig. \ref{fig: stft} depicts the processing method - Short-Time Fourier Transform.
A sliding Hanning window (e.g., $256$ points) with a stride of half of the window is used to transform the raw signals progressively.
Between two windows, the overlapping time points make sure that no information is lost by preserving the signals on the windows' boundaries.
In each window, we apply FFT to convert the signal from the time domain to the frequency domain to generate a spectrum.

\begin{equation}
    \mathbf{S}_m(t, f)=\int^{(t+1)w}_{(t-1)w}\mathbf{B}_m(\tau)e^{-j2\pi f\tau} d\tau
    \label{eq: stft}
\end{equation}
where $m=1, 2, \dots, M$ denotes the segment index, $w$ denotes half of the STFT window size, $t=1,2, \cdots, P_m/w$ denotes the index of time windows in a segment, and $f$ is the frequency selected.

\noindent There are three benefits of using STFT. 
\begin{itemize}[leftmargin=*]
    \item Noise-Filtering. As shown in Fig.~\ref{fig: stft} where the Y-axis of the spectrogram is the frequency and the brighter the color the higher the amplitude, the main energy (the brighter part) is focused on the operating frequency of the victim model, which is $150MHz$ in our case.  
    The rest frequencies have relatively lower energy.
    Thus, we can select $15$ bands around the operating frequency out of $256$ bands and filter out other lower-energy components, which increases the signal-noise ratio (SNR) of the remaining EM frequencies.
    \item Dimension-Reduction. The number of points in the spectrogram is reduced by $w$ times in the time domain compared to the raw EM segment, which makes the follow-on model learning capture the temporal patterns easier. 
    \item 2-D image. Compared to 1-D raw EM traces, the spectrogram naturally fits CNN classifiers and kernels, which not only provides time and frequency information but also the change of the spectrums along the time.  
    More details will be presented in Section~\ref{sec: exp: spectrogram} that the EM classifiers indeed exploit both time and frequency information in the spectrograms for classification.    
\end{itemize}

%

\subsection{EM Classifiers} \label{sec: EM classifiers}
\begin{figure}[t]
    \centering
    \includegraphics[width=0.95\linewidth]{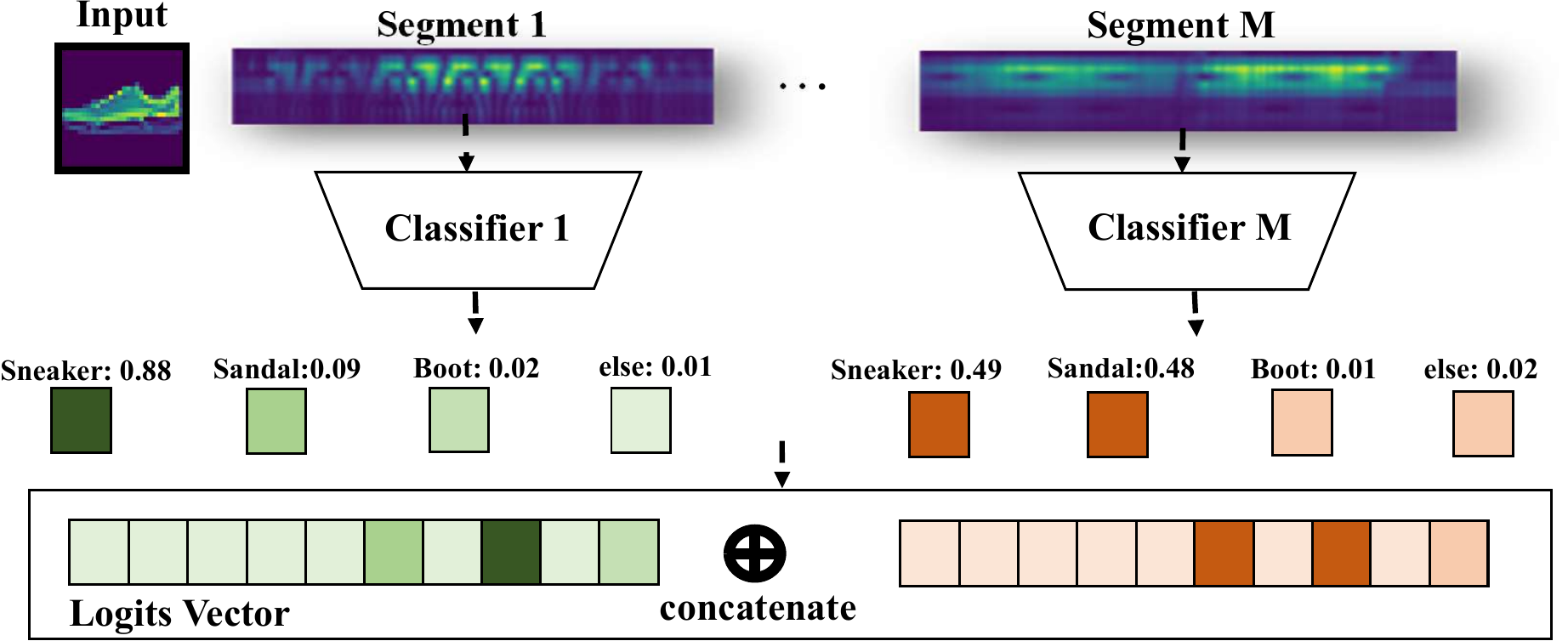}
    \caption{EM-based classifiers and logits output}
    \Description{EM classifiers}
    \label{fig: EM classifiers}\yf{the fonts for many classes are hard to see. use the same color for the text font}
    \vspace{-5mm}
\end{figure}

\ry{First paragraph: generally introduce the structure of EM classifiers}
As aforementioned, the EM trace can leak class-specific computation and activation information during the inference process for an input sample $Im_i$.
We train a DNN classifier on each segment of EM spectrogram, with all benign samples in the EM training dataset, as depicted in Fig.~\ref{fig: EM classifiers}.

\ry{In the second paragraph, we further explain some features of EM classifier, different segments shows different features, so we will concatenate the logits from all segments.}
After that, we will concatenate all the EM classifiers' outputs into one logit vector for all the benign samples that belong to one class. This is based on the observation that for one input image, although all the constituent EM classifiers give out the same class prediction, their logit vectors may differ significantly, which may carry finer-grained feature/semantic information.
We use the experimental results on the Fashion MNIST dataset as an example. 
As shown in Fig.~\ref{fig: EM classifiers}, for an input image \textit{Sneaker}, the classifier on the first spectrogram gives out the correct prediction of \textit{Sneaker}, with a $0.88$ confidence score. 
However, Classifier M on the $M^{th}$ spectrogram, although also gives out the correct prediction, has much lower confidence as $0.49$. 
One explanation is that the victim model processes the semantic information layer by layer to get a final discriminative output,
where various EM segments correspond to the computation on different layers and different features.
Further, according to our experimental results, it is not necessary for every EM classifier to correctly identify the image class to be useful for the later anomaly detector.
One segment of the model's internal operations can focus on some features that are not informative enough to identify the image class, and the corresponding EM classifier may not result in a high score/logit for the class.
However, such information (this particular segment is not informative enough to distinguish image classes) can still be helpful when adversarial samples cause a different pattern, such as resulting in a high confidence score for a class instead of among the outputs of this segment EM classifier.  
It is the \textit{deviation} of the logits from benign ones that contribute to anomaly detection.
The concatenated logits vector, $\boldsymbol{l}= concat(\mathbf{L}_1, ...,\mathbf{L}_M)$, 
provides a pattern of how segments of internal model operations correlate to the class identification, 
regardless of high or low, which resembles an entire inference process across layers of the victim DNN model.

\if false
We take the vector of logits of each classifier, and concatenate them into a longer vector.
Since each classifier  $\mathcal{C}_m$ only takes one spectrogram as input,  it tries to  predict the output class based on part of computational and activation features (e.g., of one layer).  \yf{what is the reason not to train one classifier with all batch spectrogram stitched together as the input?  For any choice you take, you need to justify it.  Also, why not do averaging across the logits instead of concatenating them together?  Concatenating logits seems weird}
Different classifiers for different segments may lead to a big difference in the logit vector even though their output classes are the same.
For instance, as Figure \ref{fig: EM classifiers} shows, for Classifier 1 the Cat class has a $0.9$ confidence score, but for Classifier M it only has $0.5$, even though both are the maximum score among all the classes.
We use the ensemble method by connecting the logits from all segments into a long vector $\boldsymbol{l}_n= concat(\mathbf{L}_1, ...,\mathbf{L}_M)$,
where $n$ denotes the original output class of the testing sample,
 and the length of $\boldsymbol{l}_n$ is $M\times N$.\yf{check notations here. you will have many $\boldsymbol{l}_n$ so it should be a set instead of a single one?}
\fi

\subsection{Anomaly Detection Models} \label{sec: anomaly detection}
\begin{figure}[t]
    \includegraphics[width=0.85\linewidth]{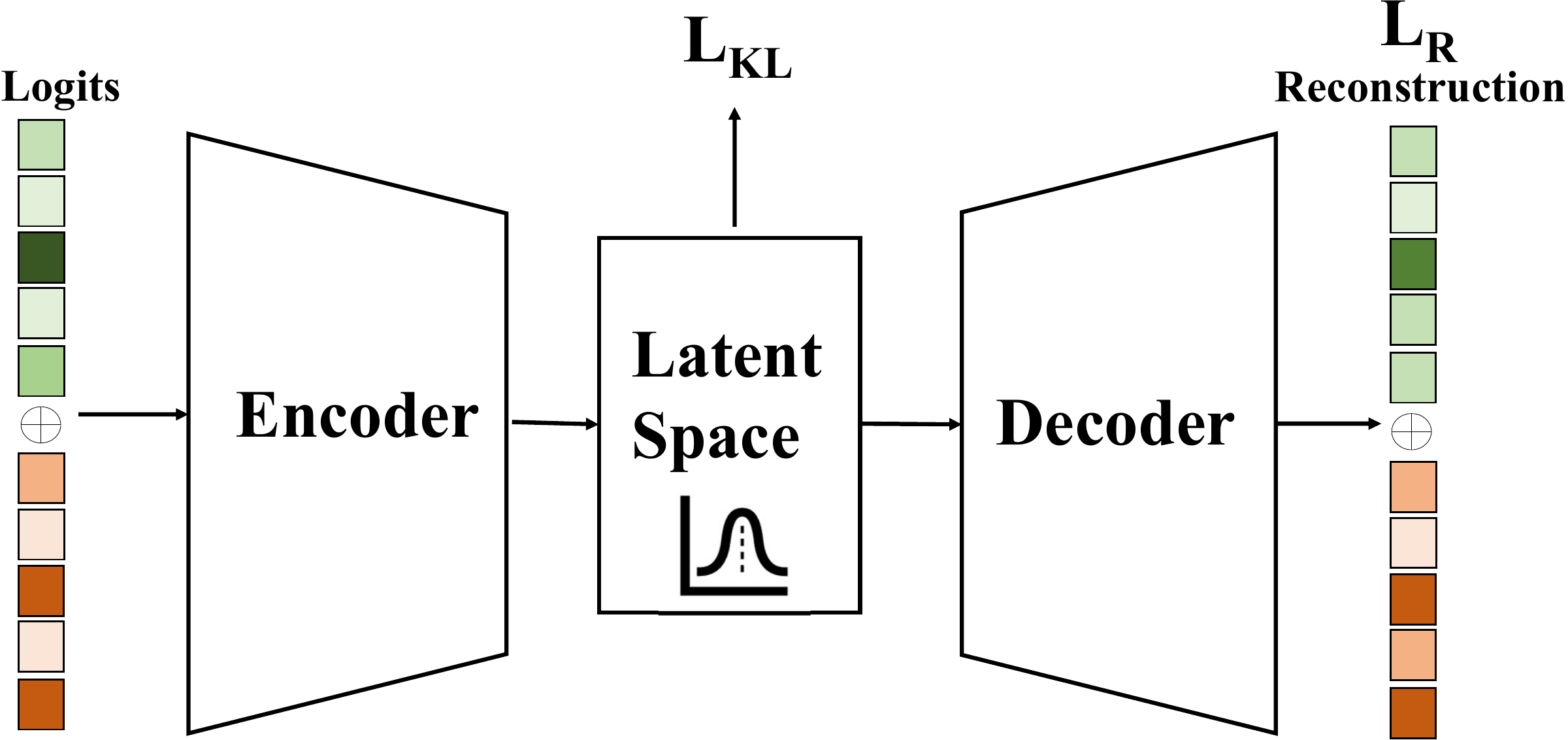}
    \Description{Illustration of Variational AutoEncoder}
    \caption{The structure of anomaly detector and VAE}
    \label{fig: vae}
    \vspace{-3mm}
\end{figure}

\ry{In this section, we talk about VAE, the main idea we want to show is that VAE can reconstruct logits value and distinguish OOD samples based on the total loss.}
The concatenated logits vector $\boldsymbol{l}$ of benign inputs from the same class are likely to be similar.
Therefore, we can build OOD detector to distinguish adversarial samples which very likely do not fall into the target class (claimed by the victim model). We select a  reconstruct-based detector using Variational AutoEncoder (VAE).
The structure of VAE is shown in Fig.~\ref{fig: vae},
with an encoder followed by a decoder, where the middle layer is the latent-space representation. 
The encoder and decoder of our VAE each contain four fully-connected layers.
The loss of the VAE includes a latent-space regularizer loss $\ell_{KL}$ in addition to the encoder-decoder's reconstruction loss $\ell_{recon}$.
The $\ell_{KL}$ measures the Kullback-Leibler divergence of the latent space when fitted to some distribution assumption, which is a Gaussian distribution in our case.
The $\ell_{recon}$ measures the difference between the output and the input of the autoencoder. 
When fitting the VAE, we utilize ADAM optimizer to minimize the total loss $\ell_{total} = \ell_{recon} + \lambda \ell_{KL}$, where $\lambda$ is a constant.

The VAE total loss $ell_{total}$ during inference can be used to detect OOD samples.
When inferring benign samples, the EM classifiers' logits will match the prediction output, which fits the pre-trained VAE model with a lower loss.  
However, when the victim model is attacked by an adversarial sample that misleads the prediction to be a different target class, the EM signals will be Out-of-Distribution.
Therefore, the EM classifiers' logits will also be Out-of-Distribution leading to a higher VAE loss.
By selecting a VAE loss threshold based on the validation of benign samples, one can detect adversarial samples with a controllable false positive rate.

\section{Experiments and Evaluations} \label{sec: experiments}
In this section, we present the experimental setup and evaluation results.

\subsection{Experiment Setup}\label{sec: setup}
\begin{figure}[t]
    \subcaptionbox{Trace Collection Setup}[.5\linewidth][c]{%
    \includegraphics[width=\linewidth]{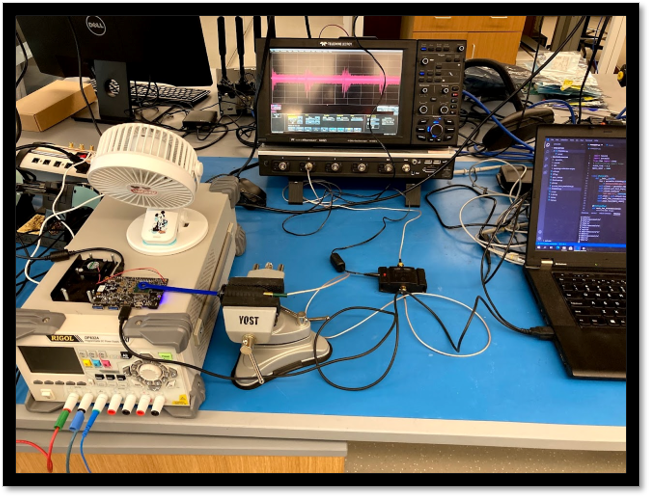}}\quad
    \subcaptionbox{Confusion Matrix}[.46\linewidth][c]{%
    \includegraphics[width=\linewidth]{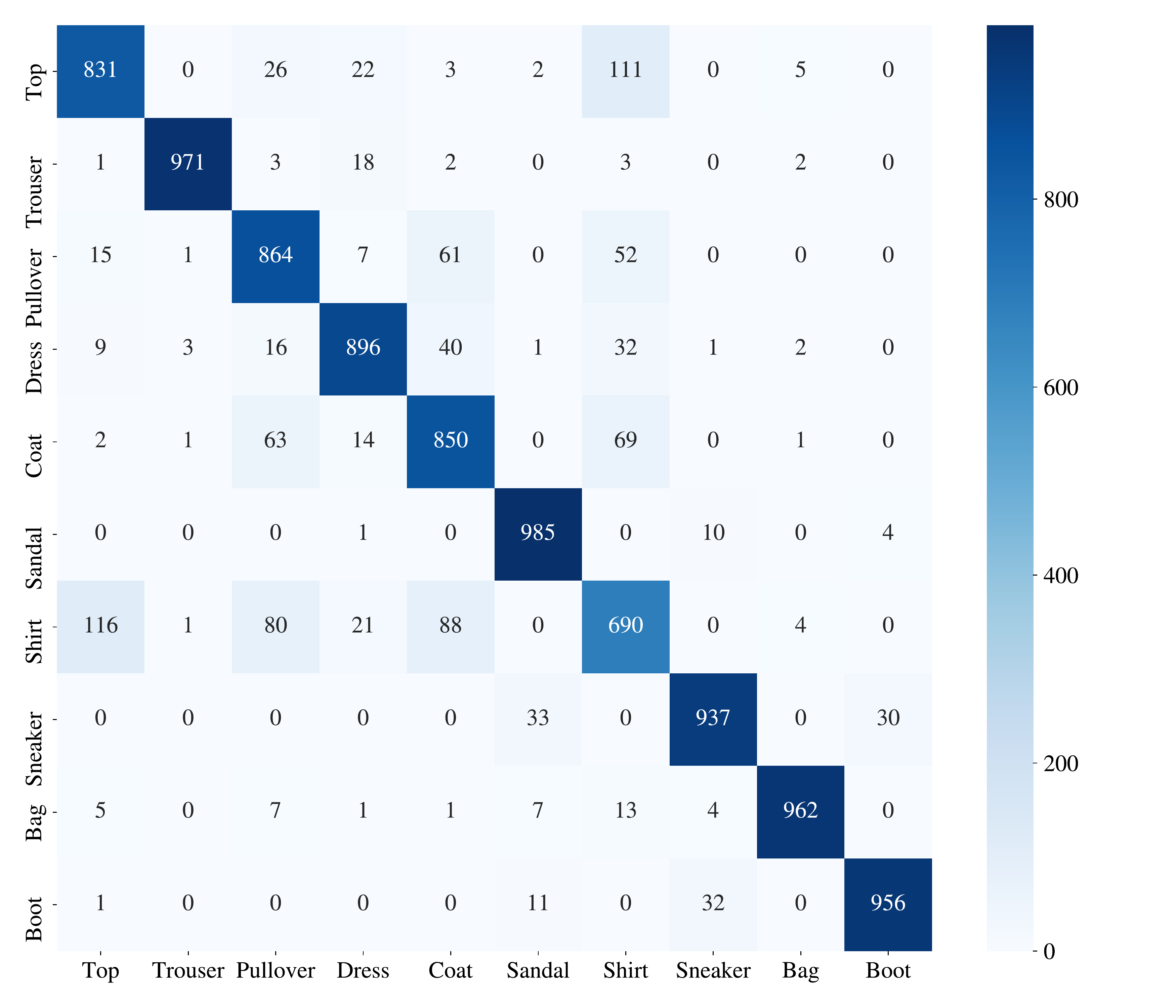}}
    \Description{Trace collection setup and confusion matrix of Fashion MNIST}
    \caption{Collection Setup and Victim's Confusion Matrix}
    \label{fig: device}
    \vspace{-3mm}
\end{figure}

\noindent\textbf{EM Trace Collection:}
The device under test (DUT) is a Xilinx DPU~\cite{ZynqDPUv92:online} running on an Ultra96-V2
board~\cite{Ultra96-V2}, a multi-processor System-on-Chip with ARM cores and Xilinx Zynq UltraScale+ FPGA.
The board runs the official PYNQ image v2.5 from the vendor AVNET
\footnote{B1600, which supports up to 1600 multiplications and additions per clock cycle.}~\cite{PYNQ}.
The EM probe is PBS set 2 with a pre-amplifier~\cite{probe}.
We use a Lecroy  WR640Zi oscilloscope~\cite{LeCroy2022Jan} to collect EM traces. 
Fig.~\ref{fig: device}(a) shows a picture of our trace collection setup.
The control and monitoring workstation first sends a command via SSH to the DUT with the pre-trained DNN model (bitstream) deployed, and the DUT loads an input image and starts executing inference.
Meanwhile, the oscilloscope is triggered to capture the EM trace until the DUT finishes execution.
Then the trace is streamed to the workstation for storage and processing.
The collected dataset of EM traces is used to train the EM classifiers and the anomaly detectors.
The training is performed on a server, with an AMD Ryzen 9 3900X 12-Core processor, 32 GB of RAM, and one Nvidia GTX TITAN GPU card.

\noindent\textbf{Datasets and Victim Models:}
We start from a LeNet-5 convolutional neural network on Fashion MNIST to evaluate our \name framework. 
We also experiment with a robust LeNet-5 retrained with adversarial examples.
Furthermore, we evaluate our framework on a larger VGG model over the colored CIFAR-10 dataset. 
The Fashion MNIST dataset is representative of computer vision tasks suitable for edge devices such as FPGA accelerators and mobile systems.
It consists of a training set of $60,000$ examples and a test set of $10,000$ samples, which are $28\times 28$ grayscale images, labeled into $10$ classes.
The LeNet-5 CNN achieves a $91.2\%$ prediction accuracy on the dataset~\cite{CNNfmnist}.
The confusion matrix of the LeNet-5 on Fashion MNIST is given in Fig.~\ref{fig: device}(b).
Note that among the ten classes, the model is more likely to misclassify Class \textit{Shirt} (the $6^{th}$ row in the confusion matrix) to other three classes, \textit{T-shirt}, \textit{Pullover}, and \textit{Coat}, due to similar features.
This lower classification rate for these classes will affect the performance of our adversarial detector accordingly, analyzed in detail in Section~\ref{sec: exp: detection performance}.
CIFAR-10 is more complex, consisting of $60,000$ $32\times 32$ color images in $10$ classes.
We adopt a larger VGG-like model and obtain a $90.5\%$ testing accuracy.
We randomly divide these datasets into a training subset ($60\%$), a validation subset ($20\%$), and a testing subset ($20\%$).


\begin{figure}[t]
    \centering
    \begin{minipage}{\linewidth}
        \subcaptionbox*{Origin}[.16\linewidth][c]{%
        \includegraphics[width=\linewidth]{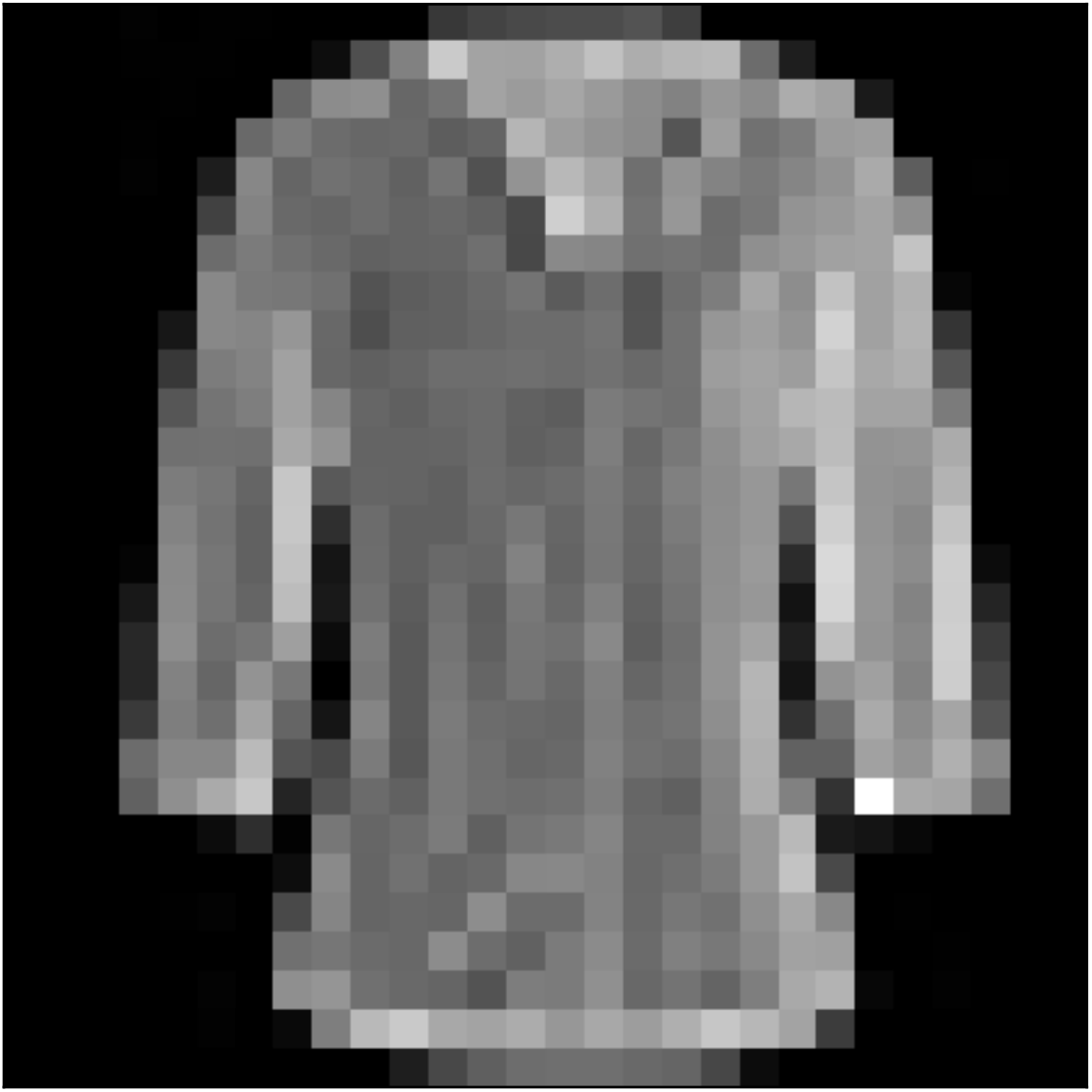}}
        \subcaptionbox*{PGD $L_1$}[.16\linewidth][c]{%
        \includegraphics[width=\linewidth]{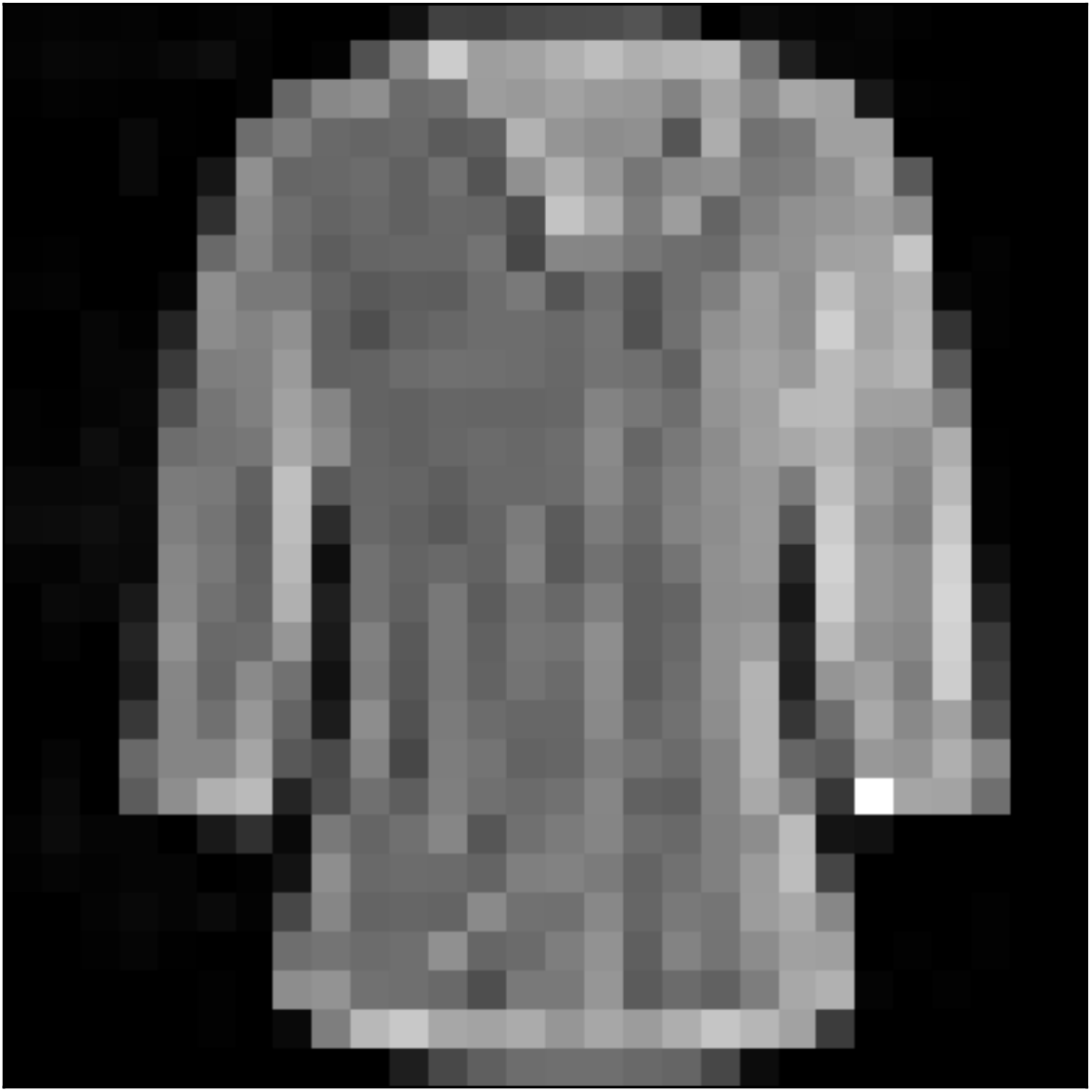}}
        \subcaptionbox*{PGD $L_2$}[.16\linewidth][c]{%
        \includegraphics[width=\linewidth]{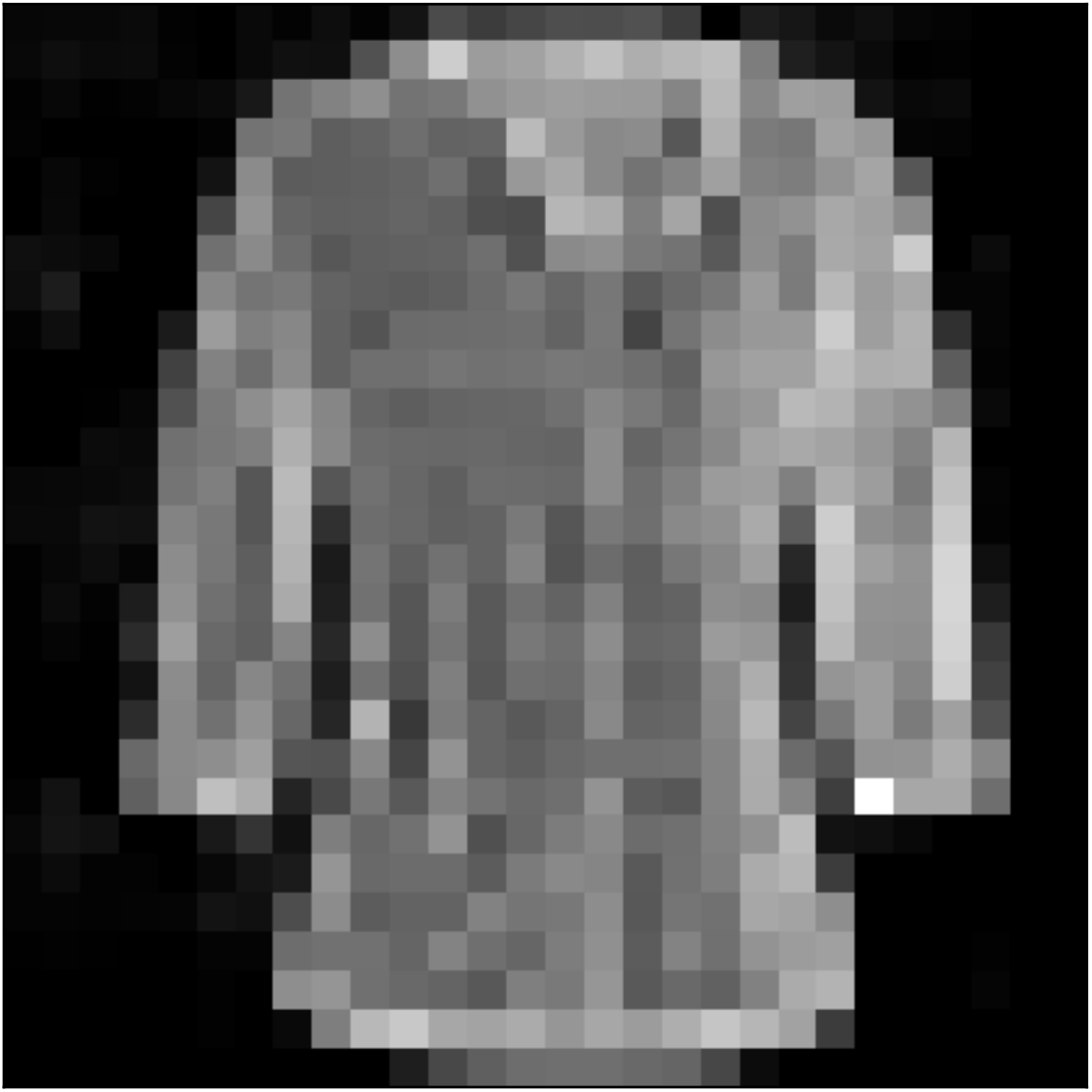}}
        \subcaptionbox*{PGD $L_{inf}$}[.16\linewidth][c]{%
        \includegraphics[width=\linewidth]{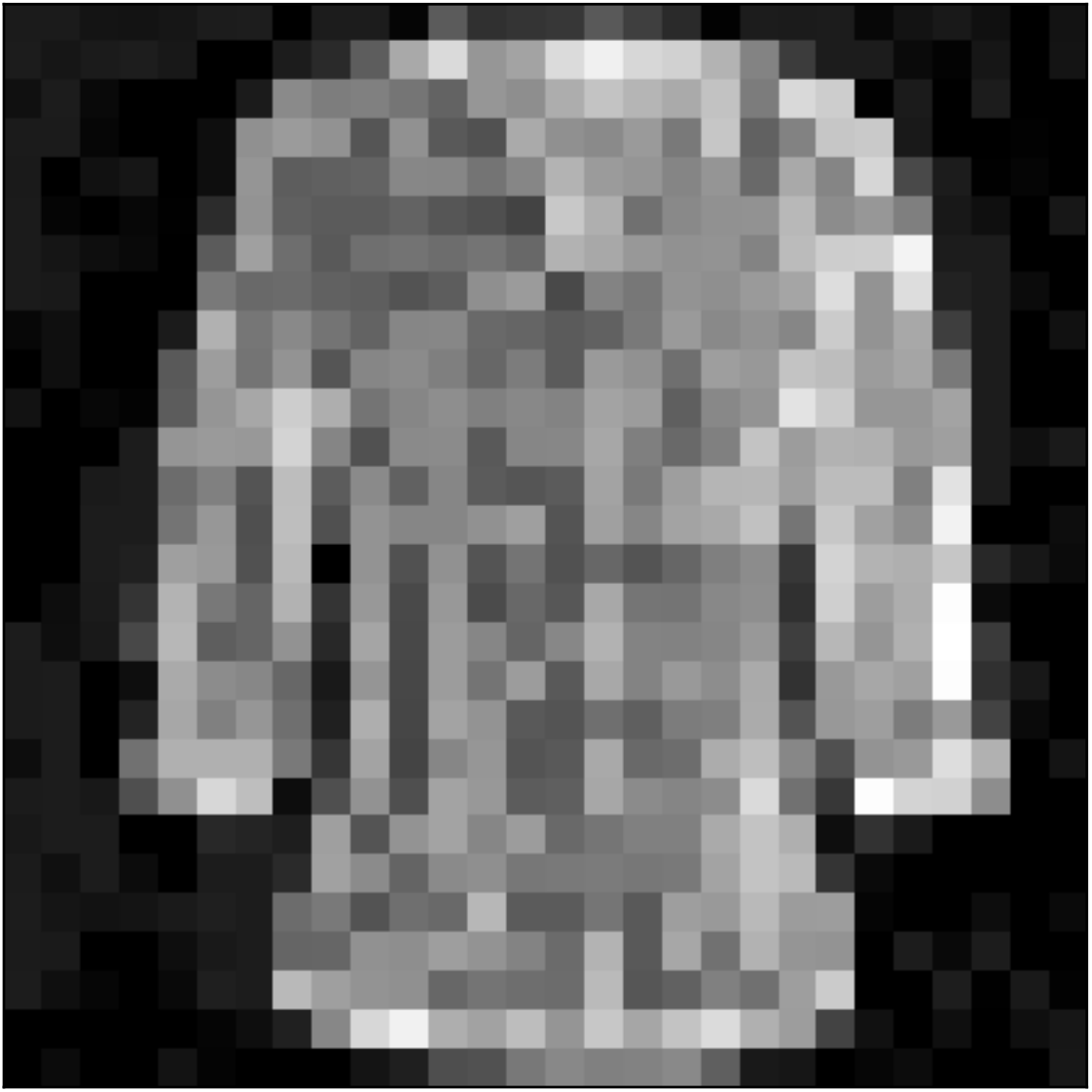}}
        \subcaptionbox*{CW $L_2$}[.16\linewidth][c]{%
        \includegraphics[width=\linewidth]{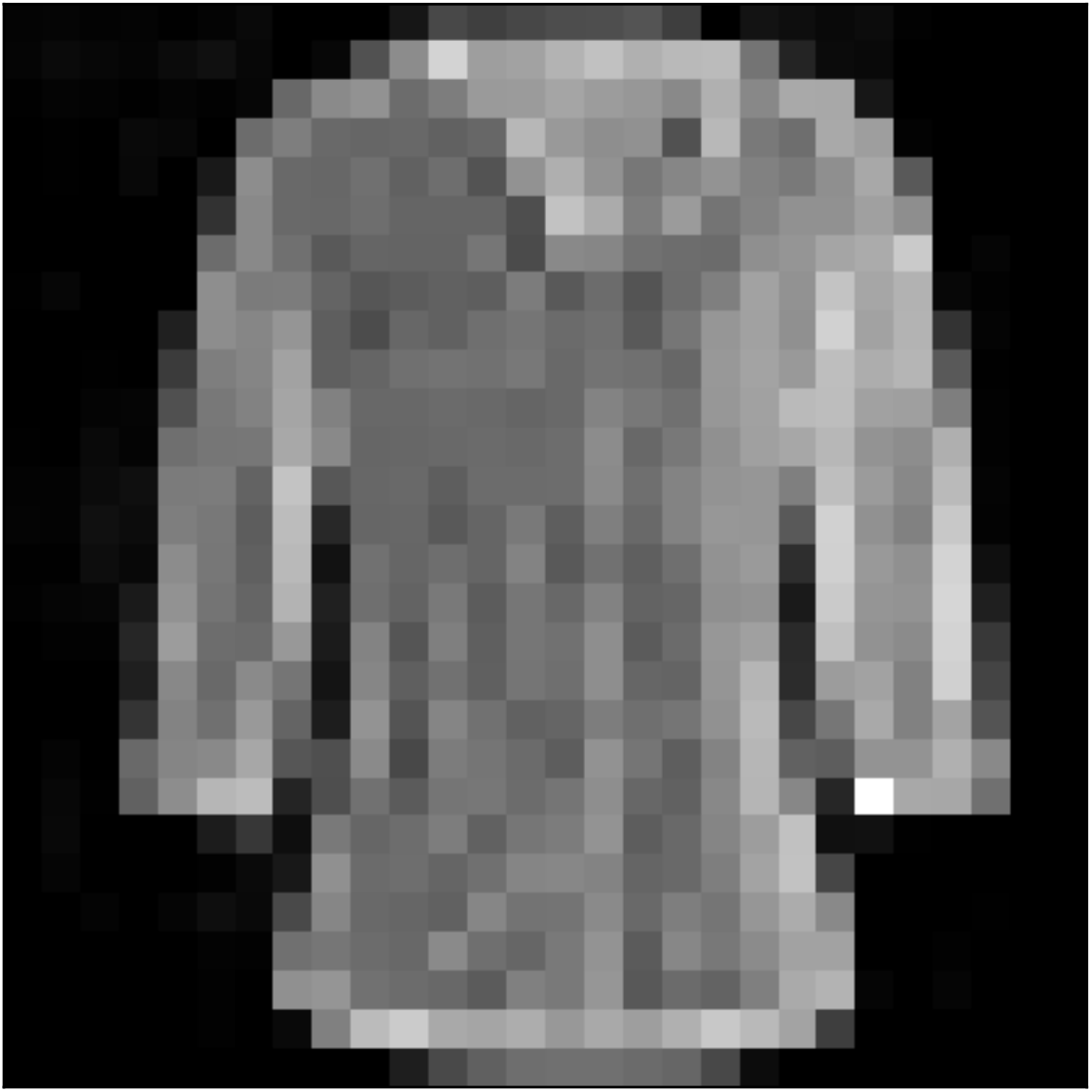}}
        \subcaptionbox*{DF $L_2$}[.16\linewidth][c]{%
        \includegraphics[width=\linewidth]{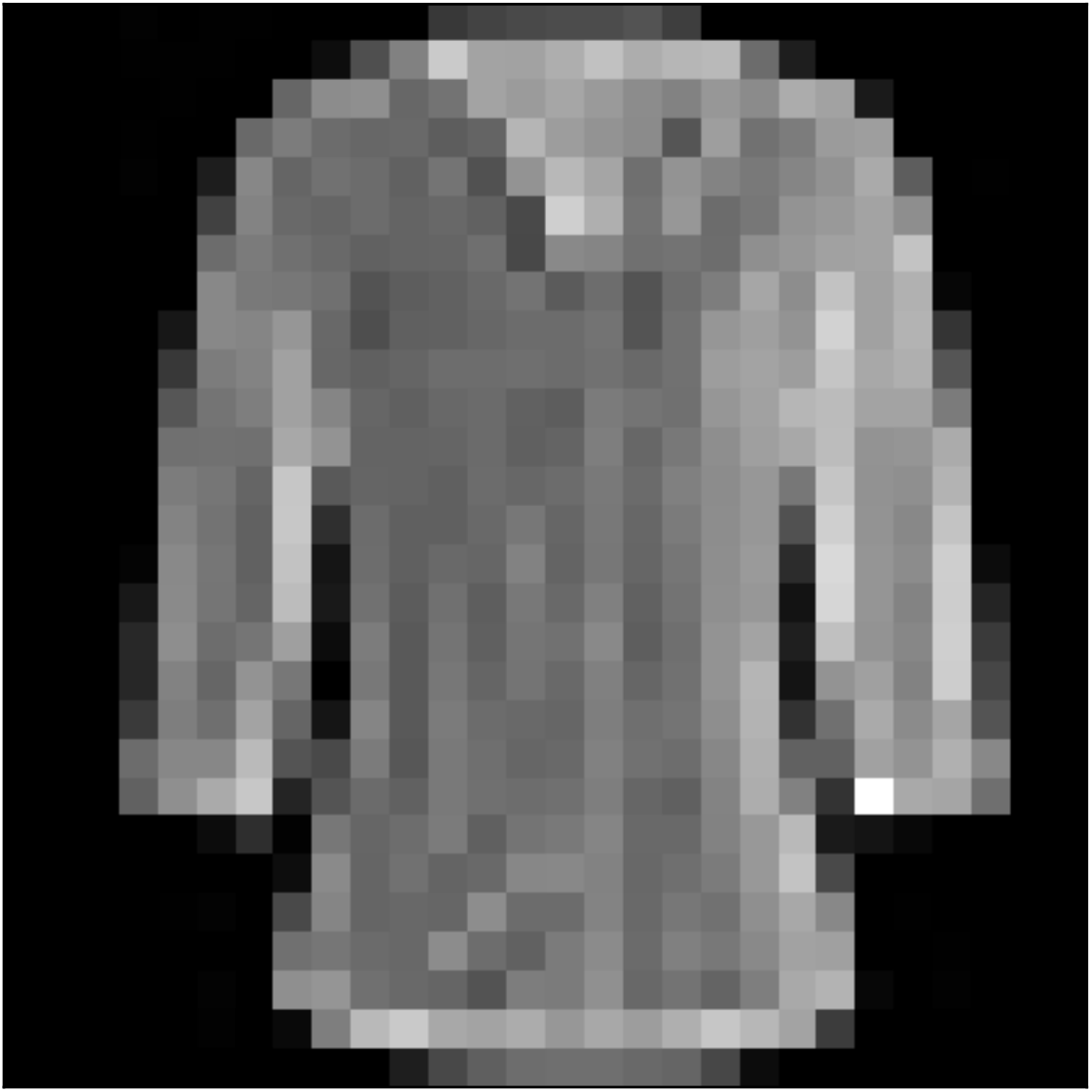}}
        \Description{Adversarial images of different attacks on Fashion MNIST.(\textit{T-shirt} to \textit{trouser})}
        \caption{Adversarial images (\textit{T-shirt} to \textit{trouser})}
        \label{fig: adversarial-examples}
    \end{minipage}
    \vspace{-3mm}
\end{figure}

\noindent\textbf{Adversarial Attacks:}
Our adversarial detector can detect a wide range of adversarial examples with EM emanations.
We employ three state-of-the-art adversarial attack methods discussed before in Section~\ref{sec:adversarial}: CW (targeted), PGD (targeted), and DeepFool (untargeted).
For PGD attacks, we test different distance measurements: $L_1$, $L_2$, and $L_{inf}$ to evaluate the model robustness against various distance losses.
For CW and DeepFool attacks, we test the commonly used $L_2$ measurements.
All the attack implementations are from the Foolbox library with commonly-used parameters~\cite{rauber2017foolbox}. 
For targeted attacks,
we consider a general attack model where the targeted label  (misclassification) can be any of the incorrect classes.
When evaluating the adversarial detector performance, 
we sample the examples to different adversarial classes for both targeted and untargeted attacks. 
For each class, we select $9,000$ adversarial samples equally distributed among the rest $9$ source classes.
Fig.~\ref{fig: adversarial-examples} shows one image from the source class of \textit{shirt} and corresponding adversarial images generated by various attack methods to a target class of \textit{trouser}. 
When evaluating CIFAR-10, we present the results of PGD L1 attacks due to the page limit.
\yf{Did you use L1 or L0?  be consistent throughout}
\ry{I use L1 attack}

\begin{figure}[t]
    \centering
    \begin{minipage}{\linewidth}
        \includegraphics[width=\linewidth]{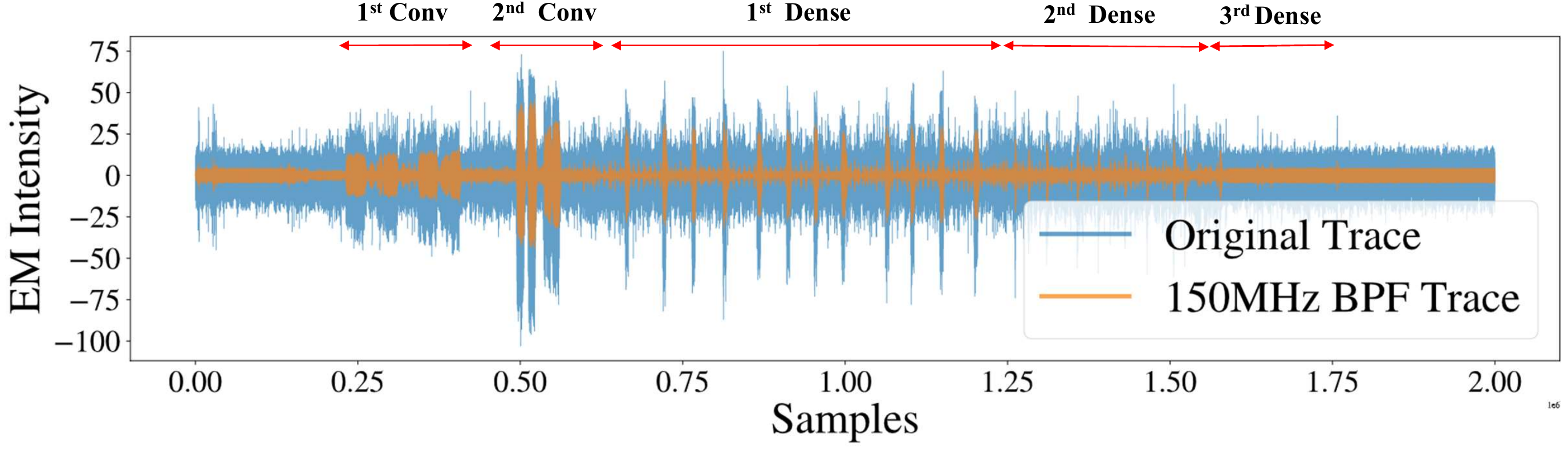}
        \Description{The raw traces and filtered traces of EM signal}
        \caption{EM trace (blue) and BPF-filtered trace (yellow)}
        \label{fig: raw-trace}
    \end{minipage}
    \vspace{-3mm}
\end{figure}

\noindent\textbf{Detection Evaluation Metrics:}
We evaluate two main components of the detector, EM classifiers and anomaly detectors, with two metrics: testing accuracy and F1-score.
Note that our training is only on benign examples while the detection (inference) is on unknown benign or adversarial samples.
In practice, the number of adversarial samples is far less than the benign ones.
Due to this imbalance, we use F1-score \cite{FScoreDe2:online}, $F_{em}$, to measure the classification performance.
\begin{equation}
    F_{em} = \frac{TP}{TP + \frac{1}{2}(FP+FN)}
    \label{eq: F1-score}
\end{equation}
where $TP$, $FP$, $FN$ are the true positive (adversarial detection rate), false positive and false negative ratio of the prediction results, respectively.
We plot Precision-Recall Curve (PR curve) to show the trade-off between detection precision and recall.
Just like ROC AUC~\cite{ROCAUC}, the Precision-Recall Area Under Curve (PR AUC) Score can be used for comparison between different detection settings.

\noindent\textbf{Baseline Comparison:}
\ry{In this section, we discuss about the comparison of \name with four baseline methods}
To the best of our knowledge, \name is the first hardware-based adversarial detector.  
It captures the contradiction between semantic EM signals and the victim output under a `black-box' setting.
We compare our method with four prior software-based detection methods, Kernel Density Estimation (KDE)~\cite{feinman2017detecting}, Network Invariant Checking (NIC)~\cite{ma2019nic}, Feature Squeezing (FS)~\cite{xu2017feature} and MagNet~\cite{meng2017magnet},  against PGD and CW $L_2$ adversarial samples.\yf{You shouldn't repeat explaining each of the prior work.  They should have been summarized in Section 2.2.  How is this section connected with 2.2? Are all the prior work inconsistency detection based}
\ry{I remove the detail description here, I want to emphasize the input }
However, these baseline methods aren't all `black-box': KDE and NIC requires the model's intermediate outputs; FS and MagNet requires testing inputs.
The metric we use for comparison is Detection Rate (DR) when the FP rate is controlled at $10\%$. 

\subsection{EM Trace Processor}\label{sec: exp: evaluate traces}
\ry{In this part, I want to give a general analysis of the EM traces.}
\subsubsection{EM traces and segmentation}
Fig.~\ref{fig: raw-trace} shows an example EM trace for one benign image inference, where the blue curve is the original trace with sample point as the $x$-axis and EM leakage intensity as the $y$-axis.
We apply a bandpass filter (BPF) with a center frequency 150MHz to reduce the noise and obtain a clearer signal (yellow trace).
After BPF, the high-intensity segments will be clean enough and can be easily partitioned.
Among the segments in Fig.~\ref{fig: raw-trace}, the first $6$ are long segments (more than $30,000$ sample points)  and the following ones are shorter (less than $10,000$ sample points).
We infer that the longer segments come from the first two convolutional layers, which utilize more Processing Element (PE) for parallel computation.
The rest shorter segments come from the dense layers, which run faster and turn out to be less informative.
In real applications, the detector has a black-box view of the model and has no information about which layer the segment comes from, but can automatically process the trace with BPF and partitioning.


\begin{figure}[t]
    \subcaptionbox{Time-domain}[.45\linewidth][c]{%
    \includegraphics[width=\linewidth]{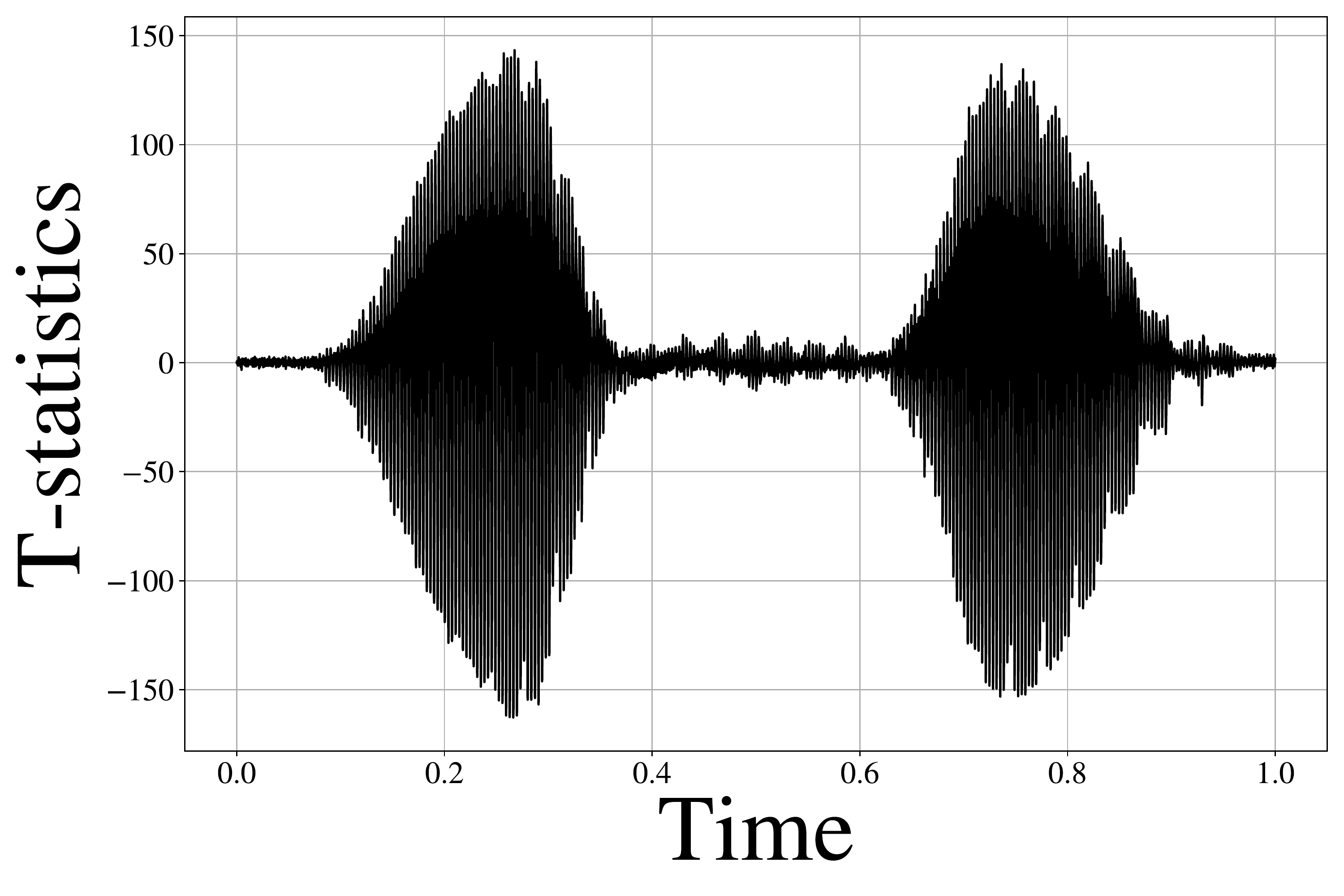}}\quad
    \subcaptionbox{Frequency-domain}[.45\linewidth][c]{%
    \includegraphics[width=\linewidth]{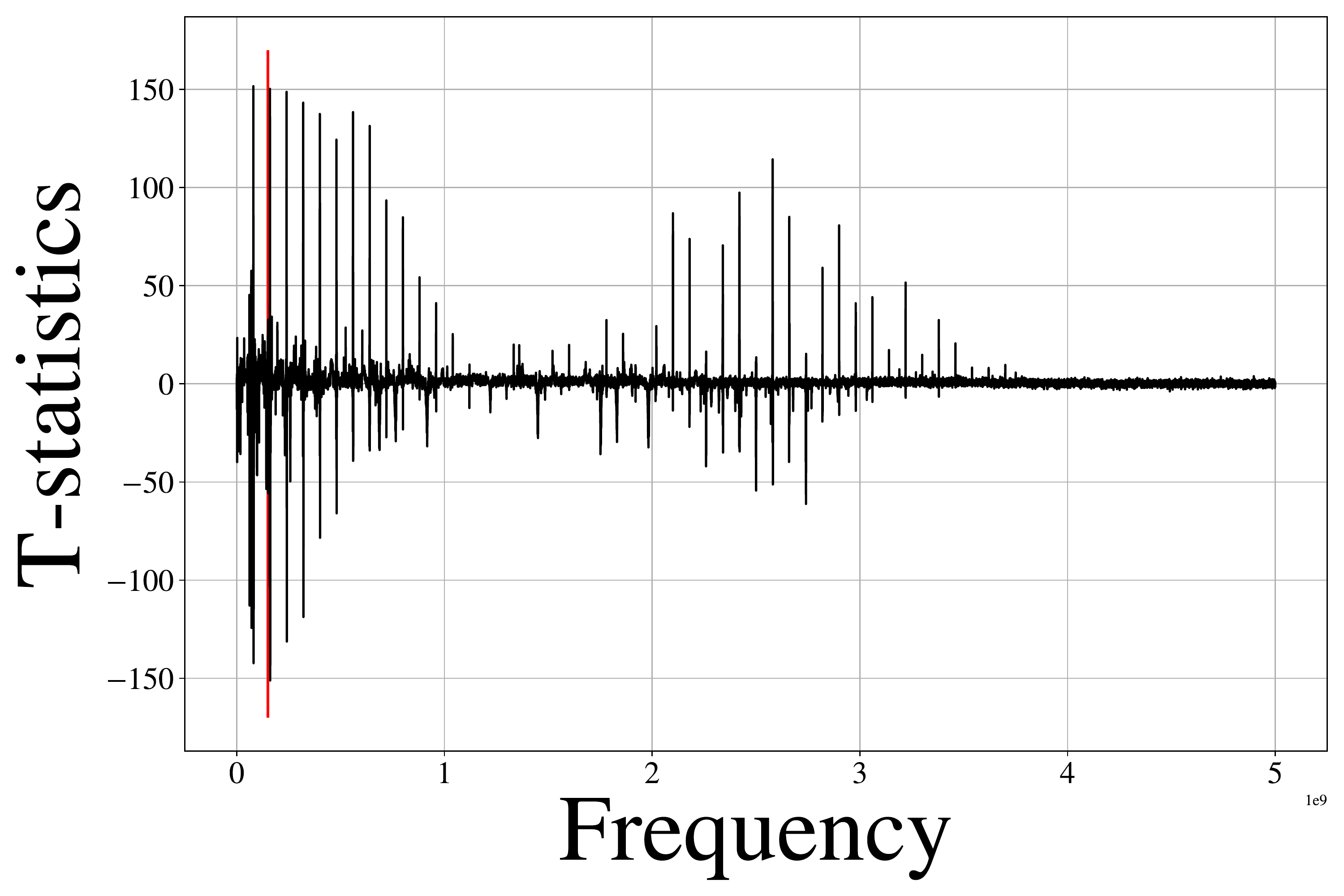}}\quad\\
    \subcaptionbox{Spectrogram T-scores}[\linewidth][c]{%
    \includegraphics[width=0.95\linewidth]{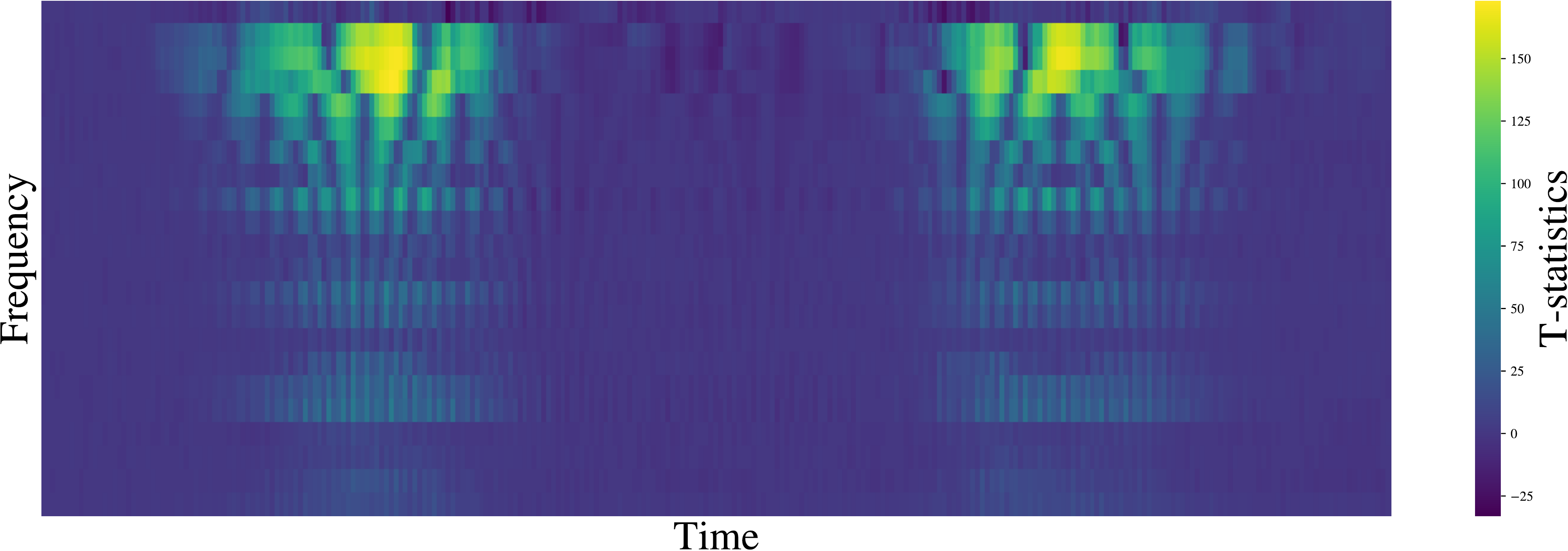}}\quad
    \caption{(a) T-statistics of time-domain traces; (b) T-statistics of frequency-domain spectrums, the red line marks the operating frequency $150$MHz;
    (c)T-statistics of spectrograms, the intensity of each point represents the T value.}
    \label{fig: Ttest}
    \Description{T-test results of the trace}
    \vspace{-3mm}
\end{figure}

\ry{In this part, I want to give some insight about class information of EM traces}
\subsubsection{EM Spectrograms} \label{sec: exp: spectrogram}
Class-related features/signals in the EM traces have to be preserved to build a highly accurate EM classifier.
We show that the spectrograms generated by our STFT data processing method outperform both the time-domain traces and the simple frequency-domain spectrum (after applying the fast-fourier-transformation (FFT) on the entire time-domain EM trace).
To localize and detect class-related signals, we run the victim model on two classes of input images and collect EM traces. We applied the student T-test across the two class-datasets, on three kinds of data representations of EM traces: spectrograms, the original time-domain traces, and the frequency-domain spectrums.
The T-test statistically tests the average difference between two groups of data.
A large absolute value of T-statistics on a point indicates that the traces of two different classes differ significantly here, thus this location contains a strong class-specific signal.

Fig.~\ref{fig: Ttest} presents the T-statistics results for Class 0 and Class 1.
The T value is shown on the y-axis, while the X axis shows the time for the time-domain trace in Fig.~\ref{fig: Ttest} (a) and frequency for the frequency-domain spectrum in Fig.~\ref{fig: Ttest} (b).
For 2-D spectrogram T scores shown in Fig.~\ref{fig: Ttest} (c), the X axis is the time, Y axis is the frequency while the T value is represented by the intensity on the heatmap.
Intuitively the spectrogram depicts the time-varying spectrum, while the frequency-domain spectrum just presents average frequency components.
When comparing Fig.~\ref{fig: Ttest} (a) and (c), we can view each row of Fig.~\ref{fig: Ttest} (c) as a constituent  component of Fig.~\ref{fig: Ttest} (a).
By filtering the bottom rows and only keeping the rows with high intensity (near the top), we are filtering irrelevant noise with low T-values.
When comparing Fig.~\ref{fig: Ttest} (b) and (c), we can view each column of Fig.~\ref{fig: Ttest} (c) as a spectrum for a short time window, and the spectrum is varying along the time.
The energy (high intensity) focuses on the frequency band near the top of Fig.~\ref{fig: Ttest} (c) (i.e., the beginning frequencies of  Fig.~\ref{fig: Ttest} (b)), which is around the operating frequency of the DUT.

Fig.~\ref{fig: Ttest} also shows the peak absolute value of T-statistics of the spectrogram is $173.14$ compared to $162.79$ and $151.64$ for the time-domain traces and frequency-domain spectrums, respectively.
We can conclude that the spectrogram contains more signals for classification than the other two forms of data.
In our experiment, we only select $15$ frequency bands of the spectrogram around the device operating frequency and discard other bands. 
This bandpass filtering is effective de-noising. 
As spectrogram preserves both the frequency and time information, its 2-D form resembles an image and suits CNN classification naturally.

\subsection{Evaluation of EM Classifiers} \label{sec: exp: EM classifiers}

Table~\ref{tab: EM-classifier-results} gives the performance of EM classifiers (we use VGG-11 models) 
for LeNet-5 on MNIST. 
For the original raw trace in Fig.~\ref{fig: raw-trace}, we analyze the first $18$ segments corresponding to computations for the two convolutional layers and the first dense layer of the victim model,
while the last two layers leak less information (with lower intensity and shorter time).
We process each of the segments separately with STFT and build a classifier, locating the most prominent class-specific information.

\begin{table*}[t]
    \centering
    \caption{The EM classifiers' performance for each segment}
    \label{tab: EM-classifier-results}
    \begin{tabularx}{\textwidth}{cXXXX|XX|XXXXXXXX}
         \toprule
        Layer Type &  \multicolumn{6}{c}{Convolutional Layers} & \multicolumn{8}{c}{Fully-connected Layers} \\
        \hline
        Index &  \multicolumn{4}{c|}{$1^{\text{st}}$  Layer} & \multicolumn{2}{c|}{$2^{\text{nd}}$  Layer} &\multicolumn{8}{c}{$3^{\text{rd}}$  Layer}\\
        \hline
        Segment  & 0 & 1 & 2 & 3 & 4 & 5 & 6 & 7 & 8 & 9 & 10 & 11 & 12-16 & 17\\
        \hline
        Accuracy & 0.68 & 0.70 & 0.64 & 0.56 & 0.68 & 0.68 & 0.23 & 0.38 & 0.37 & 0.10 & 0.33 & 0.33 & 0.10 & 0.59 \\
        F1-score & 0.65 & 0.67 & 0.61 & 0.52 & 0.66 & 0.66 & 0.18 & 0.35 & 0.32 & 0.02 & 0.26 & 0.26 & 0.02 & 0.57\\
        \hline
    \end{tabularx}
 \end{table*}

For most of the segments, the EM classifier does extract some model execution information of that segment, where the class prediction accuracies based only on EM traces range from $23\%$ to $70\%$.
Some segments (the 12th-16th) from dense layers reveal little information, resulting in only $10\%$ accuracy (the same as a random guess among $10$ classes).
We choose to only use the strong signals from convolutional layers for adversarial detection in the next part without much degradation.

For different classes, the amount of information carried in each segment also varies.
As an example, we separate the first three-segment EM classification performances by the output class labels and report them in Table~\ref{tab: resultofB0}.
For instance, for the $8^\text{th}$ class, Segment 0 contains most information;
for Class 5, Segment 1 contains most information;
for Class 1, Segment 2.
Different segments of DNN execution focus on different semantic features.
Since the most informative feature varies from one output class to another, the most informative EM segment also varies.

\begin{table*}[t]
    \centering
       \caption{The EM classifiers' classification report for Segment 0, 1, 2}
    \label{tab: resultofB0}
    \begin{tabularx}{\textwidth}{c|XXXXXXXXXXX}
        \toprule
        \multicolumn{2}{c}{Class}  & 0 & 1 & 2 & 3 & 4 & 5 & 6 & 7 & 8 & 9\\
        \hline
        \multirow{ 2}{*}{Segment 0}
        & Accuracy & 0.58 & 0.92 & 0.55 & 0.55 & 0.49 & 0.67 & 0.35 & 0.64 & \textbf{0.89} & 0.83 \\
        & F1-score & 0.56 & 0.81 & 0.38 & 0.65 & 0.50 & 0.63 & 0.37 & 0.72 & \textbf{0.91} & 0.77 \\
        \hline
        \multirow{ 2}{*}{Segment 1}
        & Accuracy & 0.70 & 0.81 & 0.71 & 0.69 & 0.40 & \textbf{0.85} & 0.37 & 0.91 & 0.81 & 0.85 \\
        & F1-score & 0.72 & 0.86 & 0.53 & 0.71 & 0.42 & \textbf{0.90} & 0.04 & 0.82 & 0.84 & 0.88 \\
        \hline
        \multirow{ 2}{*}{Segment 2}
        & Accuracy & 0.59 & \textbf{0.95} & 0.47 & 0.65 & 0.35 & 0.82 & 0.47 & 0.65 & 0.62 & 0.80 \\
        & F1-score & 0.68 & \textbf{0.92} & 0.53 & 0.68 & 0.40 & 0.70 & 0.08 & 0.74 & 0.61 & 0.82 \\
        \hline
    \end{tabularx}
    \vspace{-3mm}
 \end{table*}

To visualize the utilization of varying parts of EM segments, we use the GradCAM~\cite{selvaraju2016grad} on our EM classifiers.
The results are shown in Appendix~\ref{sec: gradCAM}.
For different classes, the benign inputs generally activate different neurons within a segment, and our EM classifiers capture the patterns.
When an adversarial example does not activate these neurons, it results in different output logits for the EM classifier.

\yf{at the end of Section 4.3, you need to add a line to discuss your experiments when changing the window and stride size of STFT.  how many you have tried? What is the impact? NO figure is fine, but need to be stated}


\begin{figure}[t]
    \subcaptionbox{Target 1 VAE loss}[.40\linewidth][c]{%
    \includegraphics[width=\linewidth]{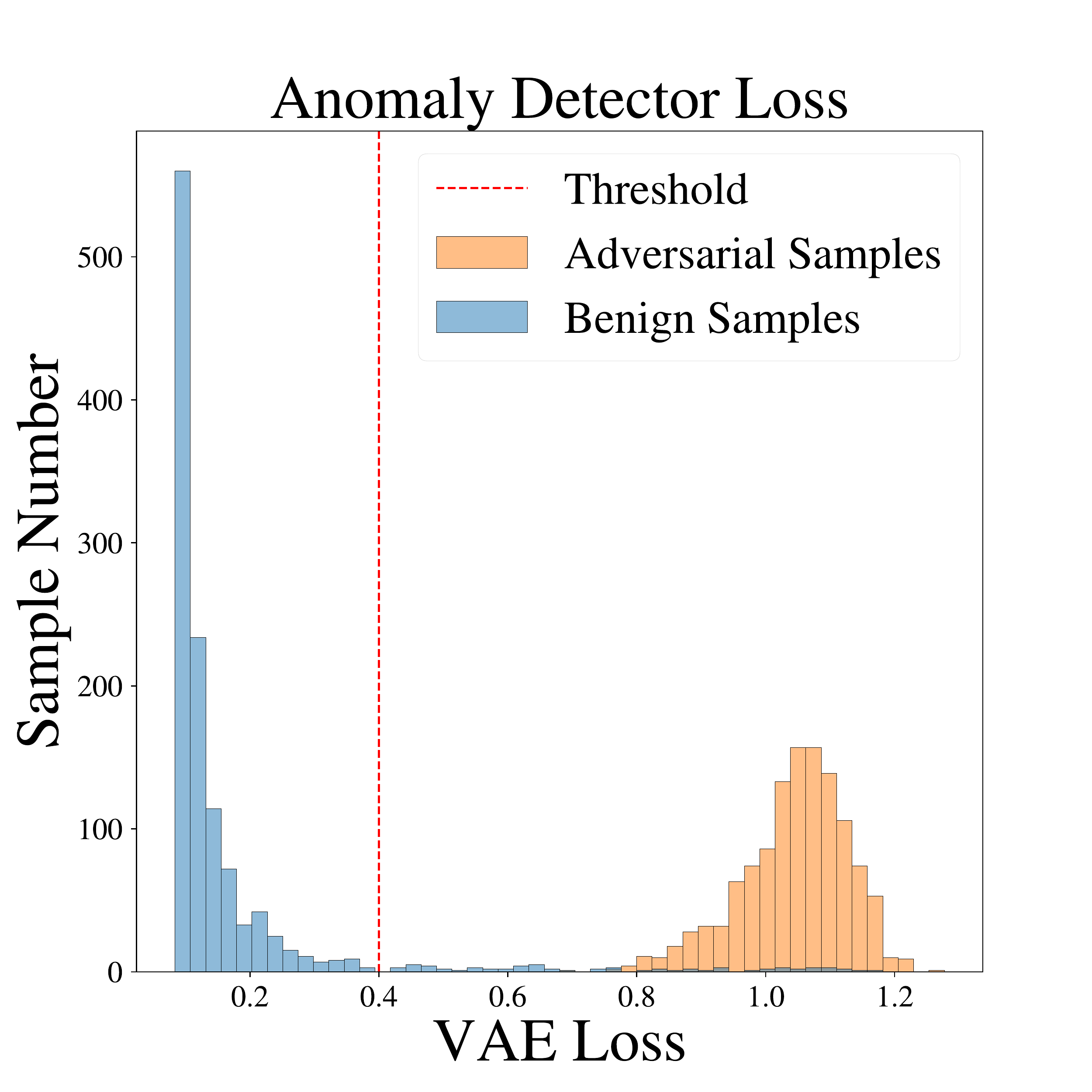}}\quad
    \subcaptionbox{PR curves for classes}[.48\linewidth][c]{%
    \includegraphics[width=\linewidth]{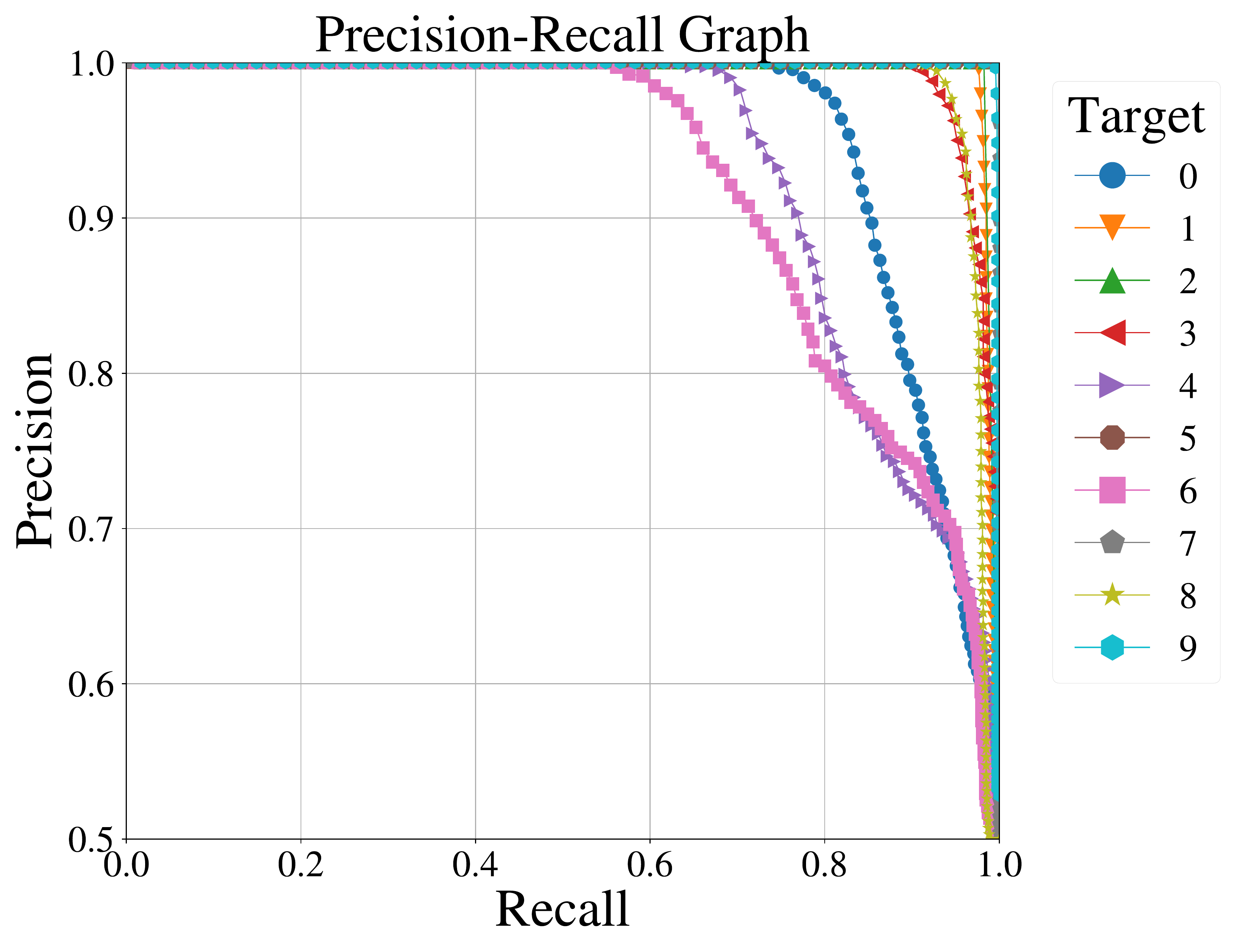}}
    \Description{Class1 anomaly detection and PR curve}
    \caption{VAE Loss of Target 1 and PRcurves for ten targets}\yf{Class 1 means the source class or target class. be accurate!}
    \label{fig: loss and pr}
    \vspace{-3mm}
\end{figure}

\subsection{Evaluation of Anomaly Detector} \label{sec: exp: detection performance}

We concatenate the logits from EM classifiers for all segments to get a logits vector reflecting the execution flow of the victim model.
The follow-on VAE extracts compressed latent features from the benign logits vectors so that the benign vectors can be reconstructed from the compressed latent features with small loss.
The adversarial examples cause different execution flows and their logits vectors can not be well reconstructed from compressed latent features.
Fig.~\ref{fig: loss and pr}(a) presents the testing reconstruction loss of the pre-trained VAE for both benign and adversarial samples, where the blue bars are for benign samples and the yellow bars are for adversarial examples from PGD L2 attack.  The red dash vertical line is an empirically selected threshold to determine whether the input is benign or adversarial.
It shows that two distributions are disjoint and adversarial examples can be easily distinguished from benign examples.
Fig.~\ref{fig: loss and pr}(b) shows the precision-recall curves of the VAE for all 10 classes.
Similar to the receiver operating characteristic curve, the PR curve shows the model performance trade-off between precisions and recalls.
The classification algorithm is desired to have both high precision and high recall.
Therefore, a larger Area Under Curve (AUC) indicates a better classifier.
Three classes, Class 0, 4, 6, have relatively worse performance, due to the original classification inaccuracy of the victim model among these classes.

In Fig.~\ref{fig: embedding}, we visualize the features of a selected adversarial sample (generated by PGD $L_2$ attack) and two benign samples (one of the source class and the other of the target class) with a 3-D embedding of their logits vectors.
The embedding demonstrates that the logits of the adversarial sample are different from both those of the source class and the target class, while relatively closer to the former (more different from the target).
Combined with the victim model output (misclassified to a target class), the anomaly detector finds that essentially the adversarial example bears more similarity to another class (the source class) than the predicted one, presenting a conflicted result and therefore capturing the discrepancy.

\aptLtoX[graphics=no, type=html]{\begin{figure}[t]
    \centering
    \begin{minipage}{.21\textwidth}
        \includegraphics[width=\linewidth]{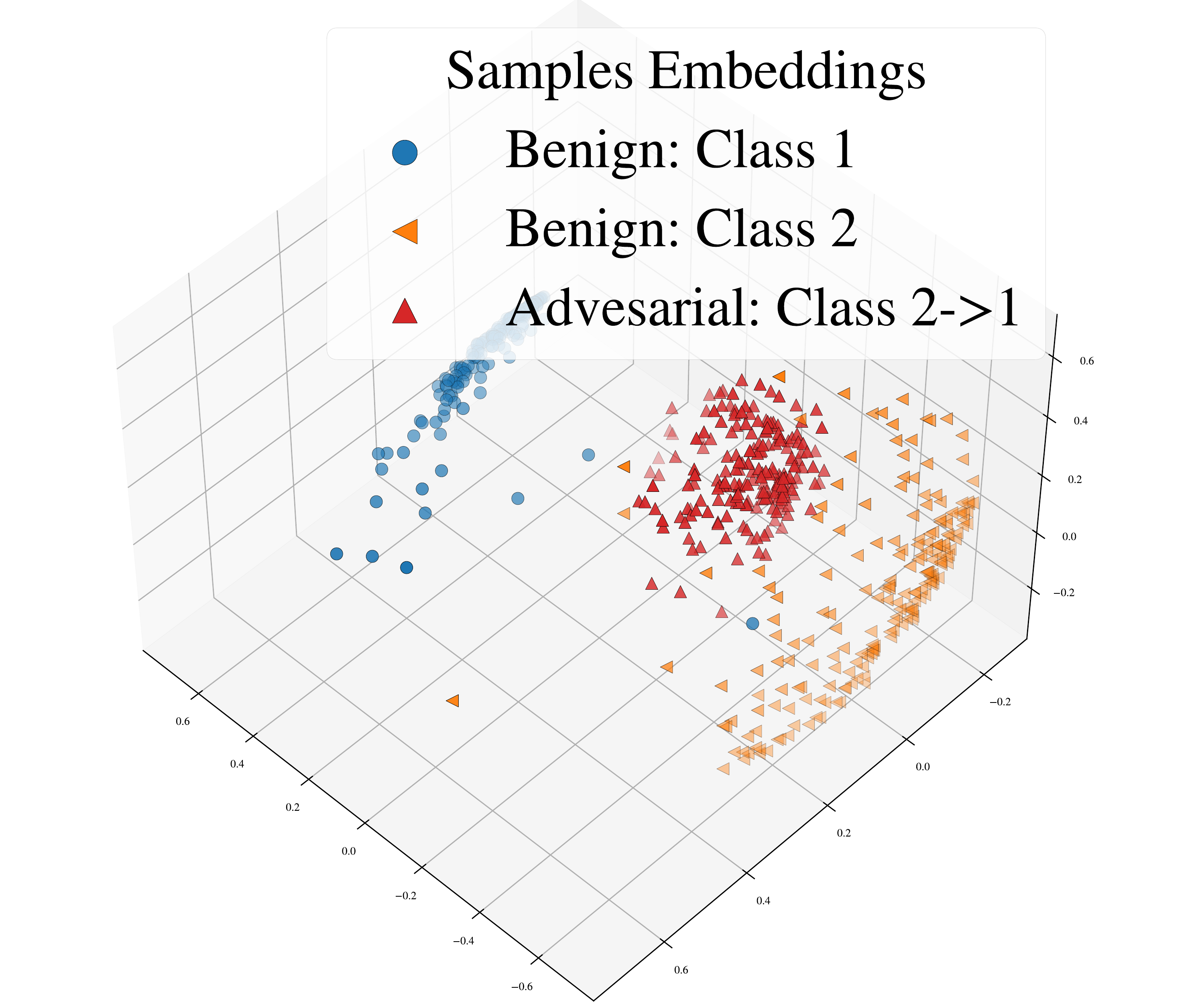}
        \caption{3D Embeddings}
        \label{fig: embedding}
        \Description{3D embeddings}
    \end{minipage}
\end{figure}}{}

\aptLtoX[graphics=no, type=html]{\begin{figure}[t]
    \begin{minipage}{.25\textwidth}
        \includegraphics[width=\linewidth]{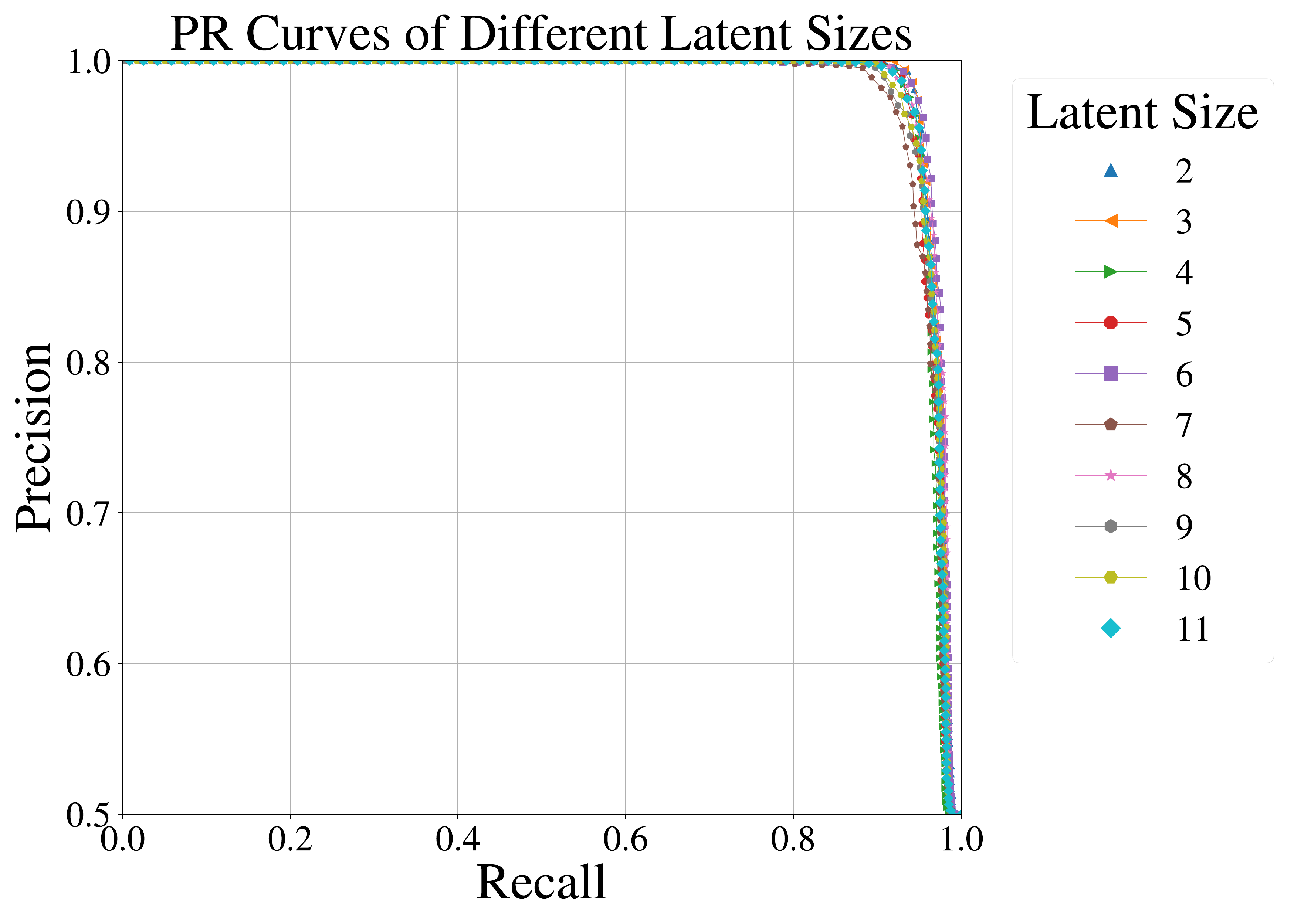}
        \caption{Latent PRcurves}
        \label{fig: pr latent}
        \Description{The Precision-Recall Curve of latent space}
    \end{minipage}
    \vspace{-3mm}
\end{figure}}{}

\aptLtoX[graphics=no, type=html]{}{\begin{figure}[t]
    \centering
    \begin{minipage}{.21\textwidth}
        \includegraphics[width=\linewidth]{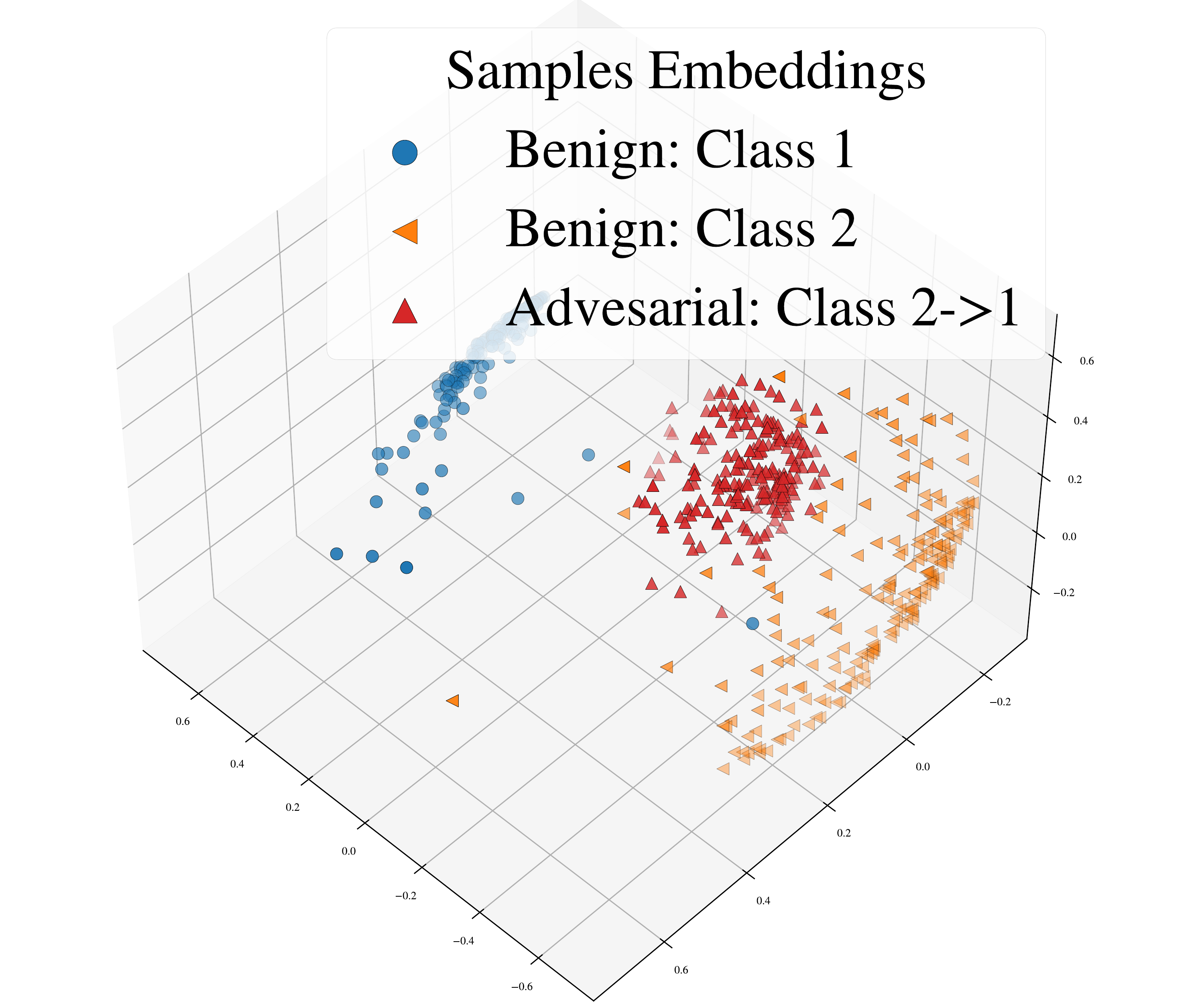}
        \caption{3D Embeddings}
        \label{fig: embedding}
        \Description{3D embeddings}
    \end{minipage}
    \begin{minipage}{.25\textwidth}
        \includegraphics[width=\linewidth]{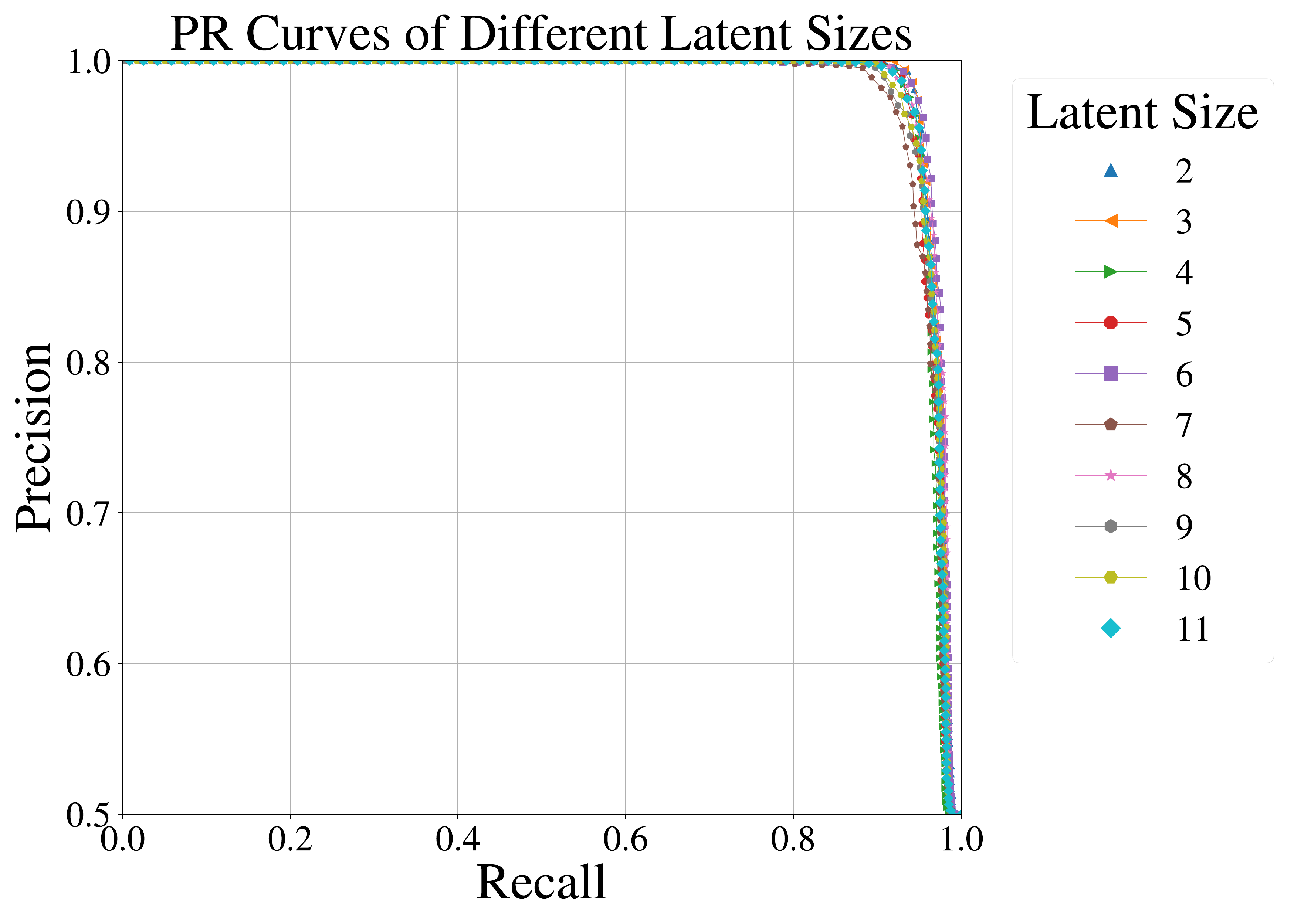}
        \caption{Latent PRcurves}
        \label{fig: pr latent}
        \Description{The Precision-Recall Curve of latent space}
    \end{minipage}
    \vspace{-3mm}
\end{figure}}

\subsection{Overhead and Delay}\label{sec: exp: overhead}
We measure the overhead and delay of the EMShepherd detection framework.
As the detector is outside of the victim model and the system, it will not affect the victim model execution at all.
It detects an adversarial example within $169$ milliseconds after the victim model finishes execution on the experimental platform.
The delay is composed of the average processing time of EM traces ($10$ ms), EM classifiers inference ($128$ ms) and anomaly detector execution ($31$ ms).
The processing time can be reduced by running the EM classifiers for different trace segments in parallel.  It can be further reduced by running the detection along with the measurements in a pipelined fashion - starting processing a segment as soon as it is measured while the victim model is still executing the next segment. 

\subsection{Impact of the Detector Parameters} \label{sec: exp: ablation}
In this section, we evaluate the impact of different experimental settings on the performance of our adversarial detector. 
For instance, the trace sampling frequency and sliding window size of STFT will impact the classification accuracy of EM classifiers,
and the structure of the anomaly detector (such as the latent space size) may affect the generalizability of VAE.
Different attack methods or the distance measures used in attacks will also lead to different detection results.

\noindent\textbf{EM Trace Sampling Frequency:}
\ry{Add results about different original trace sampling rate}
The sampling frequency of EM traces has an effect on the information density for the classifiers.
We adjust the oscillator's sampling frequency from $500$MHz to $20$GHz and report the EM classification accuracy of the first segment in Table~\ref{tab: sampling frequency}.
Note that according to the Nyquist-Shannon sampling theorem, the minimum sampling frequency should be larger than $300$ MHz (two times of operating frequency).
The results show that a sampling frequency above 2GHz is sufficient to achieve good classification accuracy.\yf{I remember there is some question about the high sampling frequency.  Do you have some segment that doesn't require this high sampling frequency to get good classification accuracy?}

\begin{table}[ht]
    \centering
    \caption{Classification accuracy versus sampling frequency }
    \begin{tabularx}{\linewidth}{cXXXXXX}
        \toprule
        Frequency (GHz) & 0.5 & 1 & 2 & 4 & 10 & 20 \\
        \hline
       Accuracy & 0.32 & 0.32 & 0.65 & 0.67 & 0.67 & \textbf{0.68}\\
        \hline
    \end{tabularx}
    \label{tab: sampling frequency}
    \vspace{-3mm}
\end{table}

\noindent\textbf{Sliding Window Size:}
We further compare the classifiers' accuracy with different STFT window sizes.
When the Hanning window size changes, the number of bands with most signals also changes.
Table~\ref{tab: stft-config} presents the results for different STFT configurations.
As the window size changes from 64 to 1024, the classifiers' average accuracy goes up first and then goes down, with the maximum accuracy of
67\% at the window size of 128 with the top $15$ frequency bands (above the red dash line in Fig.~\ref{fig: gradCAM} (d)) in the Appendix are kept for the follow-on classifier and anomaly detector).

\begin{table}[ht]
    \centering
      \caption{Classification accuracy vs. the sliding window size }
    \label{tab: stft-config}
    \begin{tabularx}{\linewidth}{cXXXXX}
        \toprule
        Window size & 64 & 128 & \textbf{256} & 512 & 1024\\
        \hline
        Band number & 60 & 30 & \textbf{15} & 8 & 4\\
        \hline
        Accuracy & 0.53 & 0.56 & \textbf{0.67} & 0.59 & 0.34\\
        \hline
    \end{tabularx}
\end{table}

\noindent\textbf{Latent Space Size:}
The latent space of VAE is a variable that may affect the performance of anomaly detection.
We vary the size of the latent space between 2 and 9, which reflects the model's capability to express the input data features.
The results are presented in Fig.~\ref{fig: pr latent}.
Overall the latent space size has no significant effect on anomaly detection. The space size of 6 is slightly better than others.  


\noindent\textbf{Attack Methods:}
Our anomaly detector can detect adversarial samples for a wide range of existing attacks, PGD, CW, and DeepFool, with different attack distance metrics.
We tested five attack methods, $PGD(L_1)$, $PGD(L_2)$, $PGD(L_{inf})$, $CW(L_2)$, and $DeepFool(L_2)$, 
and the adversarial detector all has a reasonable detection performance, shown in Table~\ref{tab: detection rate of vae}.

Comparing the different distances used in attacks, we observe that using an EM-based detector has better performance for the $L_1$ attack, then $L_{inf}$, and the worst is $L_2$ for PGD attacks.
For $L_1$ attack, only a few pixels are modified to make an adversarial example from the original source-class sample, as shown in Fig.~\ref{fig: adversarial-examples} (b). Although the victim model is misled to predict it to be the target class, the EM trace of the inference bears more similarity to the original class sample, distinctly different from the EM traces of the target class samples.  Therefore, it is easier to detect the adversarial.
While for $L_{inf}$ attack, many pixels are changed, randomly distributed, easily visualized in Fig.~\ref{fig: adversarial-examples} (d).
The EM trace will differ both from that of the source class and that of the target class, still caught easily by the adversarial detector. 
For $L_2$ attack, there are more pixels changed than the $L_1$ attack, but around the object in the image rather than randomly distributed like in $L_{inf}$, making the EM trace somewhat between those of the source class and target class and causing
the anomaly detector low confidence in making the prediction.

\begin{table*}[t]
    \centering
        \caption{F1-scores of the VAE Detector}\yf{should the DR be TPR?}\ad{Ruyi: check the notation consitency: Is DR just TPR and is FAR just FPR?}
    \begin{tabularx}{\textwidth}{X|X|X|X|X|X|X|X|X|X|X}
        \toprule
         Attack & 0 & 1 & 2 & 3 & 4 & 5 & 6 & 7 & 8 & 9 \\
        \hline
        PGD($L_2$) & 0.820 & 0.999  & 0.957 & 0.957 & 0.899 & 0.999 & 0.758 & 0.999 & 0.968 & 0.999 \\
        \hline
        PGD($L_{inf}$) & 0.946 & 0.999  & 0.956 & 0.981 & 0.918 & 0.999 & 0.884 & 0.999 & 0.963 & 0.999 \\
        \hline
        PGD($L_1$) & 0.948 & 0.999  & 0.957 & 0.982 & 0.924 & 0.999 & 0.910 & 0.999 & 0.968 & 0.999\\
        \hline
        CW($L_2$) & 0.797 & 0.999  & 0.957  & 0.982 & 0.924 & 0.999 &0.756& 0.999& 0.958 & 0.999 \\
        \hline
        DF($L_2$) & 0.918 & 0.999  & 0.811 & 0.963 & 0.804& 0.999& 0.756 & 0.999 & 0.968& 0.999\\
        \hline
    \end{tabularx}
    \label{tab: detection rate of vae}
\end{table*}

\noindent\textbf{Target Classes:}
Fig.~\ref{fig: loss and pr}(b) shows that the prediction F1-scores of the victim model on the three classes, 0, 4, and 6, are $0.84$, $0.83$, and $0.70$, respectively, lower than other classes with scores above $0.9$. \ad{Did the Figure changed during revision? Fig.~\ref{fig: device}(b) shows the confusion matrix with no F1-scores. }
Such inaccuracy affects the performance of our adversarial detectors.
The samples from these classes, therefore, include similar computation along a large part of the victim model execution flow, making the detection hard from the EM emanations of the execution.
Table~\ref{tab: detection rate of vae} also presents the detector performance on various adversarial target classes (columns) by different attack methods (rows) using F1-scores.
Particularly Class 0(T-shirt) and 6(Shirt) do not perform as well as other classes.
For other classes, our detection framework can detect close to $94\%$ of adversarial samples with less than a $10\%$ false positive rate.
\yf{\textcolor{red}{Why didn't you present individual FAR for each attack method?  Also the highest FAR excluding classes 0, 4, 6 is 9\%, not what you reported as 3.75\%?}}\ad{I assume that FAR is FPR, so same for all attack method. However, if overall FPR is 10\%, then average FAR across the 10 classes should be 10\%, but that is not the case in Table 3. Why?}

\if false 
\noindent\textbf{Distance Measurements:}
Comparing the different distances used in attacks, we observe that using an EM-based detector has better performance for the $L_1$ attack, then $L_{inf}$, and the worst is $L_2$ for PGD attacks.
For $L_1$ attack, only a few pixels are modified to make an adversarial example from the original source-class sample, as shown in Fig.~\ref{fig: adversarial-examples} (b). Although the victim model is misled to predict it to be the target class, the EM trace of the inference bears more similarity to the original class sample, distinctly different from the EM traces of the target class samples.  Therefore, it is easier to detect the adversarial.
While for $L_{inf}$ attack, many pixels are changed, randomly distributed, easily visualized in Fig.~\ref{fig: adversarial-examples} (d).
The EM trace will differ both from that of the source class and that of the target class, still caught easily by the adversarial detector. 
For $L_2$ attack, there are more pixels changed than the $L_1$ attack, but around the object in the image rather than randomly distributed like in $L_{inf}$, making the EM trace somewhat between those of the source class and target class and causing
the anomaly detector low confidence in making the prediction.

\ad{the above description does not seem to fit the visualization in the figure. I wrote a paragraph below.}
Figure~\ref{fig: embedding} shows the 3-D embedding of the logits traces of PGD $L_2$ attack versus benign logits traces of two classes. We can observe that the benign logits traces of two classes are well separated, indicating that the victim model inference flows are distinct for these two classes. The adversarial samples result in inference flows different from either of the benign class inference flows, and the resulting logits traces would be detected as anomaly.
\fi

\subsection{Comparison with Other Methods} \label{sec: exp: comparison}
\ry{This section focuses on Table 6, compare \name with other methods in terms of detection performance. The comparison conclusion includes 1. the reason we get a higher accuracy than others(contradiction between discriminative features and semantic features. 2. Difference between PGD and CW. 3. Finds about class.)}
Table~\ref{tab: comparison} shows our comparison between the hardware-based EMShepherd with state-of-the-art software detection methods. Note that our detector is under a stricter `black-box' scenario where only the EM traces of model execution along with the model prediction output are available. 
We control the False Positive (FP) rate under $10\%$ and evaluate the detection rates under targeted PGD $L_2$ attacks and CW $L_2$ attacks on all the  $10$ classes.
We draw three major conclusions.
\begin{itemize}[leftmargin=*]
    \item \name outperforms all baseline methods in the detection of PGD and CW attacks,
    with a $94\%$ detection rate on average.
    This demonstrates that our \name successfully captures the different computations of the model inference for benign and adversarial samples.
    \item The detection performance varies in different target classes.
    PGD attacks on Class 0, 4, and 6 cannot be easily detected, due to the relatively lower EM classification accuracies for these three classes (See Table~\ref{tab: resultofB0}). 
    \item Our detector performs consistently across the two different attacks, while the performance of other methods varies significantly for the two attacks. 
    PGD attacks can be effectively detected by MagNet, which utilizes only testing inputs (semantic information) from the victim inputs.
    On the other hand, NIC, which focuses on the execution flow, has a better performance on CW adversarial samples.
Our method is more general as it obtains both of such information from EM traces.\yf{rephrase the last line? be precise about semantic information? Semantic information of what? the input image??}
\end{itemize}

\begin{table*}[t]
    \centering
    \small
    \caption{Detection Rate(\%) when $FPR=10\%$}
    \label{tab: comparison}
    \begin{tabularx}{\linewidth}{c?llllllllll?llllllllll}
        \toprule
        \multirow{ 2}{*}{Method}
        & \multicolumn{10}{c}{PGD $L_2$ Targeted Class}
        & \multicolumn{10}{c}{CW $L_2$ Targeted Class}\\
        \cline{2-21}
        & 0 & 1 & 2 & 3 &4 & 5 & 6 & 7 & 8 & 9 & 0 & 1 & 2 & 3 &4 & 5 & 6 & 7 & 8 & 9\\
        \hline
        EM &  74.9 & \textbf{100} & \textbf{100} & \textbf{99.5} & 67.4 & \textbf{100} & 68.0 & \textbf{100} & \textbf{99.9} & \textbf{100} &
          72.4 & \textbf{100} & \textbf{100} & \textbf{100} & \textbf{100} & \textbf{100} & 65.6 & \textbf{100} & \textbf{100} & \textbf{100}\\
        \hline
        KDE & 57.5 & 53.8 & 51.9 & 42.4 & 50.6 & 57.3 & 50.7 & 56.3 & 52.8 & 65.2 &
        49.1 & 59.8 & 48.0 & 45.9 & 54.5 & 69.6 & 44.7 & 68.8 & 60.7 & 74.6 \\
        \hline
        NIC & 55.0 & 55.8 & 52.8 & 43.0 & 41.4 & 57.2 & 50.0 & 83.0 & 47.0 & 56.9 &
        \textbf{82.0} & 81.7 & 77.3 & 86.5 & 79.6 & 81.0 & \textbf{74.0} & 82.5 & 87.8 & 85.6\\
        \hline
        FS & 64.4 & 51.2 & 72.1 & 60.8 & 69.9 & 40.1 & 66.8 & 39.6 & 65.2 & 50.1 &
         69.5 & 62.0 & 68.0 & 62.8 & 79.0 & 66.3 & 66.8 & 73.9 & 67.4 & 69.5 \\
        \hline
        MagNet & \textbf{88.0} & 87.2& 81.2 & 88.2 & \textbf{82.0} & 84.5 & \textbf{87.4} & 87.2 & 88.1 & 85.9 &
        67.2 & 62.3 & 63.2 & 62.7 & 65.3 & 80.1 & 65.2 & 74.3 & 68.4 & 68.8\\
        \hline
    \end{tabularx}

\end{table*}

\yf{at the end of Section 4.4, you need to add a line to discuss your experiments when changing the latent feature space of VAE. similar to the STFT,  how many you have tried? What is the impact? NO figure is fine, but need to be stated}


\yf{\textcolor{red}{If the (a) and (b) figures are not related, better go with different figures using the minipage setup e.g.,  Fig. 14 and Fig. 15. The colors for FSGM and Original have to be distinctly different}}

\subsection{Adversarial Detection for Robust Models} \label{sec: exp: robust}
\begin{figure}[t]
    \begin{minipage}{.33\textwidth}
        \centering
        \includegraphics[width=\linewidth]{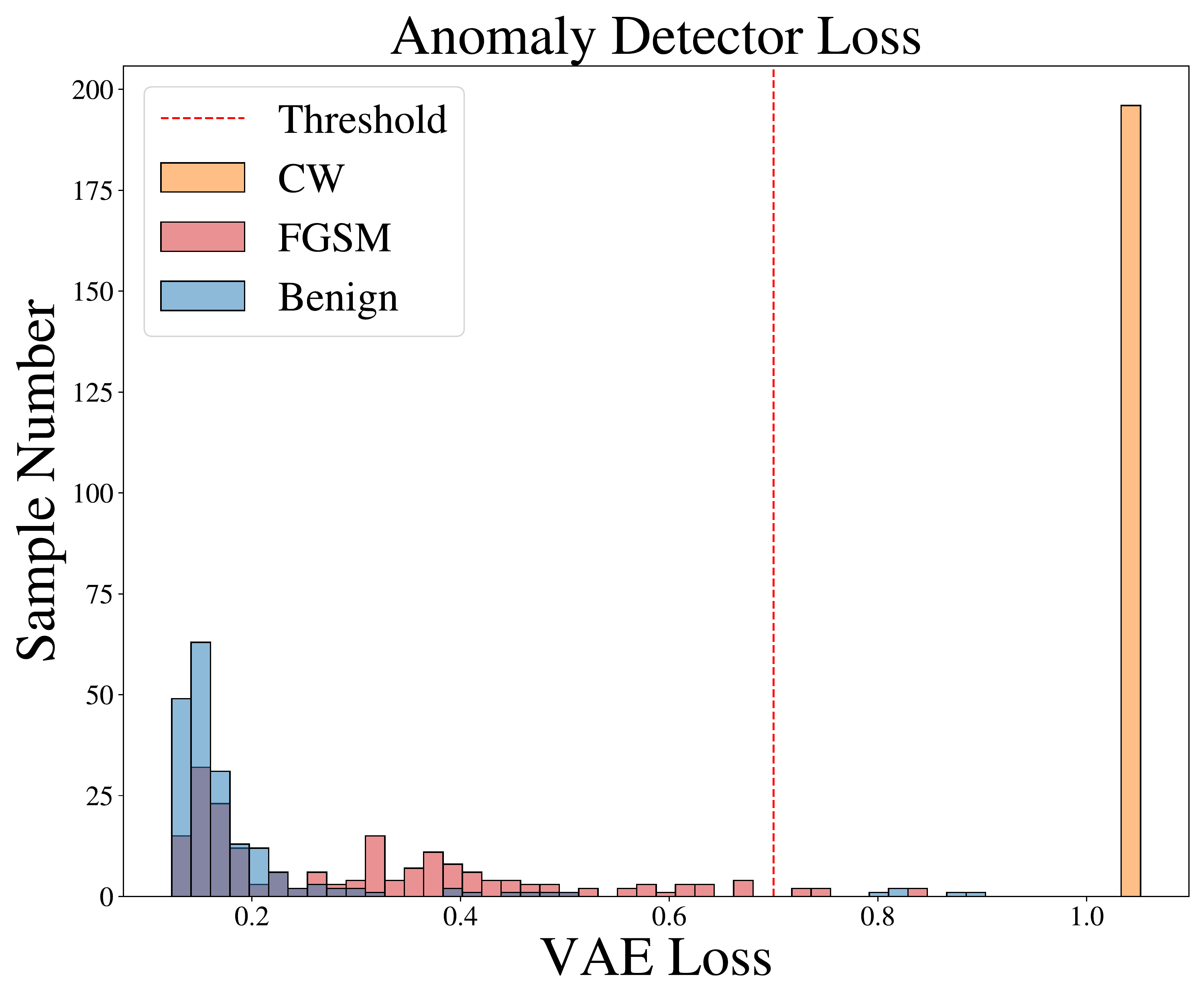}
        \caption{Robust VAE Loss}
        \label{fig: robust vae loss}
        \Description{The Anomaly detector loss of Robust Model}
    \end{minipage}
    \vspace{-3mm}
\end{figure}

The EMShepherd framework is model-agnostic and should also work for robust models enhanced with adversarial defense mechanisms, such as adversarial training.  
The robust model is only resilient to adversarial examples similar to the ones used in retraining, and may be circumvented by other unknown adversarial examples or
maliciously-designed adaptive attacks (stronger adversarial examples).  We evaluate the effectiveness of our detector on a robust model under a different adversarial attack. 
We train a robust LeNet-5 CNN model with benign samples and adversarial examples (with the correct labels) generated by the FGSM method.
Such a robust model is weak against the CW attack, while EMShepherd succeeds in detecting the CW adversarial examples.
We evaluate the detection performance on a testing dataset with benign, FGSM (regarded as noisy benign), and targeted CW examples, and the VAE loss distributions are presented in Fig.~\ref{fig: robust vae loss}.
Our findings are as follows: 
\begin{itemize}[leftmargin=*]
    \item \name can detect the stronger adversarial samples (yellow bars in Fig.~\ref{fig: robust vae loss}) and most of the benign examples correctly (blue and magenta bars).
    Using the threshold of 0.7, the DR of the targeted CW attack is $100\%$ when the FPR on unattacked samples (benign and FGSM) is $2.6\%$.
    \item It is noticeable that the FPR of FGSM samples is $5.1\%$.
    Note that although \name for the robust model is trained with only benign samples, it still correctly classifies FGSM samples, consistent with the robust model.
\end{itemize}


\yf{\textcolor{red}{elaborate the last two lines.}} \ad{I am confused by last item. Do you mean that FGSM samples mostly give correct class labels by the robust victim model, and EMShepherd correctly identify them as benign even though only trained on traces of benign not FGSM samples? Do not see why you can FGSM samples as noise?}

\subsection{Adversarial Detection for VGG Model} \label{sec: exp: cifar10}
To show the scalability of the \name framework, we further apply it to a VGG-like model on CIFAR-10 dataset and evaluate the adversarial detection performance.
The results show that our framework can cope with large victim model execution on more complex datasets.

\begin{figure*}[t]
    \centering
    \subcaptionbox{An Example of CIFAR-10 EM traces}[.76\linewidth][c]{%
    \includegraphics[width=\linewidth]{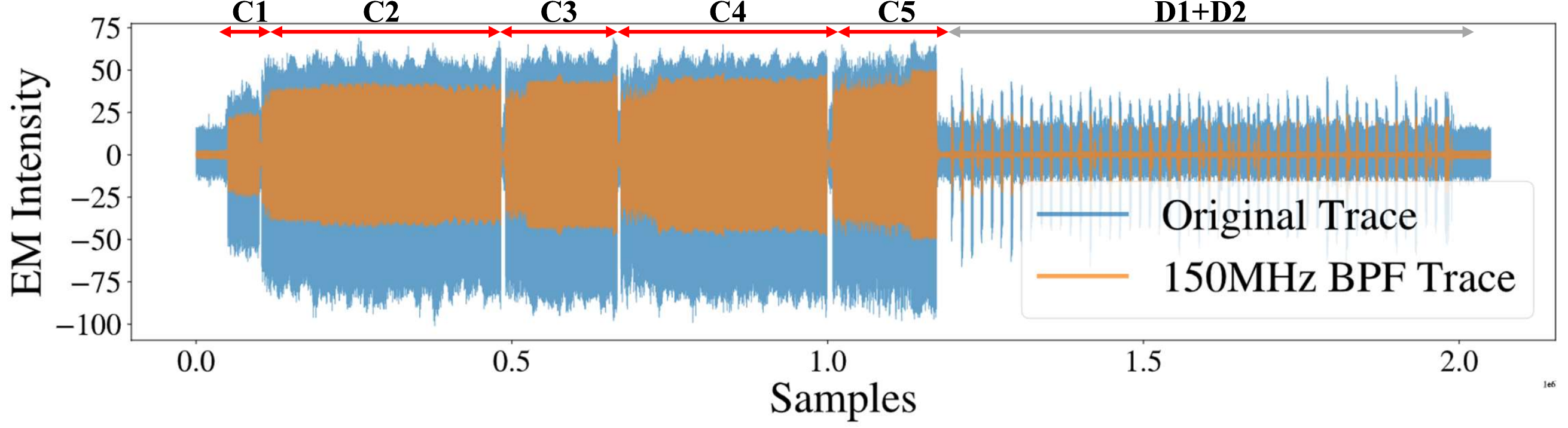}}\quad
    \subcaptionbox{Multiplications}[.22\linewidth][c]{%
    \includegraphics[width=\linewidth]{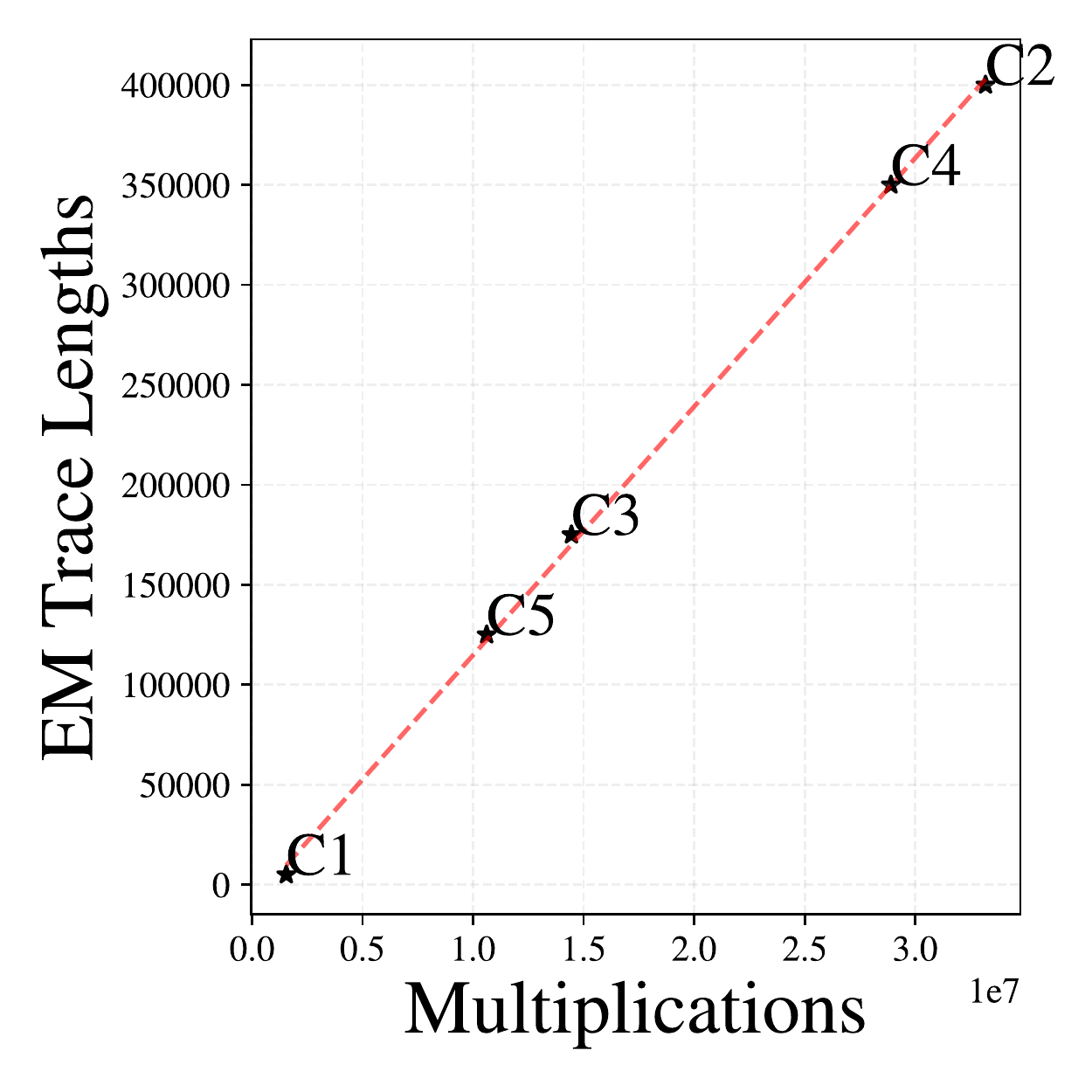}}
    \Description{An example CIFAR-10 EM trace}
    \caption{CIFAR-10 EM trace and layer operations}
    \label{fig: cifar10-trace}
    \vspace{-3mm}
\end{figure*}
\yf{\textcolor{red}{Fig.~\ref{fig: cifar10-trace}(b) figure can be represented better by a sort of correlation plot: X axis is the \# of multiplications and the Y-axis is the execution time, and you have five points in the plot with each labeled by ``layer i"}}

\noindent\textbf{EM Traces of VGG Model Execution on CIFAR-10:}
Compared with the grayscale Fashion MNIST, the CIFAR-10 dataset includes colored images used for objection detection.
The victim model and our EM trace collector both have to change accordingly.
\begin{itemize}[leftmargin=*]
    \item \textbf{Larger victim model:} The size of CIFAR-10 images is $32\times 32\times 3$, requiring more sophisticated models. Due to the limited resources on DPU, we choose a VGG-like model for implementation.
    The model includes $7$ layers: $5$ consecutive convolutional layers followed by $2$ dense layers, which achieves testing accuracy $90.5\%$ on CIFAR-10 ($93.6\%$ by the benchmark VGG-16~\cite{vgg}).
    It uses Tensorflow2 building of Ultra96 with a working frequency of $150$MHz.
    \item \textbf{Lower EM sampling frequency:} Due to the increase of execution time, the length of CIFAR-10 EM traces are longer than the Fashion MNIST one.
    Fig.~\ref{fig: cifar10-trace}(a) shows an example CIFAR-10 EM trace under a sampling frequency $1$ GHz.
    The blue part is the raw EM signal and the orange part stands for the signals after a bandpass filter at the DPU operating frequency.
    \item \textbf{Layer-wise separation:} As annotated on Fig.~\ref{fig: cifar10-trace}(a),
    the EM trace can be partitioned into $5$ convolutional layer segments (C1-C5) and $2$ dense layer segments (D1 and D2).
    We use C1-C5 for building the EM classifiers as they are for computations and easily distinguished by the data transmission segments between two computation ones.
    We find that the length of each EM segment (layer execution) is directly proportional to the number of multiplications in that layer, and the results are presented in
   Fig.~\ref{fig: cifar10-trace}(b), where C1-C5 are marked accordingly.
\end{itemize}

\noindent\textbf{CIFAR-10 Layer-wise EM Classifiers:}
Table~\ref{tab: cifar-em-accuracy} shows the classification performance of layer-wise EM classifiers on CIFAR-10.
 It shows that around half of EM traces can be correctly classified.  
 The reason for CIFAR-10 EM classifiers with lower prediction accuracies than Fashion MNIST ones is because of inherent characteristics of the datasets.
    CIFAR-10 image is 3-channel RGB while Fashion MNIST image is 1-channel Grayscale.  The VGG model used for CIFAR-10 is deeper and larger than LeNet-5.
    The resolution of the EM traces is much lower due to the lower sampling frequency and limited storage/processing capabilities.
 Comparing the different layer segments, we find that  $C_5$, the last convolutional layer has the lowest prediction accuracy.
    Because $C_5$ has a larger receptive field and a large number of kernels running in parallel, many neurons and activation are concurrent and time points on the EM traces bear low signal-to-noise ratios. 
Note although the classifiers achieve lower accuracy than the previous Fashion MNIST cases, these classifiers
 are sufficient for the follow-on anomaly detectors to catch adversarial examples, as we have analyzed it is the deviation of classifiers' logits that is the characteristics of adversarial examples, i.e., a relative value instead of absolute accuracy.

\begin{table}[ht]
    \centering
      \caption{CIFAR-10 EM Classifiers Performance}
    \label{tab: cifar-em-accuracy}
    \begin{tabularx}{\linewidth}{cXXXXX}
         \toprule
         Convolutional layer & C1 & C2 & C3 & C4 & C5\\
        \hline
        Accuracy & 0.42 & 0.46 & 0.43 & 0.49 & 0.27\\
        \hline
    \end{tabularx}
    \vspace{-0.3cm}
\end{table}

\noindent\textbf{CIFAR-10 Anomaly Detector:}
We evaluate the performance of our anomaly detector on CIFAR-10 EM classifiers.
The experimental results show that our detection framework still achieves fairly good performance on the colored CIFAR-10 dataset.
The logits from all $5$ segments (convolutional layers) are utilized as inputs for the VAE anomaly detector, against targeted PGD attacks on CIFAR-10.
Fig.~\ref{fig: cifar10 vae loss} shows the VAE loss of the vectors of logits for both benign samples and adversarial examples.
With an optimal threshold selected, the detection accuracy for the adversarial examples is close to 100\% with some false positives on the benign examples. 
 Fig.~\ref{fig: cifar10 PR Curve} shows the precision-recall curves for different classes:
    the best detection result is from target class 4 (deer) with the F1-score of $0.906$ and the worst one is class 7 (horse) with the F1-score of $0.821$.

\aptLtoX[graphics=no, type=html]{\begin{figure}[t]
    \centering
    \begin{minipage}{.22\textwidth}
        \includegraphics[width=\linewidth]{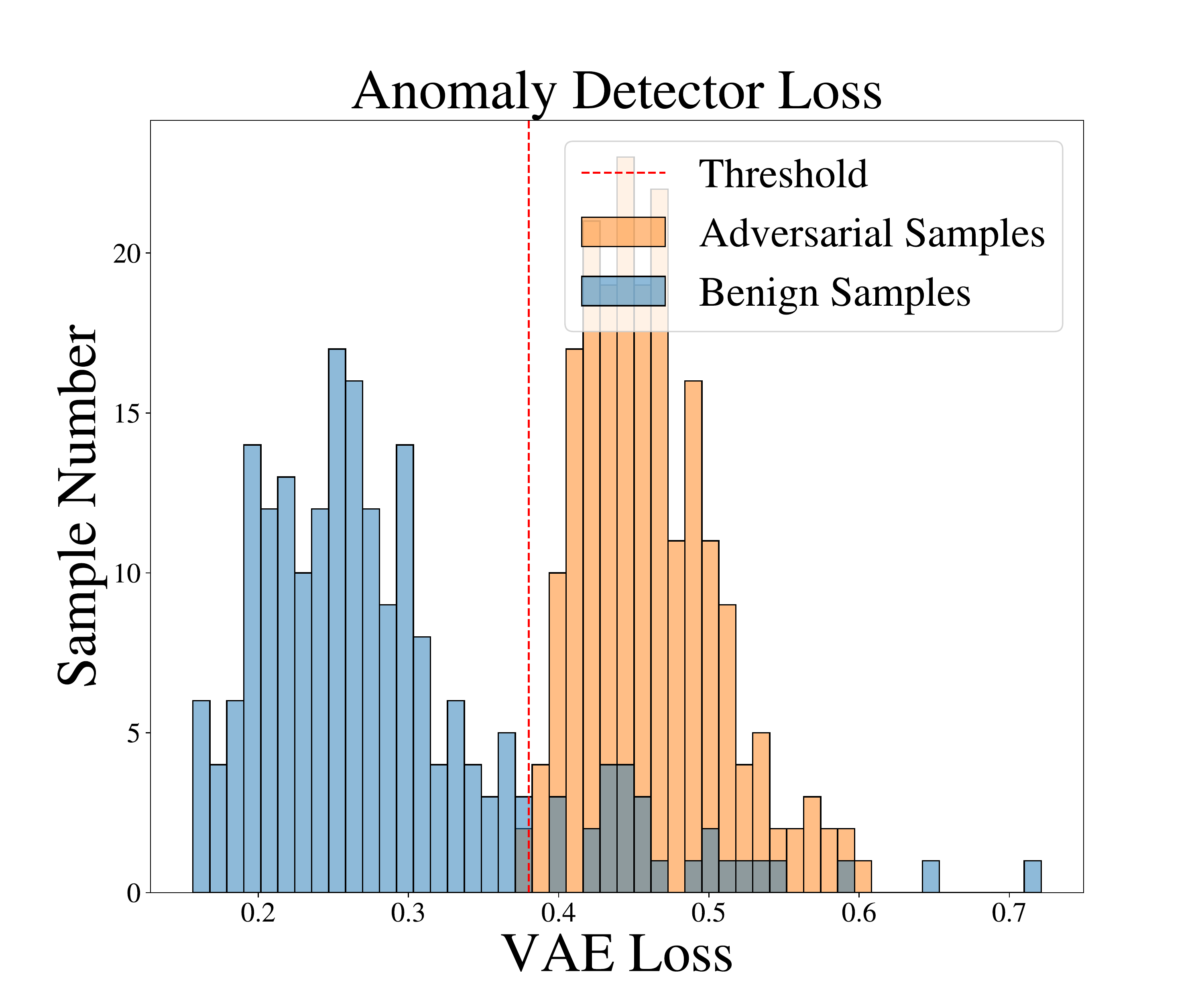}
        \caption{CIFAR-10 VAE Loss}
        \label{fig: cifar10 vae loss}
        \Description{The Anomaly detector loss of CIFAR10 Dataset}
    \end{minipage}
\end{figure}}{}

\aptLtoX[graphics=no, type=html]{\begin{figure}[t]
    \begin{minipage}{.23\textwidth}
        \includegraphics[width=\linewidth]{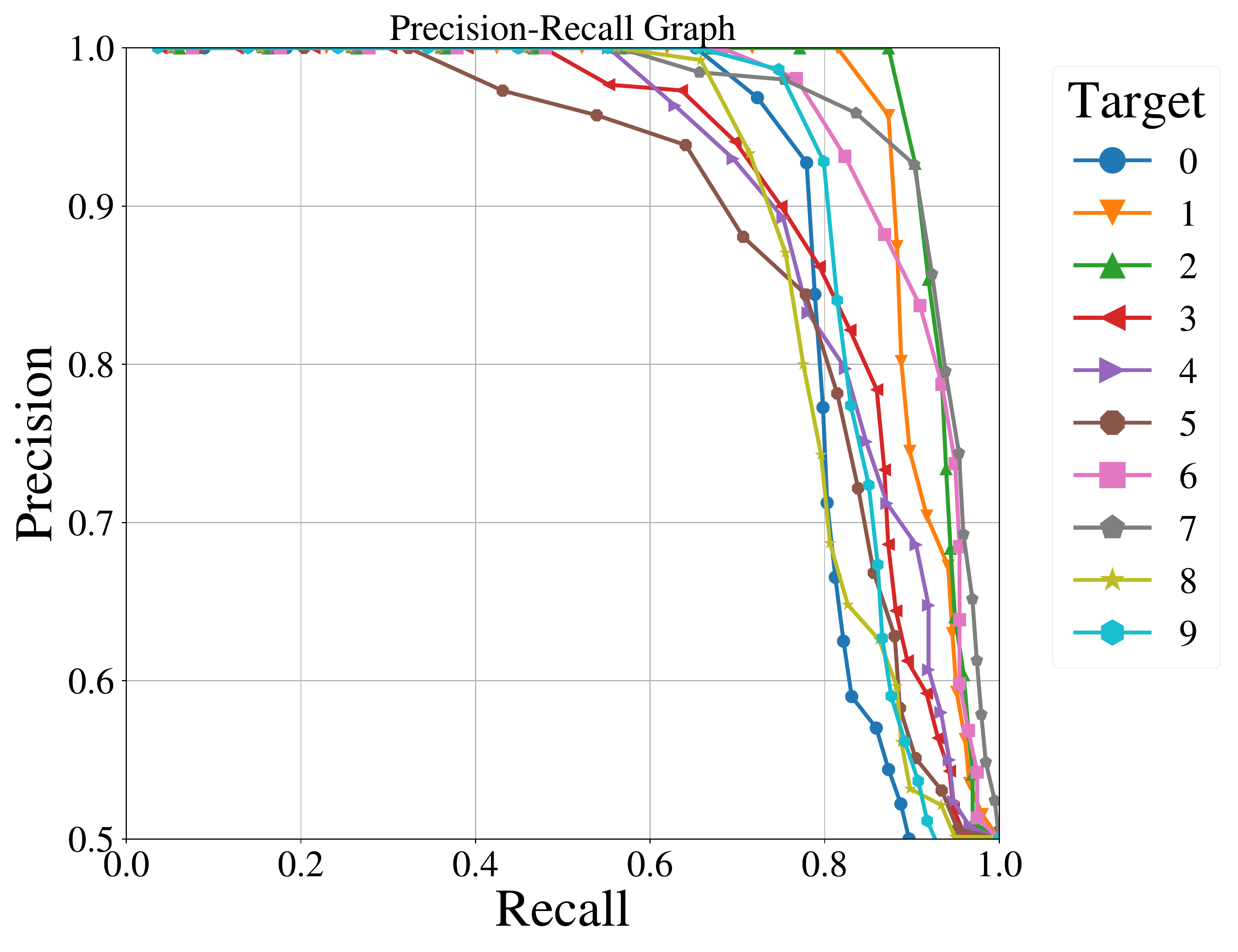}
        \caption{CIFAR PRcurves}
        \label{fig: cifar10 PR Curve}
        \Description{The Precision-Recall Curve of CIFAR10 Dataset}
    \end{minipage}
    \vspace{-3mm}
\end{figure}}{}

\aptLtoX[graphics=no, type=html]{}{\begin{figure}[t]
    \centering
    \begin{minipage}{.22\textwidth}
        \includegraphics[width=\linewidth]{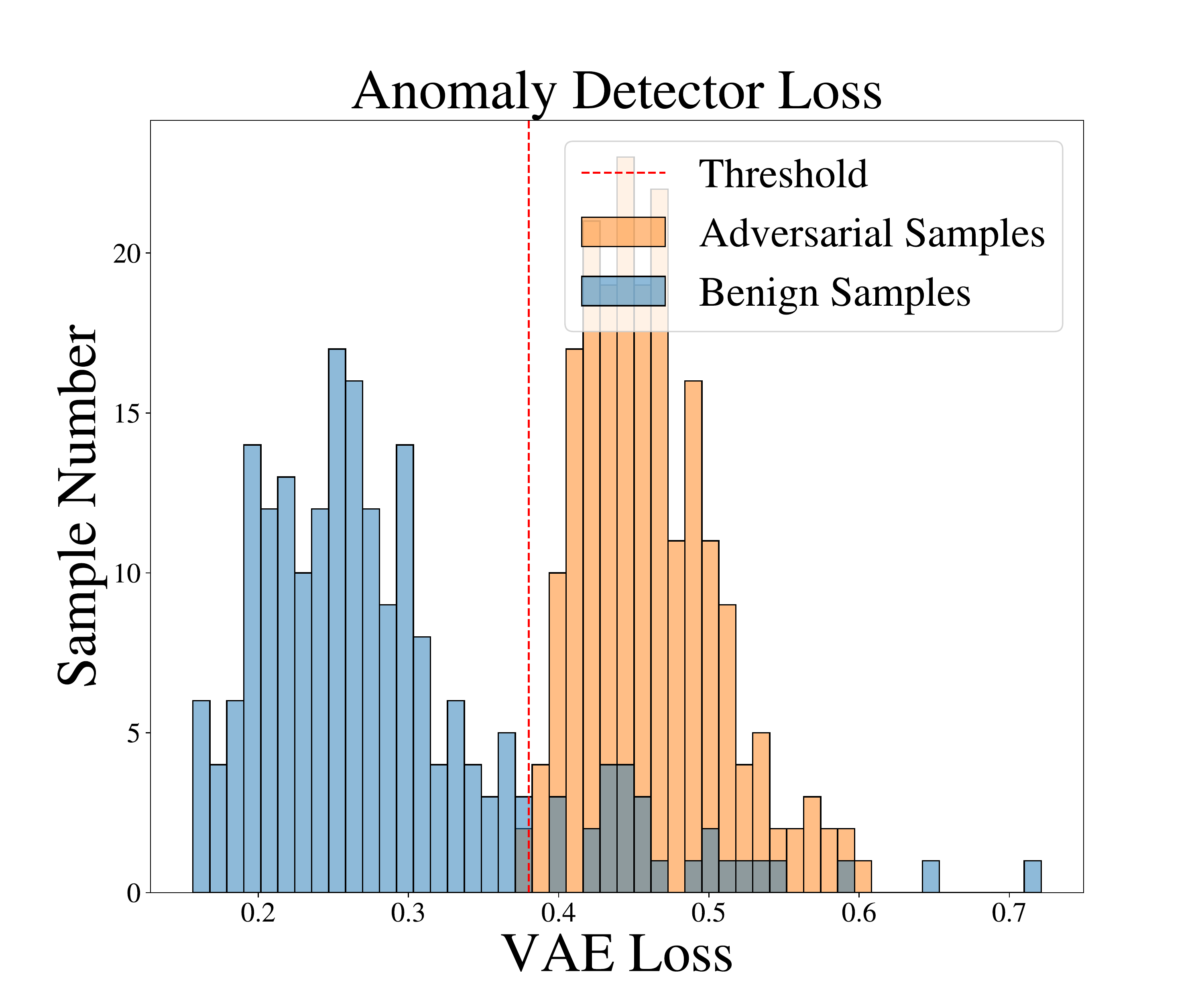}
        \caption{CIFAR-10 VAE Loss}
        \label{fig: cifar10 vae loss}
        \Description{The Anomaly detector loss of CIFAR10 Dataset}
    \end{minipage}
    \begin{minipage}{.23\textwidth}
        \includegraphics[width=\linewidth]{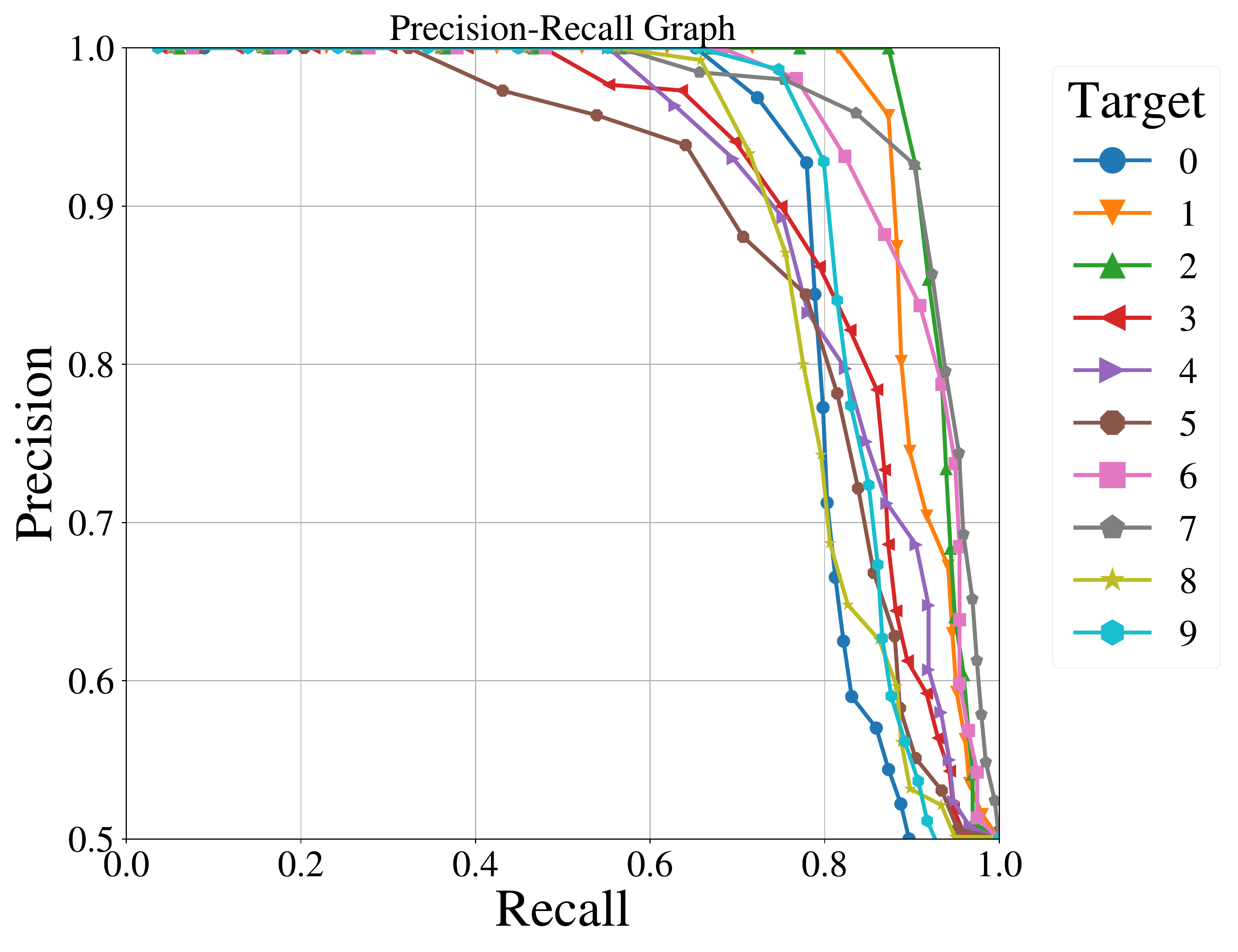}
        \caption{CIFAR PRcurves}
        \label{fig: cifar10 PR Curve}
        \Description{The Precision-Recall Curve of CIFAR10 Dataset}
    \end{minipage}
    \vspace{-3mm}
\end{figure}}

\section{Conclusion and Future Work} \label{sec: conclusion}
In this work, we propose a novel adversarial detection framework, EMShepherd, leveraging the EM side-channel of model execution. 
EM traces embody rich input (class)-dependent inference information, well suited for classification and anomaly detection. 
Our framework extracts EM feature invariants for different classes and use them for unsupervised anomaly detection. 
The adversarial detector can be deployed as an air-gapped, third-party, PnP system in the proximity of the victim system in operation. It is totally passive and noninvasive without probing the model execution or retraining the model. 
The performance of our black-box adversarial detector is comparable to the state-of-the-art software-based white-box detection method, but has a much broader and more general application to diverse DNN implementations and applications.

Our future work will adapt the framework for the detection of more attacks, such as Trojan attacks, backdoor attacks, and data poisoning attacks. 
The EM side-channel leakage of deep learning engines during execution can be further leveraged for more applications, e.g., membership inference attacks where the input categories are reverse engineered. 

\if false 
\noindent\textbf{EM Side Channel Attack} 
Previous studies on DNN EM side-channel attacks explore the possibility of recovering the model structures and parameters~\cite{yu2020deepem}. 
In addition, we also find that EM emanation may also contain category dependency.
It is possible to explore data confidentiality attacks such as membership inference attacks~\cite{shokri2017membership}.

\noindent\textbf{Adaptive Adversaries}
Software-based detection methods should consider adaptive adversarial attacks when the attacker realizes the detection framework and adjusts inputs to fool the detector. 
Such attacks may encounter difficulties when targets at EMShepred, where the adaptive adversarial samples are too complex to craft. 
Future researchers can study the insights of how embedding devices leak computation patterns via EM for effective adaptive attacks.

\noindent\textbf{Data Poisoning Attacks} 
The trojan attack is another insidious variant of data poisoning attacks, which exploits a backdoor to DNN models to misclassify any inputs signed with the trojan trigger. 
EMShepred can be extended to other data poisoning attacks on DNN to build a comprehensive attack defense system.
\fi

\appendix

\section{\name GradCAM}
\label{sec: gradCAM}

\begin{figure}[htb]
    \centering
    \subcaptionbox{Spectrogram0}[.315\linewidth][c]{%
    \includegraphics[width=\linewidth]{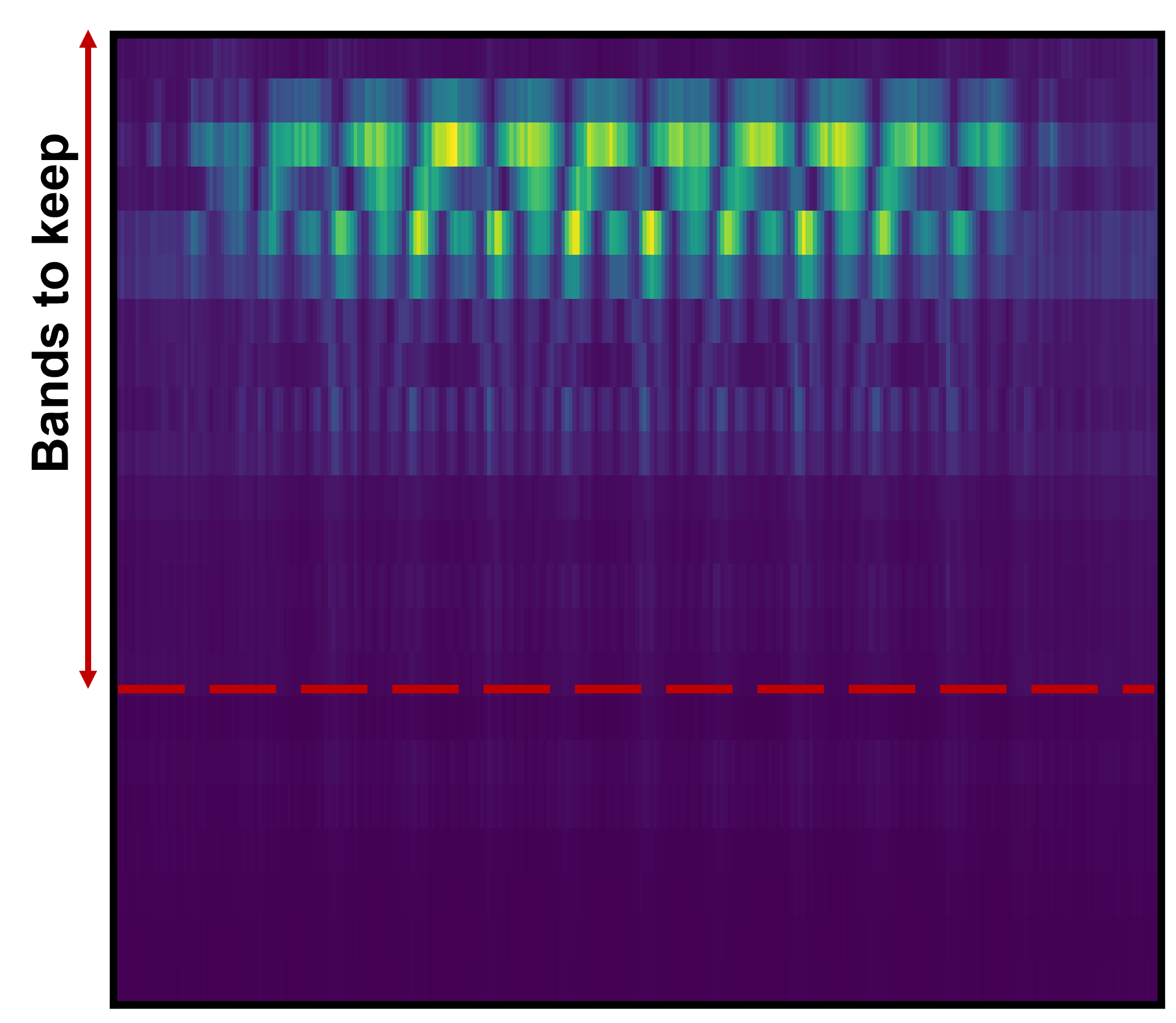}}\quad
    \subcaptionbox{Spectrogram1}[.295\linewidth][c]{%
    \includegraphics[width=\linewidth]{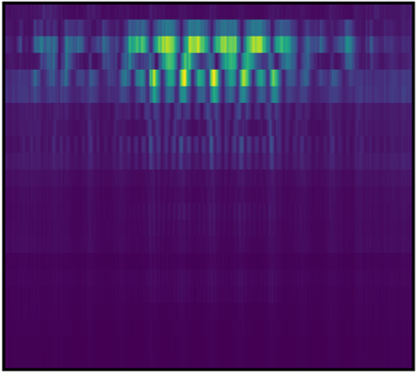}}\quad
    \subcaptionbox{Spectrogram5}[.3\linewidth][c]{%
    \includegraphics[width=\linewidth]{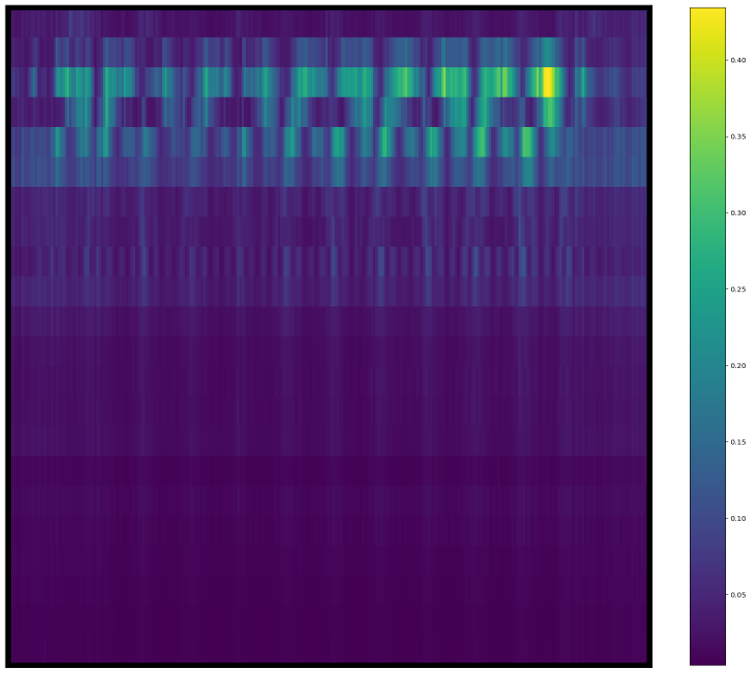}}\quad
    \subcaptionbox{GradCAM0}[.31\linewidth][c]{%
    \includegraphics[width=\linewidth]{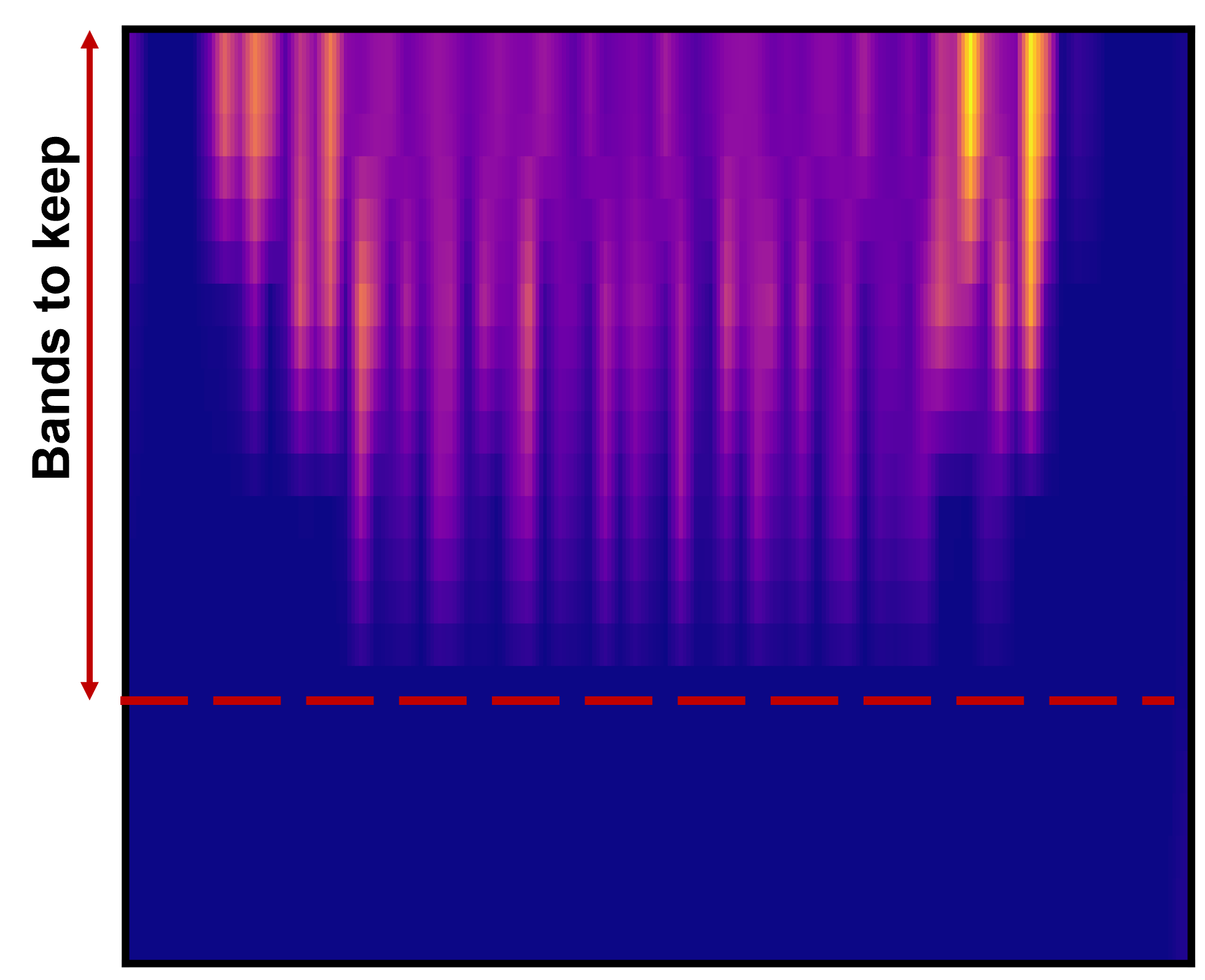}}\quad
    \subcaptionbox{GradCAM1}[.28\linewidth][c]{%
    \includegraphics[width=\linewidth]{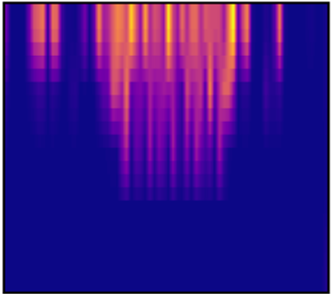}}\quad
    \subcaptionbox{GradCAM5}[.32\linewidth][c]{%
    \includegraphics[width=\linewidth]{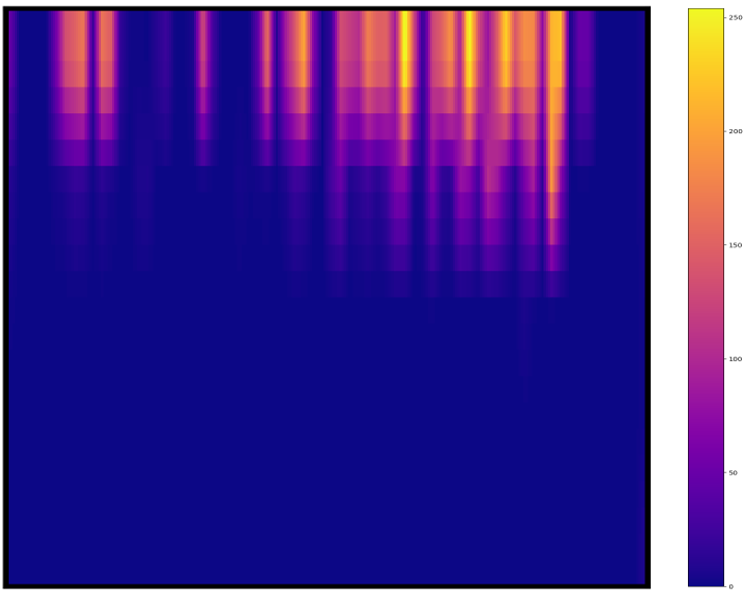}}\quad
    \caption{The average spectrogram and GradCAM results of the first EM segment from Class 0, 1, 5}
    \Description{The sample spectrogram and GradCAM reuslts of the first batch from Class 0, 1, 5.
    The figure(a)(b)(c) are the average spectrogram of samples from corresponding classes.
    The figure(d)(e)(f) are those desired classes' GradCAM results, indicating which part of image the model is looking to for make the decision.
    The model focus on the high light envelope of spectrograms to make its decision.}
    \label{fig: gradCAM}
    \vspace{-0.3cm}
\end{figure}

The GradCAM uses the gradient information flowing into the last convolutional layer of the classifier to assign importance values to each neuron for a particular decision of interest.
\yf{exaplain GradCAM.}
Fig.~\ref{fig: gradCAM} (a)-(c) present the average spectrogram of the first EM segment for Class 0, 1, and 5, respectively, \yf{which segment?}
and Figure~\ref{fig: gradCAM} (d)-(f) show the corresponding coarse GradCAM localization (red heatmaps).
The GradCAM heatmaps illustrate the sensitivity (magnitude of gradient) of the EM classification on the input neurons, 
which are the pixels on the spectrogram here.
The most sensitive (bright) parts are where the EM classifier focuses on when making a decision.
For example, the primary signal of Class 1 is from the middle part of the spectrogram as shown in Fig.~\ref{fig: gradCAM} (b), and correspondingly GradCAM in Fig.~\ref{fig: gradCAM} (e) shows that our classifier also emphasizes the central part.
The GradCAM heatmaps also guide the defender to improve the EM segments' pre-processing by selecting the significant frequency bands.

\balance

\bibliographystyle{ACM-Reference-Format}
\bibliography{references}


\begin{thebibliography}{66}


\ifx \showCODEN    \undefined \def \showCODEN     #1{\unskip}     \fi
\ifx \showDOI      \undefined \def \showDOI       #1{#1}\fi
\ifx \showISBNx    \undefined \def \showISBNx     #1{\unskip}     \fi
\ifx \showISBNxiii \undefined \def \showISBNxiii  #1{\unskip}     \fi
\ifx \showISSN     \undefined \def \showISSN      #1{\unskip}     \fi
\ifx \showLCCN     \undefined \def \showLCCN      #1{\unskip}     \fi
\ifx \shownote     \undefined \def \shownote      #1{#1}          \fi
\ifx \showarticletitle \undefined \def \showarticletitle #1{#1}   \fi
\ifx \showURL      \undefined \def \showURL       {\relax}        \fi
\providecommand\bibfield[2]{#2}
\providecommand\bibinfo[2]{#2}
\providecommand\natexlab[1]{#1}
\providecommand\showeprint[2][]{arXiv:#2}

\bibitem[FSc(2019)]%
        {FScoreDe2:online}
 \bibinfo{year}{2019}\natexlab{}.
\newblock \bibinfo{title}{F-Score Definition | DeepAI}.
\newblock
  \bibinfo{howpublished}{\url{https://deepai.org/machine-learning-glossary-and-terms/f-score}}.
\newblock
\newblock
\shownote{(Accessed on 01/23/2022)}.


\bibitem[int(2020)]%
        {intelrapl}
 \bibinfo{year}{2020}\natexlab{}.
\newblock \bibinfo{title}{Running Average Power Limit Energy Reporting}.
\newblock
  \bibinfo{howpublished}{\url{https://www.intel.com/content/www/us/en/developer/articles/technical/software-security-guidance/advisory-guidance/running-average-power-limit-energy-reporting.html}}.
\newblock


\bibitem[Zyn(2020)]%
        {ZynqDPUv92:online}
 \bibinfo{year}{2020}\natexlab{}.
\newblock \bibinfo{title}{Zynq DPU v3.2 IP Product Guide}.
\newblock
  \bibinfo{howpublished}{\url{https://www.xilinx.com/support/documentation/ip_documentation/dpu/v3_2/pg338-dpu.pdf}}.
\newblock
\newblock
\shownote{(Accessed on 01/23/2022)}.


\bibitem[Bib(2021)]%
        {BibEntry2021Dec}
 \bibinfo{year}{2021}\natexlab{}.
\newblock \bibinfo{title}{{TensorFlow}}.
\newblock
\newblock
\urldef\tempurl%
\url{https://www.tensorflow.org}
\showURL{%
\tempurl}
\newblock
\shownote{[Online; accessed 24. Jan. 2022]}.


\bibitem[PYN(2021)]%
        {PYNQ}
 \bibinfo{year}{2021}\natexlab{}.
\newblock \bibinfo{title}{{Welcome to Ultra96-PYNQ{'}s documentation!
  {\ifmmode---\else\textemdash\fi} Ultra96-PYNQ v2.6 documentation}}.
\newblock
\newblock
\urldef\tempurl%
\url{https://ultra96-pynq.readthedocs.io/en/latest}
\showURL{%
\tempurl}
\newblock
\shownote{[Online; accessed 23. Jan. 2022]}.


\bibitem[pro(2022)]%
        {probe}
 \bibinfo{year}{2022}\natexlab{}.
\newblock \bibinfo{title}{{Probe Set PBS 2 (incl. Preamplifier)}}.
\newblock
\newblock
\urldef\tempurl%
\url{https://aaronia-shop.com/products/probe-set-pbs-2-incl-preamplifier}
\showURL{%
\tempurl}
\newblock
\shownote{[Online; accessed 23. Jan. 2022]}.


\bibitem[Ult(2022)]%
        {Ultra96-V2}
 \bibinfo{year}{2022}\natexlab{}.
\newblock \bibinfo{title}{{Ultra96-V2 {$\vert$} Avnet Boards}}.
\newblock
\newblock
\urldef\tempurl%
\url{https://www.avnet.com/wps/portal/us/products/avnet-boards/avnet-board-families/ultra96-v2}
\showURL{%
\tempurl}
\newblock
\shownote{[Online; accessed 23. Jan. 2022]}.


\bibitem[Bib(2022)]%
        {BibEntry2022Jan}
 \bibinfo{year}{2022}\natexlab{}.
\newblock \bibinfo{title}{{Vitis AI}}.
\newblock
\newblock
\urldef\tempurl%
\url{https://www.xilinx.com/products/design-tools/vitis/vitis-ai.html}
\showURL{%
\tempurl}
\newblock
\shownote{[Online; accessed 24. Jan. 2022]}.


\bibitem[Agrawal et~al\mbox{.}(2002a)]%
        {agrawal2002side}
\bibfield{author}{\bibinfo{person}{Dakshi Agrawal}, \bibinfo{person}{Bruce
  Archambeault}, \bibinfo{person}{Josyula~R Rao}, {and} \bibinfo{person}{Pankaj
  Rohatgi}.} \bibinfo{year}{2002}\natexlab{a}.
\newblock \showarticletitle{The {EM} side-channel (s)}. In
  \bibinfo{booktitle}{\emph{Int. Workshop on Cryptographic Hardware \& Embedded
  Systems}}. \bibinfo{pages}{29--45}.
\newblock


\bibitem[Agrawal et~al\mbox{.}(2002b)]%
        {emsca}
\bibfield{author}{\bibinfo{person}{D. Agrawal}, \bibinfo{person}{B.
  Archambeault}, \bibinfo{person}{J.~R. Rao}, {and} \bibinfo{person}{P.
  Rohatgi}.} \bibinfo{year}{2002}\natexlab{b}.
\newblock \showarticletitle{The EM side-channels}. In
  \bibinfo{booktitle}{\emph{Int. WkShp on Cryptographic Hardware \& Embedded
  Systems}}.
\newblock


\bibitem[Batina et~al\mbox{.}(2019)]%
        {batina2019csi}
\bibfield{author}{\bibinfo{person}{Lejla Batina}, \bibinfo{person}{Shivam
  Bhasin}, \bibinfo{person}{Dirmanto Jap}, {and} \bibinfo{person}{Stjepan
  Picek}.} \bibinfo{year}{2019}\natexlab{}.
\newblock \showarticletitle{$\{$CSI$\}$$\{$NN$\}$: Reverse Engineering of
  Neural Network Architectures Through Electromagnetic Side Channel}. In
  \bibinfo{booktitle}{\emph{{USENIX} Security Symp.}}
  \bibinfo{pages}{515--532}.
\newblock


\bibitem[Bhagoji et~al\mbox{.}(2017)]%
        {bhagoji2017dimensionality}
\bibfield{author}{\bibinfo{person}{Arjun~Nitin Bhagoji},
  \bibinfo{person}{Daniel Cullina}, {and} \bibinfo{person}{Prateek Mittal}.}
  \bibinfo{year}{2017}\natexlab{}.
\newblock \showarticletitle{Dimensionality reduction as a defense against
  evasion attacks on machine learning classifiers}.
\newblock \bibinfo{journal}{\emph{arXiv preprint arXiv:1704.02654}}
  \bibinfo{volume}{2} (\bibinfo{year}{2017}), \bibinfo{pages}{1}.
\newblock


\bibitem[Bhatnagar et~al\mbox{.}(2017)]%
        {CNNfmnist}
\bibfield{author}{\bibinfo{person}{Shobhit Bhatnagar},
  \bibinfo{person}{Deepanway Ghosal}, {and} \bibinfo{person}{Maheshkumar~H
  Kolekar}.} \bibinfo{year}{2017}\natexlab{}.
\newblock \showarticletitle{Classification of fashion article images using
  convolutional neural networks}. In \bibinfo{booktitle}{\emph{Int. Conf. on
  Image Information Processing (ICIIP)}}. \bibinfo{pages}{1--6}.
\newblock


\bibitem[Bojarski et~al\mbox{.}(2016)]%
        {bojarski2016end}
\bibfield{author}{\bibinfo{person}{Mariusz Bojarski}, \bibinfo{person}{Davide
  Del~Testa}, \bibinfo{person}{Daniel Dworakowski}, \bibinfo{person}{Bernhard
  Firner}, \bibinfo{person}{Beat Flepp}, \bibinfo{person}{Prasoon Goyal},
  \bibinfo{person}{Lawrence~D Jackel}, \bibinfo{person}{Mathew Monfort},
  \bibinfo{person}{Urs Muller}, \bibinfo{person}{Jiakai Zhang},
  {et~al\mbox{.}}} \bibinfo{year}{2016}\natexlab{}.
\newblock \showarticletitle{End to end learning for self-driving cars}.
\newblock \bibinfo{journal}{\emph{arXiv preprint arXiv:1604.07316}}
  (\bibinfo{year}{2016}).
\newblock


\bibitem[Carlini and Wagner(2017a)]%
        {carlini2017adversarial}
\bibfield{author}{\bibinfo{person}{Nicholas Carlini} {and}
  \bibinfo{person}{David Wagner}.} \bibinfo{year}{2017}\natexlab{a}.
\newblock \showarticletitle{Adversarial examples are not easily detected:
  Bypassing ten detection methods}. In \bibinfo{booktitle}{\emph{ACM Workshop
  on Artificial Intelligence \& Security}}. \bibinfo{pages}{3--14}.
\newblock


\bibitem[Carlini and Wagner(2017b)]%
        {carlini2017towards}
\bibfield{author}{\bibinfo{person}{Nicholas Carlini} {and}
  \bibinfo{person}{David Wagner}.} \bibinfo{year}{2017}\natexlab{b}.
\newblock \showarticletitle{Towards evaluating the robustness of neural
  networks}. In \bibinfo{booktitle}{\emph{IEEE Symp. on Security \& Privacy}}.
  IEEE, \bibinfo{pages}{39--57}.
\newblock


\bibitem[Chari et~al\mbox{.}(2002)]%
        {chari2002template}
\bibfield{author}{\bibinfo{person}{Suresh Chari}, \bibinfo{person}{Josyula~R
  Rao}, {and} \bibinfo{person}{Pankaj Rohatgi}.}
  \bibinfo{year}{2002}\natexlab{}.
\newblock \showarticletitle{Template attacks}. In
  \bibinfo{booktitle}{\emph{Int. Workshop on Cryptographic Hardware \& Embedded
  Systems}}. Springer, \bibinfo{pages}{13--28}.
\newblock


\bibitem[Chmielewski and Weissbart(2021)]%
        {chmielewski2021reverse}
\bibfield{author}{\bibinfo{person}{{\L}ukasz Chmielewski} {and}
  \bibinfo{person}{L{\'e}o Weissbart}.} \bibinfo{year}{2021}\natexlab{}.
\newblock \showarticletitle{On reverse engineering neural network
  implementation on gpu}. In \bibinfo{booktitle}{\emph{International Conference
  on Applied Cryptography and Network Security}}. Springer,
  \bibinfo{pages}{96--113}.
\newblock


\bibitem[Das et~al\mbox{.}(2019)]%
        {das2019x}
\bibfield{author}{\bibinfo{person}{Debayan Das}, \bibinfo{person}{Anupam
  Golder}, \bibinfo{person}{Josef Danial}, \bibinfo{person}{Santosh Ghosh},
  \bibinfo{person}{Arijit Raychowdhury}, {and} \bibinfo{person}{Shreyas Sen}.}
  \bibinfo{year}{2019}\natexlab{}.
\newblock \showarticletitle{{X-DeepSCA}: Cross-device deep learning side
  channel attack}. In \bibinfo{booktitle}{\emph{Proc. Design Automation Conf.}}
  \bibinfo{pages}{1--6}.
\newblock


\bibitem[Duchi et~al\mbox{.}(2011)]%
        {sgd}
\bibfield{author}{\bibinfo{person}{John Duchi}, \bibinfo{person}{Elad Hazan},
  {and} \bibinfo{person}{Yoram Singer}.} \bibinfo{year}{2011}\natexlab{}.
\newblock \showarticletitle{Adaptive subgradient methods for online learning
  and stochastic optimization.}
\newblock \bibinfo{journal}{\emph{Journal of machine learning research}}
  \bibinfo{volume}{12}, \bibinfo{number}{7} (\bibinfo{year}{2011}).
\newblock


\bibitem[Feinman et~al\mbox{.}(2017)]%
        {feinman2017detecting}
\bibfield{author}{\bibinfo{person}{Reuben Feinman}, \bibinfo{person}{Ryan~R
  Curtin}, \bibinfo{person}{Saurabh Shintre}, {and} \bibinfo{person}{Andrew~B
  Gardner}.} \bibinfo{year}{2017}\natexlab{}.
\newblock \showarticletitle{Detecting adversarial samples from artifacts}.
\newblock \bibinfo{journal}{\emph{arXiv preprint arXiv:1703.00410}}
  (\bibinfo{year}{2017}).
\newblock


\bibitem[Forsyth and Ponce(2011)]%
        {forsyth2011computer}
\bibfield{author}{\bibinfo{person}{David Forsyth} {and} \bibinfo{person}{Jean
  Ponce}.} \bibinfo{year}{2011}\natexlab{}.
\newblock \bibinfo{booktitle}{\emph{Computer vision: A modern approach.}}
\newblock \bibinfo{publisher}{Prentice hall}.
\newblock


\bibitem[Foster et~al\mbox{.}(2014)]%
        {foster2014machine}
\bibfield{author}{\bibinfo{person}{Kenneth~R Foster}, \bibinfo{person}{Robert
  Koprowski}, {and} \bibinfo{person}{Joseph~D Skufca}.}
  \bibinfo{year}{2014}\natexlab{}.
\newblock \showarticletitle{Machine learning, medical diagnosis, and biomedical
  engineering research-commentary}.
\newblock \bibinfo{journal}{\emph{Biomedical engineering online}}
  \bibinfo{volume}{13}, \bibinfo{number}{1} (\bibinfo{year}{2014}),
  \bibinfo{pages}{1--9}.
\newblock


\bibitem[Ganin et~al\mbox{.}(2016)]%
        {ganin2016domain}
\bibfield{author}{\bibinfo{person}{Yaroslav Ganin}, \bibinfo{person}{Evgeniya
  Ustinova}, \bibinfo{person}{Hana Ajakan}, \bibinfo{person}{Pascal Germain},
  \bibinfo{person}{Hugo Larochelle}, \bibinfo{person}{Fran{\c{c}}ois
  Laviolette}, \bibinfo{person}{Mario Marchand}, {and} \bibinfo{person}{Victor
  Lempitsky}.} \bibinfo{year}{2016}\natexlab{}.
\newblock \showarticletitle{Domain-adversarial training of neural networks}.
\newblock \bibinfo{journal}{\emph{J. Machine Learning Research}}
  \bibinfo{volume}{17}, \bibinfo{number}{1} (\bibinfo{year}{2016}),
  \bibinfo{pages}{2096--2030}.
\newblock


\bibitem[Goodfellow et~al\mbox{.}(2014)]%
        {goodfellow2014explaining}
\bibfield{author}{\bibinfo{person}{Ian~J Goodfellow}, \bibinfo{person}{Jonathon
  Shlens}, {and} \bibinfo{person}{Christian Szegedy}.}
  \bibinfo{year}{2014}\natexlab{}.
\newblock \showarticletitle{Explaining and harnessing adversarial examples}.
\newblock \bibinfo{journal}{\emph{arXiv preprint arXiv:1412.6572}}
  (\bibinfo{year}{2014}).
\newblock


\bibitem[Grosse et~al\mbox{.}(2017)]%
        {grosse2017statistical}
\bibfield{author}{\bibinfo{person}{Kathrin Grosse}, \bibinfo{person}{Praveen
  Manoharan}, \bibinfo{person}{Nicolas Papernot}, \bibinfo{person}{Michael
  Backes}, {and} \bibinfo{person}{Patrick McDaniel}.}
  \bibinfo{year}{2017}\natexlab{}.
\newblock \showarticletitle{On the (statistical) detection of adversarial
  examples}.
\newblock \bibinfo{journal}{\emph{arXiv preprint arXiv:1702.06280}}
  (\bibinfo{year}{2017}).
\newblock


\bibitem[He et~al\mbox{.}(2015)]%
        {he2015deep}
\bibfield{author}{\bibinfo{person}{Kaiming He}, \bibinfo{person}{Xiangyu
  Zhang}, \bibinfo{person}{Shaoqing Ren}, {and} \bibinfo{person}{Jian Sun}.}
  \bibinfo{year}{2015}\natexlab{}.
\newblock \bibinfo{title}{Deep Residual Learning for Image Recognition}.
\newblock
\newblock
\showeprint[arxiv]{1512.03385}~[cs.CV]


\bibitem[Howard et~al\mbox{.}(2017)]%
        {howard2017mobilenets}
\bibfield{author}{\bibinfo{person}{Andrew~G. Howard}, \bibinfo{person}{Menglong
  Zhu}, \bibinfo{person}{Bo Chen}, \bibinfo{person}{Dmitry Kalenichenko},
  \bibinfo{person}{Weijun Wang}, \bibinfo{person}{Tobias Weyand},
  \bibinfo{person}{Marco Andreetto}, {and} \bibinfo{person}{Hartwig Adam}.}
  \bibinfo{year}{2017}\natexlab{}.
\newblock \showarticletitle{MobileNets: Efficient Convolutional Neural Networks
  for Mobile Vision Applications}.
\newblock \bibinfo{journal}{\emph{aXiv:1704.04861}} (\bibinfo{year}{2017}).
\newblock


\bibitem[Huang and Ling(2005)]%
        {ROCAUC}
\bibfield{author}{\bibinfo{person}{Jin Huang} {and} \bibinfo{person}{Charles~X
  Ling}.} \bibinfo{year}{2005}\natexlab{}.
\newblock \showarticletitle{Using AUC and accuracy in evaluating learning
  algorithms}.
\newblock \bibinfo{journal}{\emph{IEEE Transactions on knowledge and Data
  Engineering}} \bibinfo{volume}{17}, \bibinfo{number}{3}
  (\bibinfo{year}{2005}), \bibinfo{pages}{299--310}.
\newblock


\bibitem[Jiang et~al\mbox{.}(2021)]%
        {jiang2021layercam}
\bibfield{author}{\bibinfo{person}{Peng-Tao Jiang}, \bibinfo{person}{Chang-Bin
  Zhang}, \bibinfo{person}{Qibin Hou}, \bibinfo{person}{Ming-Ming Cheng}, {and}
  \bibinfo{person}{Yunchao Wei}.} \bibinfo{year}{2021}\natexlab{}.
\newblock \showarticletitle{Layercam: Exploring hierarchical class activation
  maps for localization}.
\newblock \bibinfo{journal}{\emph{IEEE Transactions on Image Processing}}
  \bibinfo{volume}{30} (\bibinfo{year}{2021}), \bibinfo{pages}{5875--5888}.
\newblock


\bibitem[Kingma and Ba(2014)]%
        {adam}
\bibfield{author}{\bibinfo{person}{Diederik~P Kingma} {and}
  \bibinfo{person}{Jimmy Ba}.} \bibinfo{year}{2014}\natexlab{}.
\newblock \showarticletitle{Adam: A method for stochastic optimization}.
\newblock \bibinfo{journal}{\emph{arXiv preprint arXiv:1412.6980}}
  (\bibinfo{year}{2014}).
\newblock


\bibitem[Kocher et~al\mbox{.}(1999)]%
        {kocher1999differential}
\bibfield{author}{\bibinfo{person}{Paul Kocher}, \bibinfo{person}{Joshua
  Jaffe}, {and} \bibinfo{person}{Benjamin Jun}.}
  \bibinfo{year}{1999}\natexlab{}.
\newblock \showarticletitle{Differential power analysis}. In
  \bibinfo{booktitle}{\emph{Annual Int. Cryptology Conf.}} Springer,
  \bibinfo{pages}{388--397}.
\newblock


\bibitem[Kurakin et~al\mbox{.}(2016)]%
        {kurakin2016adversarial}
\bibfield{author}{\bibinfo{person}{Alexey Kurakin}, \bibinfo{person}{Ian
  Goodfellow}, \bibinfo{person}{Samy Bengio}, {et~al\mbox{.}}}
  \bibinfo{year}{2016}\natexlab{}.
\newblock \bibinfo{title}{Adversarial examples in the physical world}.
\newblock
\newblock


\bibitem[LeCroy(2022)]%
        {LeCroy2022Jan}
\bibfield{author}{\bibinfo{person}{Teledyne LeCroy}.}
  \bibinfo{year}{2022}\natexlab{}.
\newblock
\newblock
\urldef\tempurl%
\url{http://cdn.teledynelecroy.com/files/pdf/waverunner_6_zi_datasheet.pdf}
\showURL{%
\tempurl}
\newblock
\shownote{[Online; accessed 23. Jan. 2022]}.


\bibitem[Li and Li(2017)]%
        {li2017adversarial}
\bibfield{author}{\bibinfo{person}{Xin Li} {and} \bibinfo{person}{Fuxin Li}.}
  \bibinfo{year}{2017}\natexlab{}.
\newblock \showarticletitle{Adversarial examples detection in deep networks
  with convolutional filter statistics}. In \bibinfo{booktitle}{\emph{Proc.
  IEEE Int. Conf. on Computer Vision}}. \bibinfo{pages}{5764--5772}.
\newblock


\bibitem[Lipp et~al\mbox{.}(2021)]%
        {Lipp2021Platypus}
\bibfield{author}{\bibinfo{person}{Moritz Lipp}, \bibinfo{person}{Andreas
  Kogler}, \bibinfo{person}{David Oswald}, \bibinfo{person}{Michael Schwarz},
  \bibinfo{person}{Catherine Easdon}, \bibinfo{person}{Claudio Canella}, {and}
  \bibinfo{person}{Daniel Gruss}.} \bibinfo{year}{2021}\natexlab{}.
\newblock \showarticletitle{{PLATYPUS: Software-based Power Side-Channel
  Attacks on x86}}. In \bibinfo{booktitle}{\emph{2021 IEEE Symposium on
  Security and Privacy (SP)}}. IEEE.
\newblock


\bibitem[Liu et~al\mbox{.}(2018)]%
        {liu2018towards}
\bibfield{author}{\bibinfo{person}{Xuanqing Liu}, \bibinfo{person}{Minhao
  Cheng}, \bibinfo{person}{Huan Zhang}, {and} \bibinfo{person}{Cho-Jui Hsieh}.}
  \bibinfo{year}{2018}\natexlab{}.
\newblock \showarticletitle{Towards robust neural networks via random
  self-ensemble}. In \bibinfo{booktitle}{\emph{Proceedings of the European
  Conference on Computer Vision (ECCV)}}. \bibinfo{pages}{369--385}.
\newblock


\bibitem[Ma and Liu(2019)]%
        {ma2019nic}
\bibfield{author}{\bibinfo{person}{Shiqing Ma} {and} \bibinfo{person}{Yingqi
  Liu}.} \bibinfo{year}{2019}\natexlab{}.
\newblock \showarticletitle{Nic: Detecting adversarial samples with neural
  network invariant checking}. In \bibinfo{booktitle}{\emph{Proc. Network \&
  Distributed System Security Symposium (NDSS 2019)}}.
\newblock


\bibitem[Ma et~al\mbox{.}(2018)]%
        {ma2018characterizing}
\bibfield{author}{\bibinfo{person}{Xingjun Ma}, \bibinfo{person}{Bo Li},
  \bibinfo{person}{Yisen Wang}, \bibinfo{person}{Sarah~M Erfani},
  \bibinfo{person}{Sudanthi Wijewickrema}, \bibinfo{person}{Grant Schoenebeck},
  \bibinfo{person}{Dawn Song}, \bibinfo{person}{Michael~E Houle}, {and}
  \bibinfo{person}{James Bailey}.} \bibinfo{year}{2018}\natexlab{}.
\newblock \showarticletitle{Characterizing adversarial subspaces using local
  intrinsic dimensionality}.
\newblock \bibinfo{journal}{\emph{arXiv preprint arXiv:1801.02613}}
  (\bibinfo{year}{2018}).
\newblock


\bibitem[Madry et~al\mbox{.}(2017)]%
        {madry2017towards}
\bibfield{author}{\bibinfo{person}{Aleksander Madry},
  \bibinfo{person}{Aleksandar Makelov}, \bibinfo{person}{Ludwig Schmidt},
  \bibinfo{person}{Dimitris Tsipras}, {and} \bibinfo{person}{Adrian Vladu}.}
  \bibinfo{year}{2017}\natexlab{}.
\newblock \showarticletitle{Towards deep learning models resistant to
  adversarial attacks}.
\newblock \bibinfo{journal}{\emph{arXiv preprint arXiv:1706.06083}}
  (\bibinfo{year}{2017}).
\newblock


\bibitem[Manning and Schutze(1999)]%
        {manning1999foundations}
\bibfield{author}{\bibinfo{person}{Christopher Manning} {and}
  \bibinfo{person}{Hinrich Schutze}.} \bibinfo{year}{1999}\natexlab{}.
\newblock \bibinfo{booktitle}{\emph{Foundations of statistical natural language
  processing}}.
\newblock \bibinfo{publisher}{MIT press}.
\newblock


\bibitem[Meng and Chen(2017)]%
        {meng2017magnet}
\bibfield{author}{\bibinfo{person}{Dongyu Meng} {and} \bibinfo{person}{Hao
  Chen}.} \bibinfo{year}{2017}\natexlab{}.
\newblock \showarticletitle{Magnet: a two-pronged defense against adversarial
  examples}. In \bibinfo{booktitle}{\emph{ACM SIGSAC Conf. on Computer \&
  Communications Security}}. \bibinfo{pages}{135--147}.
\newblock


\bibitem[Moosavi-Dezfooli et~al\mbox{.}(2016)]%
        {moosavi2016deepfool}
\bibfield{author}{\bibinfo{person}{Seyed-Mohsen Moosavi-Dezfooli},
  \bibinfo{person}{Alhussein Fawzi}, {and} \bibinfo{person}{Pascal Frossard}.}
  \bibinfo{year}{2016}\natexlab{}.
\newblock \showarticletitle{Deepfool: a simple and accurate method to fool deep
  neural networks}. In \bibinfo{booktitle}{\emph{Proc. IEEE Conf. Computer
  Vision \& Pattern Recognition}}. \bibinfo{pages}{2574--2582}.
\newblock


\bibitem[Parkhi et~al\mbox{.}(2015)]%
        {parkhi2015deep}
\bibfield{author}{\bibinfo{person}{Omkar~M Parkhi}, \bibinfo{person}{Andrea
  Vedaldi}, {and} \bibinfo{person}{Andrew Zisserman}.}
  \bibinfo{year}{2015}\natexlab{}.
\newblock \bibinfo{booktitle}{\emph{Deep face recognition}}.
\newblock \bibinfo{publisher}{British Machine Vision Association}.
\newblock


\bibitem[Rauber et~al\mbox{.}(2017)]%
        {rauber2017foolbox}
\bibfield{author}{\bibinfo{person}{Jonas Rauber}, \bibinfo{person}{Wieland
  Brendel}, {and} \bibinfo{person}{Matthias Bethge}.}
  \bibinfo{year}{2017}\natexlab{}.
\newblock \showarticletitle{Foolbox: A Python toolbox to benchmark the
  robustness of machine learning models}. In \bibinfo{booktitle}{\emph{Workshop
  on Reliable Machine Learning in the Wild}}.
\newblock
\urldef\tempurl%
\url{http://arxiv.org/abs/1707.04131}
\showURL{%
\tempurl}


\bibitem[Redmon et~al\mbox{.}(2016)]%
        {redmon2016look}
\bibfield{author}{\bibinfo{person}{Joseph Redmon}, \bibinfo{person}{Santosh
  Divvala}, \bibinfo{person}{Ross Girshick}, {and} \bibinfo{person}{Ali
  Farhadi}.} \bibinfo{year}{2016}\natexlab{}.
\newblock \showarticletitle{You Only Look Once: Unified, Real-Time Object
  Detection}.
\newblock \bibinfo{journal}{\emph{arXiv preprint arXiv:1506.02640}}
  (\bibinfo{year}{2016}).
\newblock


\bibitem[Richens et~al\mbox{.}(2020)]%
        {richens2020improving}
\bibfield{author}{\bibinfo{person}{Jonathan~G Richens},
  \bibinfo{person}{Ciarn~M Lee}, {and} \bibinfo{person}{Saurabh Johri}.}
  \bibinfo{year}{2020}\natexlab{}.
\newblock \showarticletitle{Improving the accuracy of medical diagnosis with
  causal machine learning}.
\newblock \bibinfo{journal}{\emph{Nature communications}} \bibinfo{volume}{11},
  \bibinfo{number}{1} (\bibinfo{year}{2020}), \bibinfo{pages}{1--9}.
\newblock


\bibitem[Selvaraju et~al\mbox{.}(2016)]%
        {selvaraju2016grad}
\bibfield{author}{\bibinfo{person}{Ramprasaath~R Selvaraju},
  \bibinfo{person}{Abhishek Das}, \bibinfo{person}{Ramakrishna Vedantam},
  \bibinfo{person}{Michael Cogswell}, \bibinfo{person}{Devi Parikh}, {and}
  \bibinfo{person}{Dhruv Batra}.} \bibinfo{year}{2016}\natexlab{}.
\newblock \showarticletitle{Grad-CAM: Why did you say that?}
\newblock \bibinfo{journal}{\emph{arXiv preprint arXiv:1611.07450}}
  (\bibinfo{year}{2016}).
\newblock


\bibitem[Shokri et~al\mbox{.}(2017)]%
        {shokri2017membership}
\bibfield{author}{\bibinfo{person}{Reza Shokri}, \bibinfo{person}{Marco
  Stronati}, \bibinfo{person}{Congzheng Song}, {and} \bibinfo{person}{Vitaly
  Shmatikov}.} \bibinfo{year}{2017}\natexlab{}.
\newblock \showarticletitle{Membership inference attacks against machine
  learning models}. In \bibinfo{booktitle}{\emph{2017 IEEE Symposium on
  Security and Privacy (SP)}}. IEEE, \bibinfo{pages}{3--18}.
\newblock


\bibitem[Simonyan and Zisserman(2014)]%
        {vgg}
\bibfield{author}{\bibinfo{person}{Karen Simonyan} {and}
  \bibinfo{person}{Andrew Zisserman}.} \bibinfo{year}{2014}\natexlab{}.
\newblock \showarticletitle{Very deep convolutional networks for large-scale
  image recognition}.
\newblock \bibinfo{journal}{\emph{arXiv preprint arXiv:1409.1556}}
  (\bibinfo{year}{2014}).
\newblock


\bibitem[Simonyan and Zisserman(2015)]%
        {simonyan2015deep}
\bibfield{author}{\bibinfo{person}{Karen Simonyan} {and}
  \bibinfo{person}{Andrew Zisserman}.} \bibinfo{year}{2015}\natexlab{}.
\newblock \bibinfo{title}{Very Deep Convolutional Networks for Large-Scale
  Image Recognition}.
\newblock
\newblock
\showeprint[arxiv]{1409.1556}~[cs.CV]


\bibitem[Srivastava et~al\mbox{.}(2014)]%
        {srivastava2014dropout}
\bibfield{author}{\bibinfo{person}{Nitish Srivastava},
  \bibinfo{person}{Geoffrey Hinton}, \bibinfo{person}{Alex Krizhevsky},
  \bibinfo{person}{Ilya Sutskever}, {and} \bibinfo{person}{Ruslan
  Salakhutdinov}.} \bibinfo{year}{2014}\natexlab{}.
\newblock \showarticletitle{Dropout: a simple way to prevent neural networks
  from overfitting}.
\newblock \bibinfo{journal}{\emph{The journal of machine learning research}}
  \bibinfo{volume}{15}, \bibinfo{number}{1} (\bibinfo{year}{2014}),
  \bibinfo{pages}{1929--1958}.
\newblock


\bibitem[Szegedy et~al\mbox{.}(2014)]%
        {szegedy2014going}
\bibfield{author}{\bibinfo{person}{Christian Szegedy}, \bibinfo{person}{Wei
  Liu}, \bibinfo{person}{Yangqing Jia}, \bibinfo{person}{Pierre Sermanet},
  \bibinfo{person}{Scott Reed}, \bibinfo{person}{Dragomir Anguelov},
  \bibinfo{person}{Dumitru Erhan}, \bibinfo{person}{Vincent Vanhoucke}, {and}
  \bibinfo{person}{Andrew Rabinovich}.} \bibinfo{year}{2014}\natexlab{}.
\newblock \showarticletitle{Going Deeper with Convolutions}.
\newblock \bibinfo{journal}{\emph{arXiv preprint arXiv:1409.4842}}
  (\bibinfo{year}{2014}).
\newblock


\bibitem[Tao et~al\mbox{.}(2018)]%
        {tao2018attacks}
\bibfield{author}{\bibinfo{person}{Guanhong Tao}, \bibinfo{person}{Shiqing Ma},
  \bibinfo{person}{Yingqi Liu}, {and} \bibinfo{person}{Xiangyu Zhang}.}
  \bibinfo{year}{2018}\natexlab{}.
\newblock \showarticletitle{Attacks meet interpretability: Attribute-steered
  detection of adversarial samples}.
\newblock \bibinfo{journal}{\emph{arXiv preprint arXiv:1810.11580}}
  (\bibinfo{year}{2018}).
\newblock


\bibitem[Tramer et~al\mbox{.}(2020)]%
        {adaptive}
\bibfield{author}{\bibinfo{person}{Florian Tramer}, \bibinfo{person}{Nicholas
  Carlini}, \bibinfo{person}{Wieland Brendel}, {and}
  \bibinfo{person}{Aleksander Madry}.} \bibinfo{year}{2020}\natexlab{}.
\newblock \showarticletitle{On adaptive attacks to adversarial example
  defenses}.
\newblock \bibinfo{journal}{\emph{Advances in Neural Information Processing
  Systems}}  \bibinfo{volume}{33} (\bibinfo{year}{2020}),
  \bibinfo{pages}{1633--1645}.
\newblock


\bibitem[Van~der Maaten and Hinton(2008)]%
        {van2008visualizing}
\bibfield{author}{\bibinfo{person}{Laurens Van~der Maaten} {and}
  \bibinfo{person}{Geoffrey Hinton}.} \bibinfo{year}{2008}\natexlab{}.
\newblock \showarticletitle{Visualizing data using t-SNE.}
\newblock \bibinfo{journal}{\emph{Journal of machine learning research}}
  \bibinfo{volume}{9}, \bibinfo{number}{11} (\bibinfo{year}{2008}).
\newblock


\bibitem[Vinogradova et~al\mbox{.}(2020)]%
        {vinogradova2020towards}
\bibfield{author}{\bibinfo{person}{Kira Vinogradova}, \bibinfo{person}{Alexandr
  Dibrov}, {and} \bibinfo{person}{Gene Myers}.}
  \bibinfo{year}{2020}\natexlab{}.
\newblock \showarticletitle{Towards interpretable semantic segmentation via
  gradient-weighted class activation mapping (student abstract)}. In
  \bibinfo{booktitle}{\emph{Proceedings of the AAAI conference on artificial
  intelligence}}, Vol.~\bibinfo{volume}{34}. \bibinfo{pages}{13943--13944}.
\newblock


\bibitem[Wang et~al\mbox{.}(2018)]%
        {wang2018defensive}
\bibfield{author}{\bibinfo{person}{Siyue Wang}, \bibinfo{person}{Xiao Wang},
  \bibinfo{person}{Pu Zhao}, \bibinfo{person}{Wujie Wen},
  \bibinfo{person}{David Kaeli}, \bibinfo{person}{Peter Chin}, {and}
  \bibinfo{person}{Xue Lin}.} \bibinfo{year}{2018}\natexlab{}.
\newblock \showarticletitle{Defensive dropout for hardening deep neural
  networks under adversarial attacks}. In \bibinfo{booktitle}{\emph{Proceedings
  of the International Conference on Computer-Aided Design}}.
  \bibinfo{pages}{1--8}.
\newblock


\bibitem[Wang et~al\mbox{.}(2019b)]%
        {wang2019protecting}
\bibfield{author}{\bibinfo{person}{Xiao Wang}, \bibinfo{person}{Siyue Wang},
  \bibinfo{person}{Pin-Yu Chen}, \bibinfo{person}{Yanzhi Wang},
  \bibinfo{person}{Brian Kulis}, \bibinfo{person}{Xue Lin}, {and}
  \bibinfo{person}{Sang~Peter Chin}.} \bibinfo{year}{2019}\natexlab{b}.
\newblock \showarticletitle{Protecting Neural Networks with Hierarchical Random
  Switching: Towards Better Robustness-Accuracy Trade-off for Stochastic
  Defenses}. In \bibinfo{booktitle}{\emph{IJCAI}}.
\newblock


\bibitem[Wang et~al\mbox{.}(2019a)]%
        {wang2019beyond}
\bibfield{author}{\bibinfo{person}{Zhibo Wang}, \bibinfo{person}{Mengkai Song},
  \bibinfo{person}{Zhifei Zhang}, \bibinfo{person}{Yang Song},
  \bibinfo{person}{Qian Wang}, {and} \bibinfo{person}{Hairong Qi}.}
  \bibinfo{year}{2019}\natexlab{a}.
\newblock \showarticletitle{Beyond inferring class representatives: User-level
  privacy leakage from federated learning}. In \bibinfo{booktitle}{\emph{IEEE
  Conf. on Computer Communications}}. IEEE, \bibinfo{pages}{2512--2520}.
\newblock


\bibitem[Xu et~al\mbox{.}(2017)]%
        {xu2017feature}
\bibfield{author}{\bibinfo{person}{Weilin Xu}, \bibinfo{person}{David Evans},
  {and} \bibinfo{person}{Yanjun Qi}.} \bibinfo{year}{2017}\natexlab{}.
\newblock \showarticletitle{Feature squeezing: Detecting adversarial examples
  in deep neural networks}.
\newblock \bibinfo{journal}{\emph{arXiv preprint arXiv:1704.01155}}
  (\bibinfo{year}{2017}).
\newblock


\bibitem[Yang et~al\mbox{.}(2022)]%
        {yang2022you}
\bibfield{author}{\bibinfo{person}{Yijun Yang}, \bibinfo{person}{Ruiyuan Gao},
  \bibinfo{person}{Yu Li}, \bibinfo{person}{Qiuxia Lai}, {and}
  \bibinfo{person}{Qiang Xu}.} \bibinfo{year}{2022}\natexlab{}.
\newblock \showarticletitle{What You See is Not What the Network Infers:
  Detecting Adversarial Examples Based on Semantic Contradiction}.
\newblock \bibinfo{journal}{\emph{arXiv preprint arXiv:2201.09650}}
  (\bibinfo{year}{2022}).
\newblock


\bibitem[Yu et~al\mbox{.}(2020)]%
        {yu2020deepem}
\bibfield{author}{\bibinfo{person}{Honggang Yu}, \bibinfo{person}{Haocheng Ma},
  \bibinfo{person}{Kaichen Yang}, \bibinfo{person}{Yiqiang Zhao}, {and}
  \bibinfo{person}{Yier Jin}.} \bibinfo{year}{2020}\natexlab{}.
\newblock \showarticletitle{{DeepEM}: Deep Neural Networks Model Recovery
  through EM Side-Channel Information Leakage}. In
  \bibinfo{booktitle}{\emph{IEEE Int. Symp. on Hardware Oriented Security \&
  Trust (HOST)}}. IEEE, \bibinfo{pages}{209--218}.
\newblock


\bibitem[Zhang et~al\mbox{.}(2021)]%
        {zhang2021stealing}
\bibfield{author}{\bibinfo{person}{Yicheng Zhang}, \bibinfo{person}{Rozhin
  Yasaei}, \bibinfo{person}{Hao Chen}, \bibinfo{person}{Zhou Li}, {and}
  \bibinfo{person}{Mohammad~Abdullah Al~Faruque}.}
  \bibinfo{year}{2021}\natexlab{}.
\newblock \showarticletitle{Stealing neural network structure through remote
  fpga side-channel analysis}.
\newblock \bibinfo{journal}{\emph{IEEE Trans. on Information Forensics \&
  Security}}  \bibinfo{volume}{16} (\bibinfo{year}{2021}),
  \bibinfo{pages}{4377--4388}.
\newblock


\bibitem[Zhou et~al\mbox{.}(2021)]%
        {zhou2021removing}
\bibfield{author}{\bibinfo{person}{Dawei Zhou}, \bibinfo{person}{Nannan Wang},
  \bibinfo{person}{Chunlei Peng}, \bibinfo{person}{Xinbo Gao},
  \bibinfo{person}{Xiaoyu Wang}, \bibinfo{person}{Jun Yu}, {and}
  \bibinfo{person}{Tongliang Liu}.} \bibinfo{year}{2021}\natexlab{}.
\newblock \showarticletitle{Removing adversarial noise in class activation
  feature space}. In \bibinfo{booktitle}{\emph{Proceedings of the IEEE/CVF
  International Conference on Computer Vision}}. \bibinfo{pages}{7878--7887}.
\newblock


\bibitem[Zhu et~al\mbox{.}(2020)]%
        {9069951}
\bibfield{author}{\bibinfo{person}{Jiang Zhu}, \bibinfo{person}{Lizan Wang},
  \bibinfo{person}{Haolin Liu}, \bibinfo{person}{Shujuan Tian},
  \bibinfo{person}{Qingyong Deng}, {and} \bibinfo{person}{Jianqi Li}.}
  \bibinfo{year}{2020}\natexlab{}.
\newblock \showarticletitle{An Efficient Task Assignment Framework to
  Accelerate DPU-Based Convolutional Neural Network Inference on FPGAs}.
\newblock \bibinfo{journal}{\emph{IEEE Access}}  \bibinfo{volume}{8}
  (\bibinfo{year}{2020}), \bibinfo{pages}{83224--83237}.
\newblock
\urldef\tempurl%
\url{https://doi.org/10.1109/ACCESS.2020.2988311}
\showDOI{\tempurl}


\end{thebibliography}
\end{document}